\documentclass{article}

\usepackage{graphicx, amsfonts}
\newcommand{\be}{\begin{equation}}
\newcommand{\ee}{\end{equation}}
\newcommand{\bea}{\begin{eqnarray}}
\newcommand{\eea}{\end{eqnarray}}
\newcommand{\bean}{\begin{eqnarray*}}
\newcommand{\eean}{\end{eqnarray*}}

\title{\bf Spacetime geometry from algebra: \\ spin foam
models \\ for non-perturbative quantum gravity}
\author{\vspace{1cm} \\ {\bf Daniele Oriti} \\ \vspace{1cm} \\ Department of Applied Mathematics and
Theoretical Physics, \\ Centre for Mathematical Sciences, University of
Cambridge, \\ Wilberforce Road, Cambridge CB3 0WA, UK \\ and \\ Girton
College, Cambridge CB3 0JG, UK \\ d.oriti@damtp.cam.ac.uk}

\begin{document}
\maketitle
\begin{abstract}
This is an introduction to spin foam models for non-perturbative quantum gravity, an approach that lies at the point of convergence of many different research areas, including loop quantum gravity, topological quantum field theories, path integral quantum gravity, lattice gauge theory, matrix models, category theory, statistical mechanics. We describe the general formalism and ideas of spin foam models, the picture of quantum geometry emerging from them, and give a review of the results obtained so far, in both the Euclidean and Lorentzian case. We focus in particular on the Barrett-Crane model for 4-dimensional quantum gravity.      
\end{abstract}
\newpage
\tableofcontents
\newpage
\section{Introduction}
The construction of a quantum theory of gravity remains probably {\it
the} issue left unsolved in the last century, in spite of a lot of
efforts and many important results obtained during an indeed long
history of works (for an account of this history see \cite{Rov0}). The
problem of a complete formulation of quantum gravity is still quite
far from being solved (for reviews of the present situation see
\cite{Hor,Rov00,Carlip}), but we can say that the last (roughly) fifteen years
have seen a considerable number of developments, principally
identifiable with the construction of superstring theory \cite{GSW,Polch}, in its
perturbative formulation and, more recently, with the understanding of
some of its non-perturbative features, with the birth of non-commutative geometry \cite{Connes, Majid, Majid2}, and with the renaissance of the
canonical quantization program in the form of loop quantum gravity \cite{Ash, RovSmo2},
based on the reformulation of classical general relativity in terms of
the Ashtekar variables \cite{Ash0}. In recent years many different approaches on
the non-perturbative and background-independent side have
been converging to the formalism of the so-called spin foams
\cite{Baez,Baez2}. 
 
We give here an introduction to and a
review of 
this new line of research, the basics of the spin foam
formalism, and focus in particular on the Barrett-Crane model for Euclidean
4-dimensional quantum gravity \cite{BC}, about which more results are available, and show what picture of quantum spacetime geometry
emerges out of it. We will concentrate on the aspects of the
subject directly related to quantum gravity, so many interesting
topics, like for example the spin foam models for discrete topological
field theories or lattice gauge theories will not be covered. Moreover
we will confine ourselves mainly to the 4-dimensional case, and to the
Euclidean signature. Something about Lorentzian spin foam models will
be said in section ~\ref{sec:lor}.

The presentation is organized as follows: we first discuss briefly spin
networks, and the role they play in quantum gravity, and in loop
quantum gravity in particular; then in section ~\ref{sec:spfo} we
introduce the basic formalism and ideas of spin foams and spin
foam models; the Barrett-Crane model is
described in detail in section ~\ref{sec:BC}, where we review its
construction and derivation, the picture of quantum geometry emerging
from it, and its relationship with gravity at the classical and
semiclassical level; in section ~\ref{sec:matrix} we show how the
Barrett-Crane model can be obtained also from a generalized matrix model in 4 dimensions,
i.e. from a field theory over a group manifold, a recent development
which led to considerable progress and insight; in section
~\ref{sec:lor} we discuss the Lorentzian version of the Barrett-Crane
model; we end
up with a discussion of the present status and prospectives of the subject.

\subsection{Many roads lead to spin foams}
Spin foam models for quantum gravity are obtained by translating the
geometric information about a (usually triangulated) 
manifold into the language of combinatorics and group theory, so that
the usual concepts of a 
metric and of metric properties are somehow emerging from them, and not regarded as 
fundamental. In fact we can identify the major conceptual contribution
coming from this new approach in the idea that a truly non-perturbative and background independent theory of quantum gravity has to be formulated in a purely quantum and algebraic way, and that all the structures we commonly use to deal with gravity and spacetime, smooth manifolds, metrics, space and time themselves, are to be considered useful only as semiclassical, macroscopic, approximate concepts. Moreover, as we will see, a spin foam
model implements in a precise way
the idea of a sum over geometries, so that we can envisage here a renaissance of the covariant and path integral program for
the quantization of gravity \cite{Haw}, but now we are 
summing over labelled 2-complexes (spin foams), i.e. collections of faces, edges and vertices combined together 
and labeled by representations of a group (or a quantum group). The
spacetime geometry comes out only of these elements, i.e. out of the
lands of algebra and combinatorics (and piecewise linear
topology). 

Notably, many roads led to spin foams. A spin foam picture emerges when considering the evolution
 in time of spin networks \cite{ReiRov,Rov4}, which were discovered to represent states of 
quantum general relativity at the kinematical level
\cite{RovSmolin,Baez3,Ash1,Ash3,Ash4}, so from the canonical side.
 Spin foam models were developed also for topological field
theories in different dimensions, including 3d quantum gravity \cite{PonzReg,T-V,CrYet,CrKauYet,BaWe},
and this represents a completely different line of research coming to
 the same formalism. In these models category theory plays a major
 role, since their whole construction can be rephrased in terms 
 of operations in the category of a Lie group (or quantum group),
 making the algebraic nature of the models manifest. A spin foam model
 can also be seen, as we will discuss in the following, as a state sum
 of the type used in statistical mechanics, giving another bridge
 between different areas. 
 A spin foam formulation can also be given
 \cite{Reis1,OePf} for lattice Yang-Mills theory; here the partition function of the spin foam model is equal to that of the lattice gauge theory and can be used to calculate the strong coupling expansion. Finally, and most important for our present
 interest, pioneered by the works of Reisenberger \cite{Reis1}, many different spin foam models have been developed for gravity 
\cite{BC,Reis1,Reis2, Reis3, Iwa1,Iwa2,BC2}, opening a new path towards a
 formulation of a non-perturbative quantum gravity theory; some of
 these models make use of methods and ideas from category theory as well \cite{CraneAlg}, and moreover can be derived using a generalization of the matrix
 models developed for 2d gravity and string theory. 

Thus, we can say that, strikingly, spin foam models lie at the point
of convergence of research areas as different as canonical quantum
gravity, topological quantum field theories, category theory, statistical mechanics
 and field theory, lattice gauge theory, path integral formulation of
quantum gravity, matrix models, to name but a few, with consequent and remarkable
cross-fertilization of ideas and techniques.

\subsection{The covariant approach to quantum gravity: summing over
geometries}
Since our main interest in this work is quantum gravity,
and spin foam models can be naturally interpreted, as we will see, as
a new form of the path integral approach to quantum gravity, the
covariant approach, we would like to give here the basic ideas of this
line of research, referring to the literature \cite{Misn, Haw, HarHaw}
for more details. 

The key idea is to represent the transition amplitude between two
gravity states as a Feynman path integral over the histories
interpolating between them. The states are given by 3-geometries
relative to 3d hypersurfaces, not necessarily connected, and the histories are 4-geometries
inducing the given 3-geometries on the hypersurfaces (see Fig.1). The weight for
each history, in the Lorentzian case, is the exponential of $i$ times the classical action, the
Einstein-Hilbert action (modulo boundary terms), as usual.

Thus,
\be
\langle h_{2}, S_{2}\mid h_{1},S_{1}\rangle\,=\,\int_{g/
g(S_{1})=h_{1}, g(S_{2})=h_{2}}\mathcal{D}g\,\exp{(i\,I_{EH}(g))}
\ee 
where, as we said, we have to specify only the given 3-geometries,
meaning only the equivalence classe of 3-metric on the boundary
$\partial M$ of the 4-manifold $M$, under diffeomorphisms mapping the
boundary into itself.

\begin{center}
\includegraphics[width=6cm]{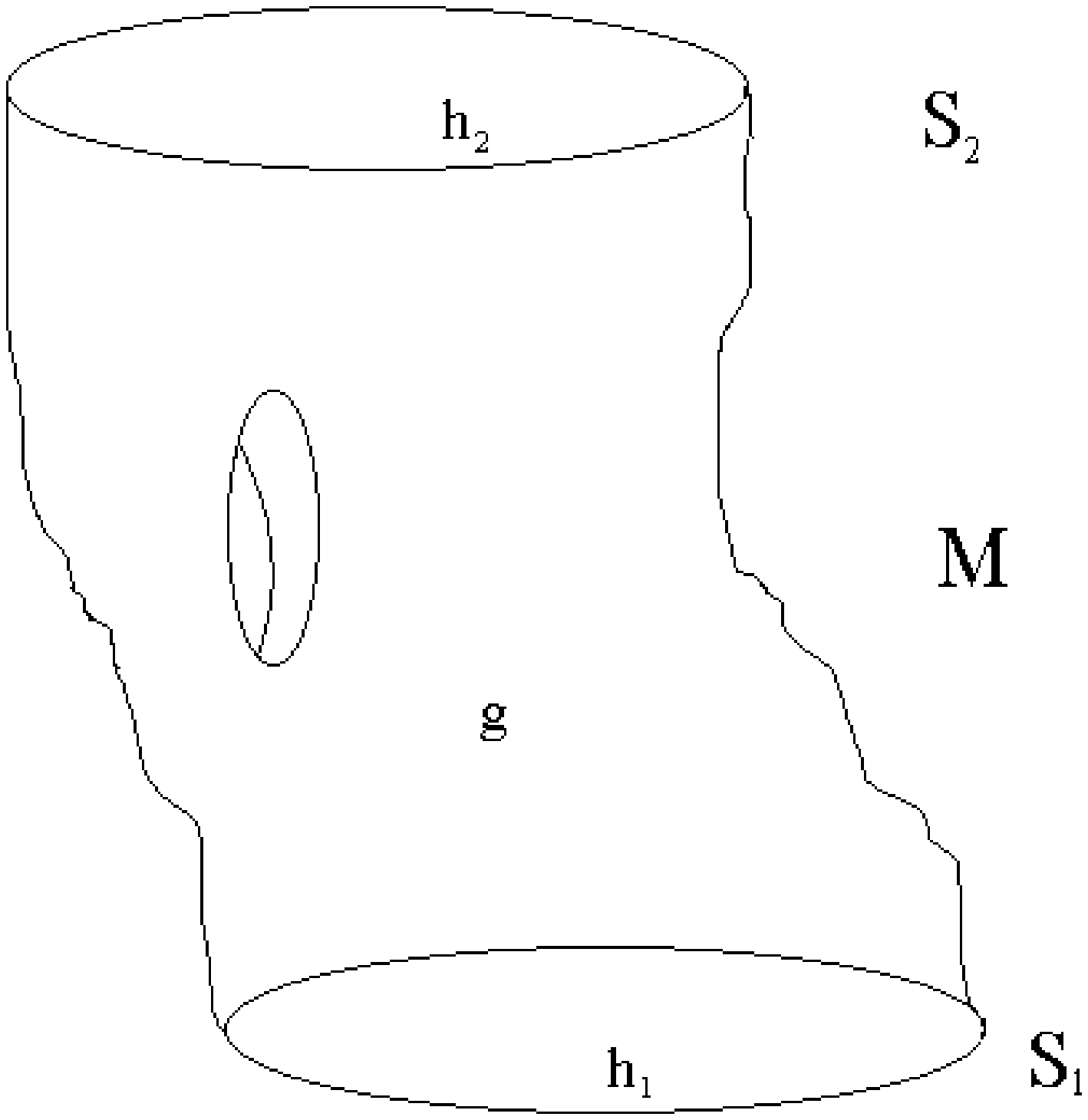}

{\small Fig.1 - Schematic representation of a 4-manifold (of genus one) with two disconnected boundary components}

\end{center}

Of course one then asks for the transition from a 3-geometry $h_{1}$
to another $h_{2}$, and then to a third $h_{3}$ to be independent on
the intermediate state:
\be
\langle h_{3}, S_{3}\mid h_{1},S_{1}\rangle\,=\,\sum_{h_{2}}\langle
h_{3}, S_{3}\mid h_{2},S_{2}\rangle\langle h_{2}, S_{2}\mid
h_{1},S_{1}\rangle
\ee
corresponding to the gluing of two manifolds $M_{1}$, with boundaries
$S_{1}$ and $S_{2}$, and $M_{2}$, with boundaries $S_{2}$ and $S_{3}$,
along the common boundary $S_{2}$.

Also, the state themselves can be defined through a path integral of
the same kind, considering now manifolds with only one boundary:

\be
\mid h_{1},S_{1}\rangle\,=\,\int_{g/
g(S_{1})=h_{1}}\mathcal{D}g\,\exp{(i\,I_{EH}(g))}
\ee  
 
Topology itself is also allowed to change, and the path integral can
be completed by a sum over different topologies, giving rise to a
foam-like structure of spacetime \cite{wheeler}, with quantum
fluctuation from one 4-metric to another for a given topology, but
also from one topology to another.

Instead of discussing the results and problems of this line of
research, for which a consistent literature is available, we now turn
to our actual subject, spin foam models for quantum gravity, to see which form these ideas take in a spin
foam context.

\section{Spin networks and the geometry of space} \label{sec:spnet}
In this section we introduce the concept of spin network, first invented by Penrose \cite{Penr}, show how it
arises from the canonical quantization program \cite{Baez2,RovSmolin,RovGaul,DePieRov}, which role it plays in
that context and how it may be thought of as giving a quantum picture
of the geometry of space (for a complete introduction to canonical
quantum general relativity, see \cite{ThieRev}). This is only one of the ways in which spin
networks turn out to be important for physics, and we refer to the
literature \cite{Baez2,S1} for other points of view. 
\subsection{Spin networks as states of the gravitational field}
We start with an ADM formulation of classical gravity in terms of
local triads $e^{i}_{a}$ (related to the 3-metric by $h_{ab}=e^{i}_{a}e^{i}_{b}$), on a 4-dimensional manifold
$\mathcal{M}$ with topology $\mathbb{R}\times M$, $M$ compact. Thus there is an
additional SU(2) local gauge symmetry (coming from the reduction of
the 4-dimensional Lorentz invariance to the 3-dimensional $M$), given by arbitrary local frame
rotations. We have $\{
E^{a}_{i},K^{i}_{a}\}$ as canonical pair on the phase space of the
theory, where $E^{i}_{a}=ee^{i}_{a}$ ($e$ is the determinant of
$e^{i}_{a}$) and $K_{a}^{i}$ is related to the extrinsic curvature by
$K^{i}_{a}=K_{ab}E^{bi}/\sqrt{h}$ with $h$ the determinant of the 3-metric
$h_{ab}$. Then given the canonical transformation
$A^{i}_{a}(x)=\omega^{i}_{a}+\beta K^{i}_{a}$, with $\omega^{i}_{a}$
being the SU(2) spin connection compatible with the triad, we arrive
at the new canonical pair of variables in the phase space,
$(A^{i}_{a}(x),E^{a}_{i})$. The new configuration variable is now the
$su(2)$-Lie algebra-valued connection 1-form on $M$, $A^{i}_{a}$, and $E^{a}_{i}$ is
the conjugate momentum. 
At the quantum level the canonical variables
will be replaced by operators acting on the state space of the theory
and the full dynamical content of general relativity will be encoded
in the action of the first-class constraints on the physical states:
the SU(2) Gauss constraint, imposing the local gauge invariance on the
states, the diffeomorphism constraint, generating 3-dimensional
diffeomorphisms on $M$, and the Hamiltonian constraint, representing
the evolution of $M$ in the (unphysical) coordinate time. The issue of
quantization is then the realization of the space of the physical
states of the theory (in general, functionals of the connection) with the correct action of the constraints on
them, i.e. they should lie in the kernel of the quantum constraint
operators. Consequently let us focus on these states.
 
Consider a graph $\Gamma_{n}$, given by a collection of n {\it links}
$\gamma_{i}$, piecewise smooth curves embedded in $M$ and meeting only
at their endpoints, called {\it vertices}, if at all. Now we can
assign group elements to each link $\gamma_{i}$ taking the holonomy or parallel transport $g_{i}=\mathcal{P}exp\int_{\gamma_{i}}A$, where the
notation means that the exponential is defined by the path ordered
series, and consequently assigning an element of $SU(2)^{n}$ to the
graph. Given now a complex-valued function $f$ of $SU(2)^{n}$, we define
the state $\Psi_{\Gamma_{n},f}(A)=f(g_{1},...,g_{n})$. The set of
states so defined (called {\it cylindrical functions}) forms a subset of the space of
smooth functions on the space of connections, on which it is possible to
define consistently an inner product \cite{Ba,AshLew}, and then
complete the space of linear combinations of cylindrical functions
in the norm induced by this inner product. In this way we obtain a
Hilbert space of states $\mathcal{H}_{aux}$. This is not the physical
space of states, $\mathcal{H}_{phys}$, 
which is instead given by the subspace of it annihilated by all the
three quantum constraints of the theory. 

An orthonormal basis in
$\mathcal{H}_{aux}$ is constructed in the following way. Consider the
graph $\Gamma_{n}$ and assign {\it irreducible} representations to the
links. Then consider the tensor product of the $k$ Hilbert spaces of
the representations associated to the $k$ links intersecting at a
{\it vertex} $v$ of the graph, fix an orthonormal basis in this space,
and assign to the vertex an element $\iota$ of the basis. The corresponding state is then
defined to be:
\be 
\Psi_{\Gamma_{n}}(A)=\otimes_{i}\rho^{j_{i}}(g_{i})\otimes_{v}\iota_{v},
\ee where the products are over all the links and all the vertices of
the graph, $\rho$ indicates the representation matrix of the group
element $g$ in the irreducible representation $j$, and the indices of
$\rho$ and $\iota$ (seen as a tensor) are suitably contracted. It is
possible to show that the states so defined for all the possible
graphs and all the possible colorings of links and vertices are
orthonormal in the previously defined inner product. Moreover, if we
define $\iota_{v}$ to be an {\it invariant} tensor, i.e. to be in the
singlet subspace in the decomposition of the tensor product of the $k$ Hilbert
spaces into irreducible parts, then the resulting quantum state
$\Psi_{\Gamma}$ is {\it invariant} under SU(2). As a result, the set
of all these states, with all the possible choices of graphs and
colorings gives an orthonormal basis for the subspace of
$\mathcal{H}_{aux}$ which is annihilated by the first of the quantum
constraint operators of the theory, namely the Gauss constraint. The
invariant tensors $\iota_{v}$ are called {\it intertwiners}, they make it
possible to couple the representations assigned to the links
intersecting at the vertex, and are given by the standard recoupling
theory of SU(2) (in this case). The colored graph above is a {\it spin
network} $S$ and defines a quantum state $\mid S\rangle$, represented in
terms of the connection by a functional $\Psi_{S}(A)=\langle A\mid
S\rangle$ of the type just described. 

We have still to consider the
diffeomorphism constraint. But this is (quite) easily taken into
account by considering the equivalence class of embedded spin networks
$S$ under the action of $Diff(M)$, called {\it s-knots}, which are
abstract, non-embedded, colored graphs (spin networks). We have to clarify what we mean by \lq\lq abstract, non-embedded" in this case. Modding out by diffeomorphisms does not leave us without any informations about the relationship between the manifold and our graphs. There is still (homotopy-theoretic) information about how loops in the graph wind around holes in the manifold, for example, or about how the edges intersect each other. What is no more defined is \lq\lq where" the graph sits inside the manifold, its location, and its metric properties (e.g. lenght of the edges,...). Each s-knot defines an element
$\mid s\rangle$
of $\mathcal{H}_{Diff}$, that is the Hilbert space of both gauge
invariant and diffeomorphism invariant states of the gravitational
field, and it can be proven that the states $(1/\sqrt{is(s)})\mid
s\rangle$, where $is(s)$ is the number of isomorphisms of the s-knot
into itself, preserving the coloring and generated by a diffeomorphism
of $M$, form an orthonormal basis for that space. 

Now we can say that we have found which structure gives the quantum
states of the gravitational field, at least at the kinematical level,
since the dynamics is encoded in the action of the Hamiltonian
constraint, that we have not yet considered. They are spin networks,
purely algebraic and combinatorial objects, defined regardless of any
embedding in the following way (here we also generalize the previous
definition to any compact group $G$)

- a spin network is a triple $(\Gamma, \rho, \iota)$ given by: a
1-dimensional oriented (with a suitable orientation of the edges or
links) complex $\Gamma$, a labelling $\rho$ of each edge $e$ of
$\Gamma$ by an
irreducible representation $\rho_{e}$ of $G$, and a labelling $\iota$
of each vertex $v$ of $\Gamma$ by an intertwiner $\iota_{v}$ mapping
(the tensor product of) the irreducible representations of the edges
incoming to $v$ to (the tensor product of) the irreducible
representations of the edges outgoing from $v$. 

A simple example of a spin network is given in Fig.2, where we indicate
also the corresponding state in the connection representation.

\begin{center}

\includegraphics[width=6cm]{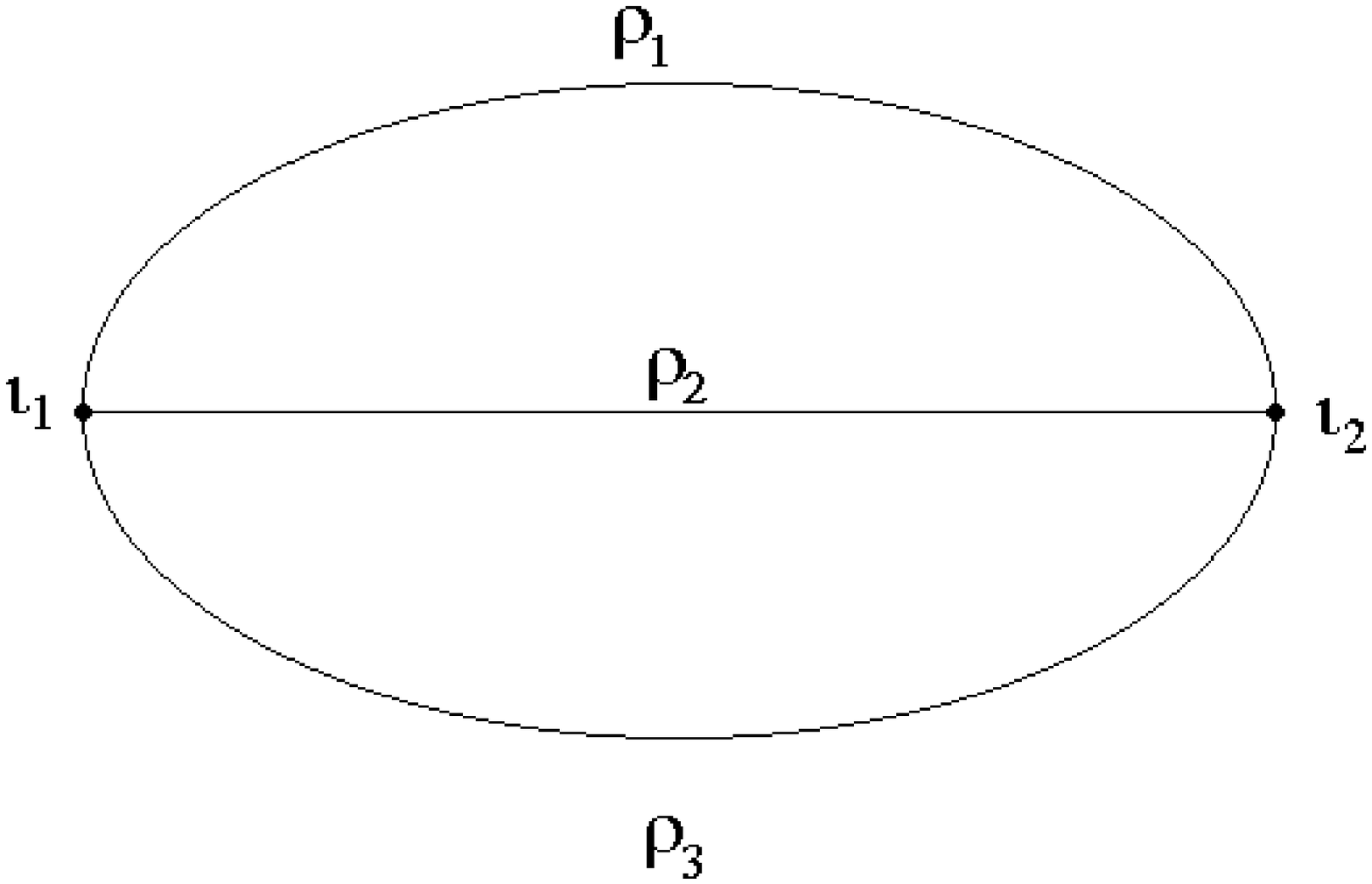}

{\small Fig.2 - A spin network whose corresponding functional is: \\
$\psi_{S}(A)=\rho_{e_{1}}(\mathcal{P}\,e^{\int_{e_{1}}A})^{a}_{b}\,\rho_{e_{2}}(
\mathcal{P}\,e^{\int_{e_{2}}A})^{c}_{d}\,\rho_{e_{3}}(
\mathcal{P}\,e^{\int_{e_{3}}A})^{e}_{f}\,(\iota_{v_{1}})_{ace}(\iota_{v_{2}})^{bdf}$}
\end{center}

These are, as we stress again,  the 3d diffeomorphism invariant
quantum states
of the gravitational field (provided we add to them homotopy-theoretic information of the kind mentioned above), and, accordingly, they do not live anywhere
in the space, but {\it define} \lq\lq the where'' itself (they can be seen
as elementary excitations of the space itself). Moreover,
we will now see that they carry the geometrical information necessary
to construct the geometry of the space in which we {\it decide} to
embed them.

\subsection{Geometry of space from spin networks} \label{sec:3geo}
Having at hand the kinematical states of our quantum theory, we want
now to construct gauge invariant operators acting on them. The
simplest example is the trace of the holonomy around a loop $\alpha$,
$\hat{T}(\alpha)=TrU(A,\alpha)$. This is a multiplicative operator,
whose action on spin network states is simply given by:
\be 
\hat{T}(\alpha)\Psi_{S}(A)\,=\,TrU(A,\alpha)\Psi_{S}(A). \ee
Since our configuration variable is the connection $A$, we then look
for a conjugate momentum operator in the form of a derivative with
respect to it, so replacing the $E^{i}_{a}$ field by the operator
$-i\delta/\delta A$. This is an operator-valued distribution so we
have to suitably smear it in order to have a well-posed operator. It
is convenient to contract it with the Levi-Civita density and to
integrate it over a surface $\Sigma$, with embedding in $M$ given by
$(\sigma^{1},\sigma^{2})\rightarrow x^{a}(\vec{\sigma})$, where
$\vec{\sigma}$ are coordinates on $\Sigma$. We then define the
operator 
\be 
\hat{E}^{i}(\Sigma)=-i\int_{\Sigma}d\vec{\sigma}\,n_{a}(\vec{\sigma})\frac{\delta}{\delta
A^{i}_{a}} \ee
where $n(\sigma)$ is the normal to $\Sigma$. This operator is well
defined but not gauge invariant. We then consider its square given by
$\hat{E}^{i}(\Sigma)\hat{E}^{i}(\Sigma)$, but we discover that its
action on a spin network state $\Psi_{S}(A)$ is gauge invariant only
if the spin network $S$, when embedded in $M$, has only one point of
intersection with the surface $\Sigma$. But now we can take a
partition $p$ of $\Sigma$ in $N(p)$ surfaces $\Sigma_{n}$ so that
$\cup\Sigma_{n}=\Sigma$ and such that each $\Sigma_{n}$ has only one point
of intersection with $S$, if any. Then the operator 
\be
\hat{\mathcal{A}}(\Sigma)\,=\,\lim_{p}\sum_{n}\sqrt{\hat{E}^{i}(\Sigma_{n})\hat{E}^{i}(\Sigma_{n})},
\ee where the limit is an infinite refinement of the partition $p$, is
well-defined, after a proper regularization, independent on the
partition chosen, and gauge invariant. Given a spin network $S$ having
(again, when embedded in $M$)
a finite number of intersections with $\Sigma$ and no nodes lying on
it (only for simplicity, since it is possible to consider any other
more general situation)(see Fig.3), the action of $\hat{\mathcal{A}}$ on the
corresponding state $\Psi_{S}(A)$ is 
\be
\hat{\mathcal{A}}(\Sigma)\Psi_{S}(A)=\sum_{i}\sqrt{j_{i}(j_{i}+1)}\Psi_{S}(A),
\ee
where the sum is over the intersection between $\Sigma$ and $S$ and
$j_{i}$ is the irreducible representation of the edge of $S$
intersecting $\Sigma$. 

\begin{center}
\includegraphics[width=6.5cm]{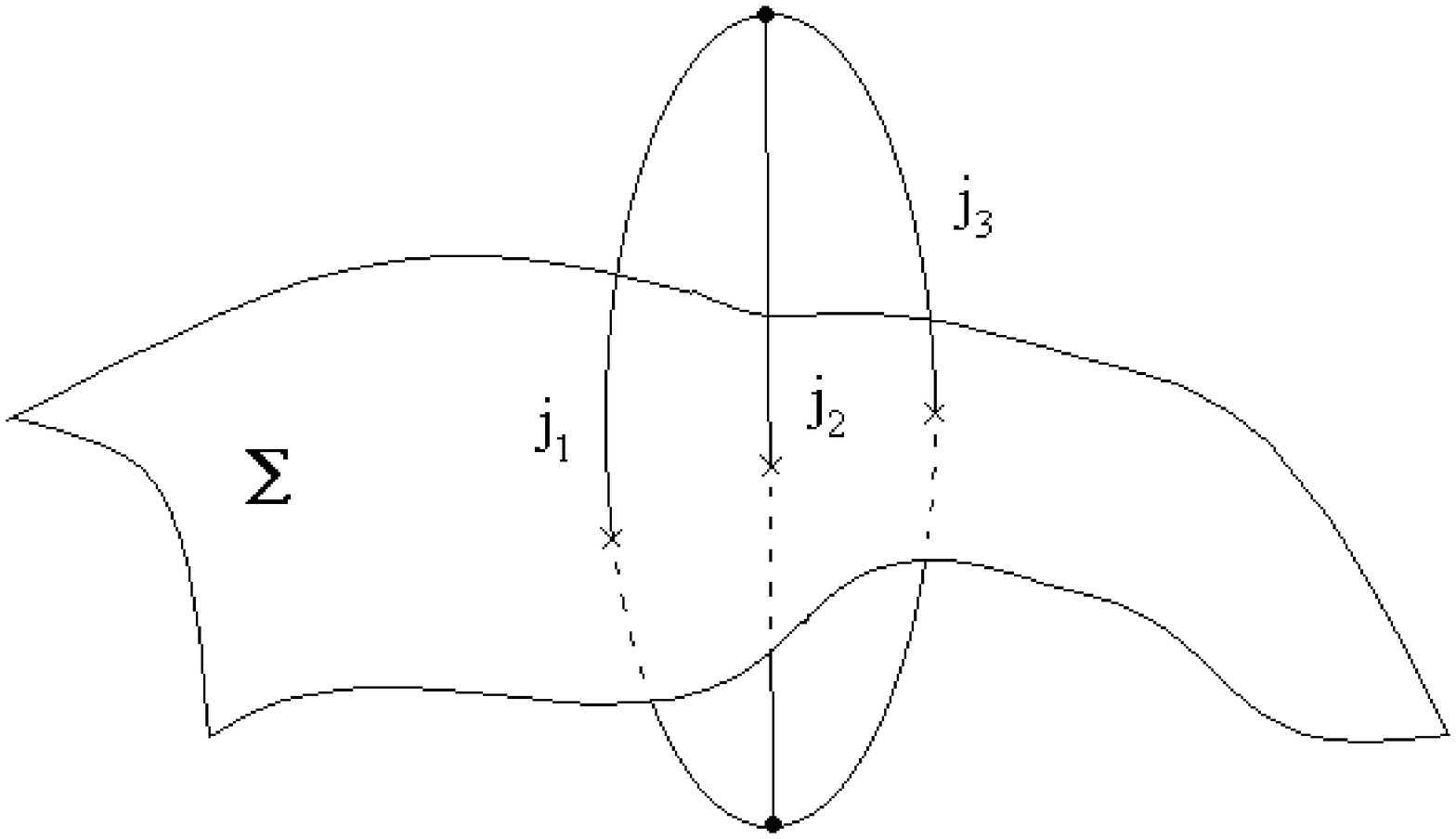}

{\small Fig.3 - A spin network embedded in a manifold and having three points of intersection with a surface $\Sigma$}

\end{center}

So the spin network states are eigenstates of
the operator with discrete eigenvalues. The crucial point is that the
operator we considered has the classical interpretation of the
area of the surface $\Sigma$. In fact its classical counterpart is
 \be
\mathcal{A}(\Sigma)\,=\,\int_{\Sigma}d^{2}\sigma\sqrt{n_{a}(\vec{\sigma})E^{ai}(\vec{x}(\vec{\sigma}))n_{b}(\vec{\sigma})E^{bi}(\vec{x}(\vec{\sigma}))}\,=\,\int_{\Sigma}d^{2}\sigma\sqrt{det(^{2}h)},\ee

which describes exactly the area of $\Sigma$ ($^{2}h$ is the 2-metric
induced by $h_{ab}$ on $\Sigma$).

This means that the area is quantized and has a discrete spectrum of
eigenvalues! Moreover we see that the \lq\lq carriers'' of this area
at the fundamental level are the edges of the spin network $S$ that
we choose to embed in the manifold. 

The same kind of procedure can be applied also to construct a
quantum operator corresponding to the volume of a 3-hypersurface and
to find that the spin network states diagonalize it as well and that
the eigenspectrum is again discrete. In this case, however, the volume is given by the vertices of the spin network
inside the hypersurface.

So we can say that an s-knot is a purely algebraic kinematical quantum state of the
gravitational field in which the vertices  give volume and the edges
give areas to the space in which we embed it.

So far we have considered only smooth embedding of our spin networks,
but the picture above works also in other contexts, such as the case of spin networks embedded in a triangulated manifold. We sketch here
how we can describe the quantum geometry of a tetrahedron in 3
dimensions as given by a spin network (for more details see
\cite{Baez,Barb}). Consider a compact, oriented, triangulated
3-manifold $\Delta$, and the complex $\Delta*$ dual to it, so having
one vertex for each tetrahedron in $\Delta$ and one edge for each face
(triangle) (see Fig.4). 

\begin{center}
\includegraphics[width=6cm]{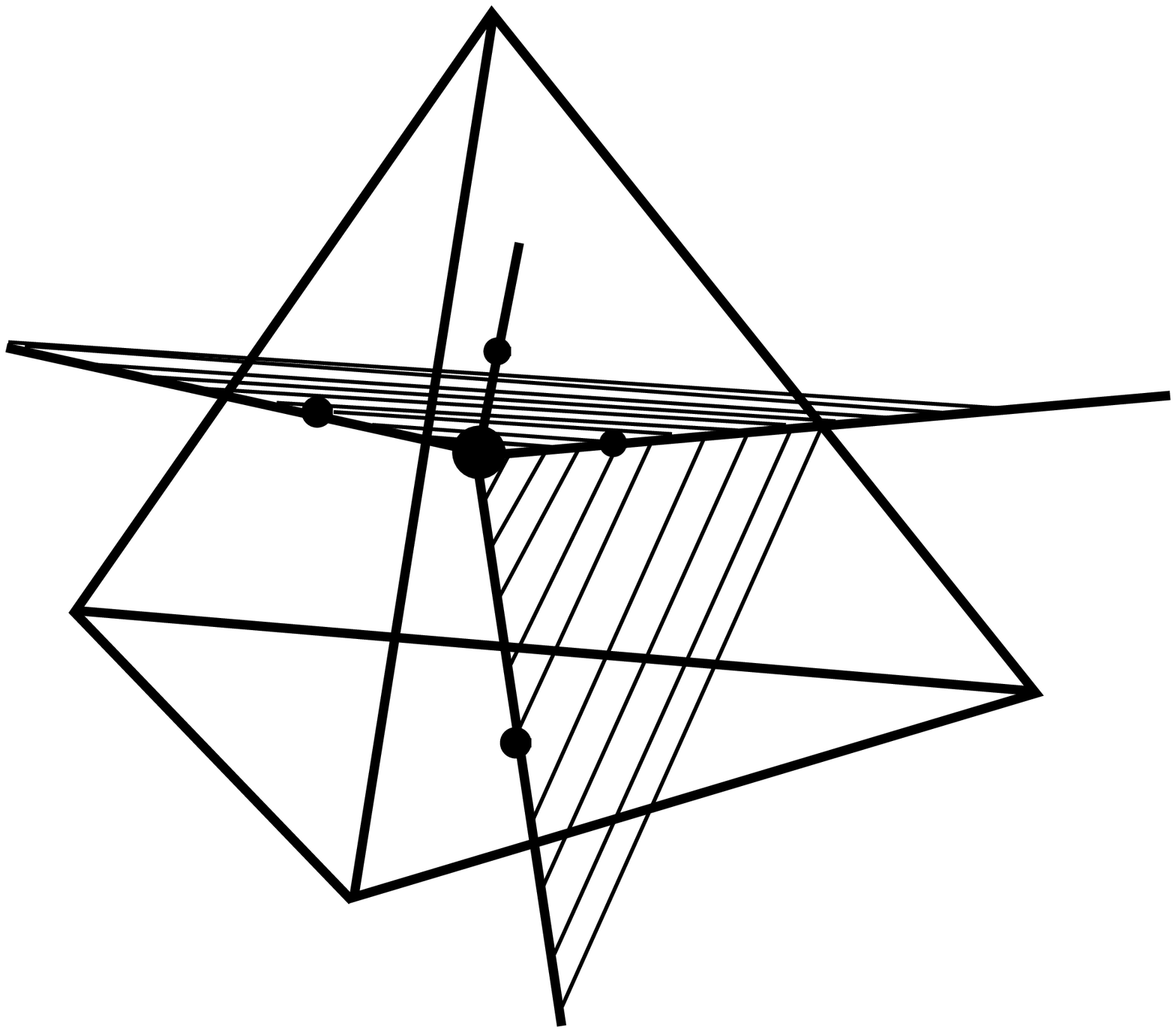}

{\small Fig.5 - The dual 2-skeleton of a tetrahedron}

\end{center}

Considering each tetrahedron in $\mathbb{R}^{3}$ with one
vertex at the origin, its classical geometry is determined
uniquely by a set of 3 vectors corresponding to the 3 edges outgoing
from that vertex. Alternatively, it can be determined by a set of 4
bivectors $E_{i}$ (i.e. elements of $\wedge^{2}\mathbb{R}^{3}$) satisfying the
{\it closure constraint} $E_{0}+E_{1}+E_{2}+E_{3}=0$. We can think of
them as assigned to the 4 triangles of the tetrahedron (by taking the
wedge product of the displacement vectors of the edges) and normal to
them and of the
constraint as simply saying that the triangles should close to form a
tetrahedron. The quantum picture goes as follows. Each bivector
corresponds uniquely to an angular momentum operator (in 3
dimensions), so an element of $SU(2)$ (using the isomorphism between
$\wedge^{2}\mathbb{R}^{3}$ and $so(3)$), and we can consider the Hilbert
space of states of a quantum bivector as given by $\mathcal{H}=\oplus j$, where $j$ indicates
the spin-j representation space of $SU(2)$. Considering the tensor
product of 4 copies of this Hilbert space, we have the following
operators acting on it:

\bean 
\hat{E}^{I}_{0}\,=\,J^{I}\otimes 1\otimes 1\otimes 1 \nonumber \\  
\hat{E}^{I}_{1}\,=\,1\otimes J^{I}\otimes 1\otimes 1 \\ 
\hat{E}^{I}_{2}\,=\,1\otimes 1\otimes J^{I}\otimes 1 \\
\hat{E}^{I}_{3}\,=\,1\otimes 1\otimes 1\otimes J^{I}
\eean
with $I=1,2,3$, and the closure constraint is given by $\sum_{i}
\hat{E}^{I}_{i}\psi=0\;\;\;\forall\psi\in\mathcal{H}^{\otimes 4}$. But
now the closure constraint indicates nothing but the invariance under
$SU(2)$ of the state $\psi$ so that {\it the Hilbert space of a
quantum tetrahedron} is given by:
\be
\mathcal{T}\,=\,\oplus_{j_{i}}\,Inv(j_{0}\otimes j_{1}\otimes
j_{2}\otimes j_{3}) \ee or in other words the sum, for all the
possible irreducible representations assigned to the triangles in the
tetrahedron, of all the possible invariant tensors of them, i.e. all
the possible intertwiners.

Moreover we can define 4 area operators
$\hat{A}_{i}=\sqrt{\hat{E}_{i}\hat{E}_{i}}$ and a volume operator
$\hat{V}=\sqrt{\mid\epsilon_{IJK}\hat{E}_{1}^{I}\hat{E}_{2}^{J}\hat{E}_{3}^{K}\mid}$,
and find that all are diagonal on $Inv(j_{0}\otimes j_{1}\otimes
j_{2}\otimes j_{3})$ (the area having eigenvalue
$\sqrt{j_{i}(j_{i}+1)}$).

But now we can think of a spin network living in the dual complex
$\Delta*$ of the triangulation, and so with one 4-valent vertex,
labelled with an intertwiner,  inside
each tetrahedron, and one edge, labelled with an irreducible
representation of $SU(2)$, intersecting exactly one triangle of the
tetrahedron. We then immediately recognize that this spin network
completely characterizes a state of the quantum tetrahedron, so a state
in $\mathcal{T}$, and gives volume to it and areas to its faces (also
matching the results from loop quantum gravity). 

Having understood what are spin networks and how they describe
the geometry of space, we can introduce spin foams and see how they
describe the geometry of {\it spacetime}.

\section{Spin foams and spin foam models} \label{sec:spfo}
\subsection{Definition and basic properties}
Let us start with the definition of what is a spin foam \cite{Baez}. It is very analogous to the definition of a spin network, but everything is one dimension higher. More precisely:
- given a spin network $\Psi=(\Gamma,\rho,\iota)$, a spin foam $F:0\rightarrow\Psi$ is a triple   $(\kappa,\tilde{\rho},\tilde{\iota})$ where: $\kappa$ is a 2-dimensional oriented complex such that $\Gamma$ borders $\kappa$, $\tilde{\rho}$ is a labeling of each face $f$ of $\kappa$ by an irreducible representation $\tilde{\rho}_{f}$ of a group $G$, and $\tilde{\iota}$ is a labelling of each edge $e$ not lying in $\Gamma$ by an intertwiner mapping
(the tensor product of) the irreducible representations of the faces
incoming to $e$ to (the tensor product of) the irreducible
representations of the faces outgoing from $e$, such that: for any edge $e'$ of $\Gamma$ $\tilde{\rho}_{f}=\rho_{e'}$ if $f$ is incoming to $e'$ and $\tilde{\rho}_{f}=\rho^{*}_{e'}$ if it is outgoing from it, and for any vertex $v$ in $\Gamma$ $\tilde{\iota}_{e}=\iota_{v}$ with appropriate dualization.

Similarly, given two disjoint spin networks $\Psi$ and $\Psi'$, a spin foam $F:\Psi\rightarrow\Psi'$ is defined to be the spin foam $F:0\rightarrow\Psi^{*}\otimes\Psi'$, where the dual of a spin network is defined to be a spin network with the same underlying 1-dimensional oriented complex with each edge labeled by dual representations $\rho^{*}_{e}$ and consequently the intetwiners appropriately dualized, while the tensor product of two spin networks is the spin network whose underlying oriented 1-complex is the disjoint union of the complexes of the original spin networks and with the labelling coming from them. 

Now some properties. A spin foam is {\it non-degenerate} if every vertex is the endpoint of at least one edge, every edge of at least one face, and every face is labelled with a non trivial irrep of $G$. We can define an equivalence relation for spin foams  so that two spin foams are equivalent if one can be obtained from the other by a sequence of the following moves: {\it affine transformation} - a spin foam $F=(\kappa, \rho, \iota)$ is obtained from $F'=(\kappa', \rho', \iota')$ by an affine transformation if there is a 1-1 affine map between cells of $\kappa$ and of $\kappa'$, preserving their orientations, and $\rho_{f}=\rho'_{\phi(f)}$ and $\iota_{e}=\iota'_{\phi(e)}$; {\it subdivision} - $F'$ is obtained from $F$ by subdivision if: the oriented complex $\kappa'$ is obtained from $\kappa$ by subdivision, when $f'$ is contained in $f$ then $\rho'_{f'}=\rho_{f}$, when $e'$ is contained in $e$ then $\iota'_{e'}=\iota_{e}$, while if $e'$ is the edge of two faces of $\kappa'$ contained in the same face $f$ of $\kappa$ then $\iota'_{e'}=1_{\rho_{f}}$; {\it orientation reversal} - $F'$ is obtained from $F$ by orientation reversal if: $\kappa$ and $\kappa'$ have the same cells but possibly different orientations, $\rho'_{f}=\rho_{f}$ or $\rho'_{f}=\rho_{f}^{*}$ if the complexes $\kappa'$ and $\kappa$ give equal or opposite orientations to the face $f$ respectively, and $\iota'_{e}=\iota_{e}$ after appropriate dualization.

We can then define {\it composition} between two (equivalence classes of) spin foams $F:\Psi\rightarrow\Psi'$ and $F':\Psi'\rightarrow\Psi''$ as follows. We choose reresentatives of $F$ and $F'$ in the same space $\mathbb{R}^{n}$ such that $\Psi'$ is the same in both and the affine maps $c,c':\Gamma'\times[0,1]\rightarrow\mathbb{R}^{n}$, by which $\Gamma'$ borders both $\kappa$ and $\kappa'$, fit together to a single affine map $C:\Gamma'\times[-1,1]\rightarrow\mathbb{R}^{n}$. Then the composite $FF'$ is defined to be the spin foam whose complex is $\kappa\cup\kappa'$, with label given by the original labels, while the edges of $\Gamma'$ are now labeled by (appropriately dualized) identity intertwiners.

It is important to note that we can define, for any given group $G$, a category
$\mathcal{F}$ in which the objects are non-degenerate spin networks
and the (non-degenerate) spin foams represent morphisms between them \cite{Baez,Baez2}. In order to have associativity of composition and unit laws to hold, in addition to the given definitions of equivalence and composition, we impose the extra equivalence relations $F(GH)\sim (FG)H$, for any spin foams $F, G, H$, and $1_{\Psi}F\sim F\sim F 1_{\Psi}$, where, for any spin network $\Psi$, $1_{\Psi}:\Psi\rightarrow\Psi$ is a left and right unit.    

We note also that a generic slice of a spin foam is a spin network so that every spin foam can be considered as a composition of \lq\lq smaller" spin foams composed along common spin networks.

Moreover, with this formulation the close analogy is apparent between the category of spin foams and the category of cobordisms used in topological quantum field theory. In fact, if we choose $G$ to be the trivial group, then a spin network is just a 1-dimensional complex, thought to represent space, and a spin foam is just a 2-dimensional complex representing spacetime. The use of a non trivial group corresponds to the addition of extra labels, representing fields, in our present case the gravitational field, i.e. the geometry.

This analogy leads us to consider spin foams as a tool for calculating transition amplitudes for the gravitational field and for formulating a quantum theory of gravity.

\subsection{Spin foam as quantum histories}
Consider an n-dimensional compact oriented cobordism $\mathcal{M}:S\rightarrow S'$ (spacetime), with $S$ and $S'$ compact oriented (n-1)-dimensional manifolds (space). A triangulation of $\mathcal{M}$ induces triangulations on $S$ and $S'$ with dual skeletons $\Gamma$ and $\Gamma'$ respectively (a dual skeleton has one vertex at the center of each (n-1)-simplex and an edge intersecting each (n-2)-simplex). We can thus consider spin networks whose underlying graph is this dual 1-skeleton (so we work with embedded spin networks). We know that the space of all the possible spin networks embedded in $S$ (or $S'$) (so for all the possible triangulations) defines the gauge invariant state space $\mathcal{H}$ (or $\mathcal{H}'$) on $S$ (or $S'$), so that time evolution is naturally given by an operator $Z(\mathcal{M}):\mathcal{H}\rightarrow\mathcal{H'}$. Since as we said spin networks are a complete basis for the state space, to characterize this operator it is enough to have the transition amplitudes $\langle\Psi'\mid Z(\mathcal{M})\mid\Psi\rangle$ between two spin networks. The idea is then to write this amplitude as a sum over all the possible spin foams going from $\Psi$ to $\Psi'$:
\be
  \langle\Psi'\mid Z(\mathcal{M})\mid\Psi\rangle\,=\,\sum_{F:\Psi\rightarrow\Psi'}\,Z(F) \label{eq:ampli}
\ee
where the sum over spin foams implies both a sum over 2-dimensional complexes (or, in the dual picture, over all the triangulations of $\mathcal{M}$ matching the graphs of $\Psi$ and $\Psi'$ on the boundary), and  a sum over all the possible labelling of the elements of them by irreps of $G$. We see that in this picture spin networks are states while spin foams represent histories of these states. So the problem is now to find the right form of the amplitude $Z(F)$ for a given spin foam $F$. Moreover, this amplitude should satisfy
\be 
Z(F')Z(F)\,=\,Z(F'F)      
\ee
where $FF':\Psi\rightarrow\Psi''$ is the spin foam obtained gluing together $F:\Psi\rightarrow\Psi'$ and $F':\Psi'\rightarrow\Psi''$ along $\Psi'$. When this happens, and the sum in (~\ref{eq:ampli}) converges in a sufficiently nice way, we have:
\be 
Z(\mathcal{M}')Z(\mathcal{M})\,=\,Z(\mathcal{M}'\mathcal{M})      
\ee
for composable cobordisms $\mathcal{M}:S\rightarrow S'$ and $\mathcal{M}':S'\rightarrow S''$.

We want to stress that even though we used a picture of spin networks and spin foams living in a triangulated manifold, this is a priori not necessary and one can instead work with abstract (non embedded) objects.

Before going on to describe how a spin foam model is constructed, we want to give some more details about how one can make precise this idea of spin foams as giving the evolution of spin networks \cite{ReiRov}. 

In the canonical approach to quantum gravity, as we said, the
(coordinate) time evolution of the gravitational field is generated by the Hamiltonian $H_{N,\vec{N}}(t)=C[N(t)]+C[\vec{N}(t)]$, i.e. it is given by the sum of the Hamiltonian constraint (function of the lapse $N(t)$)  and the diffeomorphism constraint (function of the shift $\vec{N}(t)$. The quantum operator giving evolution from one hypersurface $\Sigma_{i}$ ($t=0$) to another $\Sigma_{f}$ ($t=1$) is given by 
\be
U_{N,\vec{N}}\,=\,e^{-i\int_{0}^{1}dt  H_{N,\vec{N}}(t)}.
\ee
We then define the proper time evolution operator as:
\be U(T)\,=\,\int_{T}dN d\vec{N}\,U_{N,\vec{N}} \ee
where the integral is over all the lapses and shifts satisfying $N(x,t)=N(t)$ and $\int_{0}^{1}dt N(t)=T$, and $T$ is the proper time separation between $\Sigma_{i}$ and $\Sigma_{f}$ (this construction can be generalized to a multifingered proper time \cite{ReiRov}). Now we want to calculate the matrix elements of this operator between two spin network states (two s-knots). These are to be interpreted as transition amplitudes between quantum states of the gravitational field, and computing them is equivalent to having solved the theory imposing both the Hamiltonian and diffeomorphism constraints.   

Now \cite{ReiRov} we take a large number of hypersurfaces separated by
small intervals of coordinate time, and write
$U_{N,\vec{N}}=D[g]U_{N_{\vec{N}},0}$, where $g$ is the finite
diffeomorphism generated by the shift between the slices at $t=0$ and
$t$ and $D(g)$ is the corresponding diffeomorphism operator acting on
states (in fact we can always rearrange the coordinates to put the shift equal to zero, then compensating with a finite change of space coordinates (a diffeomorphism) at the end).
Considering $N_{\vec{N}}=N=const$ (partial gauge fixing) we can expand $U_{N,0}$ as:
\be
U_{N,0}\,=\,1\,+\,(-i)\int_{0}^{\tau}dt\,C[N(t)]\,+\,(-i)^{2}\int_{0}^{\tau}dt\int_{0}^{\tau}dt'\,C[N(t')]C[N(t)]\,+....
\ee
and compute its matrix elements as:
\bea
\langle S_{f}\mid U_{N,0}(T)\mid S_{i}\rangle = \langle S_{f}\mid S_{i}\rangle + (-i)\int_{0}^{\tau}dt\langle S_{f}\mid C[N(t)]\mid S_{i}\rangle +\nonumber \\+ (-i)^{2}\int_{0}^{\tau}dt\int_{0}^{\tau}dt'\langle S_{f}\mid C[N(t')]\mid S_{1}\rangle\langle S_{1}\mid C[N(t)]\mid S_{i}\rangle +.....
\eea
having inserted the resolution of the identity in terms of a complete set of states $\mid S_{1}\rangle$ (with the summation implicit).

Using the explicit form of the Hamiltonian constraint operator \cite{ReiRov,Tie}: 
\be
C[N]\mid S\rangle\,=\,\sum_{v,e,e',\epsilon,\epsilon'}A_{vee'\epsilon\epsilon'}(S)D_{vee'\epsilon\epsilon'}\mid S\rangle\,=\,\sum_{\alpha}A_{\alpha}(S)D_{\alpha}\mid S\rangle\,+\,h.c.
\ee
where the operator $D_{vee'\epsilon\epsilon'}$ acts on the spin network $S$ creating two new trivalent vertices $v'$ and $v''$ on the two edges $e$ and $e'$ (intersecting at the vertex $v$), connecting them with a new edge with label 1, and adding $\epsilon=\pm 1$ (and $\epsilon'=\pm 1$) to the color of the edge connecting $v$ and $v'$ (and $v$ and $v''$); the $A$s are coefficients that can be explicitely computed. Doing this and carrying out the integrals we find:
\bea 
  \lefteqn{\langle S_{f}\mid U_{N,0}(T)\mid S_{i}\rangle = \langle S_{f}\mid S_{i}\rangle +} \nonumber \\ &+&\,(-iT)\left( \sum_{\alpha\in S_{i}} A_{\alpha}(S_{i})\langle S_{f}\mid D_{\alpha}\mid S_{i}\rangle+ \sum_{\alpha\in S_{f}} A_{\alpha}(S_{f})\langle S_{f}\mid D^{\dagger}_{\alpha}\mid S_{i}\rangle\right)+\nonumber \\ &+&\frac{(-iT)^{2}}{2!}\sum_{\alpha\in S_{i}}\sum_{\alpha'\in S_{1}}A_{\alpha}(S_{i})A_{\alpha'}(S_{1})\langle S_{f}\mid D_{\alpha'}\mid S_{1}\rangle\langle S_{1}D_{\alpha}\mid S_{i}\rangle + .....
\eea
So at each order $n$ we have the operator $D$ acting $n$ times, there are $n$ factors $A$ and a finite number of terms coming from a sum over vertices, edges, and $\epsilon=\pm 1$. Moreover since the sum over intermediate states $S_{1}$ is finite, the expansion above is finite order by order.
Now we should integrate over $N$ and $\vec{N}$ to obtain the matrix
elements of $U(T)$. It happens that the integration over the lapse is
trivial, since there is no dependence on it in the integral, while the
integration over the shift is just the imposition of the
diffeomorphism constraint, so its effect is to replace the spin
networks in the matrix elements with their diffeomorphism equivalence classes. At the end we have:
    \bea 
  \langle s_{f}\mid U(T)\mid s_{i}\rangle = \langle s_{f}\mid s_{i}\rangle + (-iT)\left( \sum_{\alpha\in s_{i}} A_{\alpha}(s_{i})\langle s_{f}\mid D_{\alpha}\mid s_{i}\rangle+ \sum_{\alpha\in s_{f}} A_{\alpha}(s_{f})\langle s_{f}\mid D^{\dagger}_{\alpha}\mid s_{i}\rangle\right)+\nonumber \\ +\frac{(-iT)^{2}}{2!}\sum_{\alpha\in s_{i}}\sum_{\alpha'\in s_{1}}A_{\alpha}(s_{i})A_{\alpha'}(s_{1})\langle s_{f}\mid D_{\alpha'}\mid s_{1}\rangle\langle s_{1}D_{\alpha}\mid s_{i}\rangle + .....\label{eq:exp}
\eea
This is the transition amplitude between a 3-geometry $\mid s_{i}\rangle$ and a 3-geometry $\mid s_{f}\rangle$.
The crucial observation now is that we can associate to each term in the expansion above a 2-dimensional colored surface $\sigma$ in the manifold defined up to a 4-diffeomorphism. The idea is the following. Consider the initial hypersurfaces $\Sigma_{i}$ and $\Sigma_{f}$ and draw $s_{i}$ in the first and $s_{f}$ in the second; of course location is chosen arbitrarily, again up to diffeomorphisms, since there is no information in $s_{i}$ and $s_{f}$ about their location in spacetime. Now let $s_{i}$ slide across the manifold $\mathcal{M}$ from $\Sigma_{i}$ towards $s_{f}$ in $\Sigma_{f}$ (let it \lq\lq evolve in time"). The edges of $s_{i}$ will describe 2-surfaces, while the vertices will describe lines. Each \lq\lq spatial" slice of the 2-complex so created will be a spin network in the same s-knot (with the same combinatorial and algebraic structure), unless an \lq\lq interaction" occurs, i.e. unless the Hamiltonian constraint acts on one of these spin networks. When this happens the Hamiltonian constraint creates a spin network with an additional edge (or with one edge less) and two new vertices; this means that the action on the 2-complex described by the evolving spin network is given by a creation of a vertex in the 2-complex connected by two edges to the new vertices of the new spin network, originating from the old one by the action of the Hamiltonian constraint. So at each event in which the Hamiltonian constraint acts the 2-complex \lq\lq branches" and this branching is the elementary interaction vertex of the theory. So an $n$-th order term in the expansion (~\ref{eq:exp}) corresponds to a 2-complex with $n$ interaction vertices, so with $n$ actions of the operator $D_{\alpha}$ on the s-knot giving the 3-geometry at the \lq\lq moment" at which the action occurs. Moreover, each surface in the full 2-complex connecting in this way $s_{i}$ to $s_{f}$ can be coloured, assigning to each 2-surface (face) in it the irrep of the spin network edge that has swept it out, and to each edge in it the intertwiner of the corresponding spin network vertex. We give a picture of a second order term in Fig.5.

\begin{center}
\includegraphics[width=7cm]{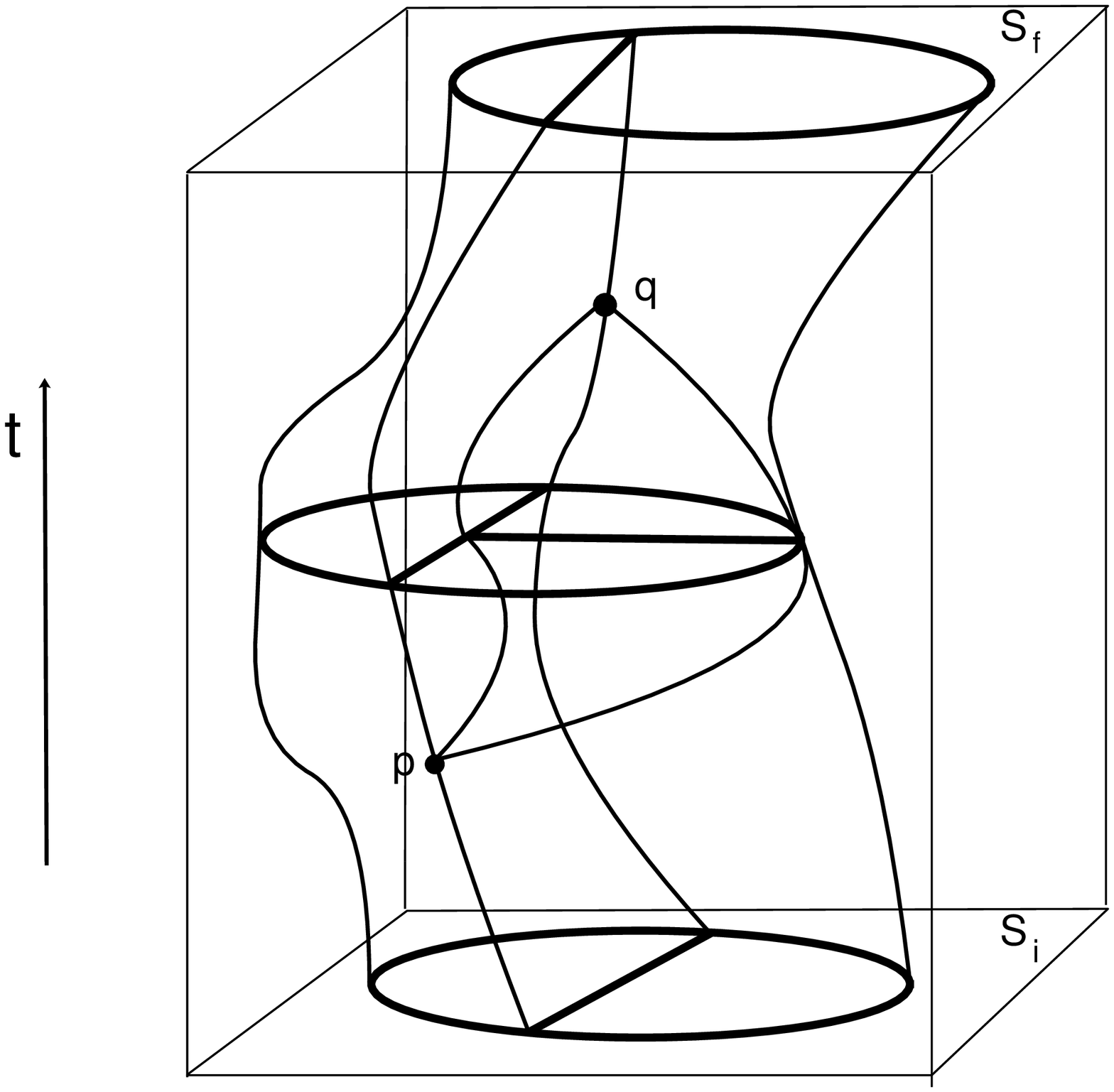}
\end{center}
{\small Fig.5 - The 2-complex correspondent to a second order term in the expansion of the amplitude} 
\vspace{0.5cm}

If we also fix an ordering of the vertices of the 2-complex, then each term in (~\ref{eq:exp}) corresponds uniquely to a 2-complex of this kind, in a diffeomorphism invariant way, in the sense that two 2-complexes correspond to the same term if and only if they are related by a 4-diffeomorphism.

Given this, we can rewrite the transition amplitudes as sums over topologically inequivalent (ordered) 2-complexes $\sigma$ bounded by $s_{i}$ and $s_{f}$, and with the weight for each 2-complex being a product over the $n(\sigma)$ vertices of the 2-complex $\sigma$ with a contribution from each vertex given by the coefficients of the Hamiltonian constraint:
\be
\langle s_{f}\mid U(T)\mid s_{i}\rangle\,=\,\sum_{\sigma}\mathcal{A}[\sigma](T)\,=\,\sum_{\sigma}\frac{(-iT)^{n(\sigma)}}{n(\sigma)!}\prod_{v}A_{\alpha}
\ee

These colored 2-complexes arising from the evolution in time of
spin networks are precisely spin foams! Moreover the structure of the
amplitude calculated in this approach is the same as the one outlined in (~\ref{eq:ampli}).

\subsection{Spin foam models for quantum gravity}
The expression for the amplitudes of the propagation operator in loop quantum gravity can be seen as an example of a spin foam model for 4-dimensional quantum gravity. More generally, a spin foam model is given by a partition function $Z$ expressed as a sum over topologically inequivalent spin foams, this meaning a sum over all the possible 2-complexes and a sum over all the possible representation of the group $G$ labelling the faces of it, with a weight for each spin foam given, in general, by a product of amplitudes for the elements of the spin foam, i.e. faces, edges, vertices:
\be
Z\,=\,\sum_{\sigma:\partial\sigma=\Psi\cup\Psi'}w(\sigma)\sum_{J}\prod_{f}A_{f}\prod_{e}A_{e}\prod_{v}A_{v} \label{eq:spinfoam} \ee
including the case in which $\Psi$ and/or $\Psi'$ are null spin networks, i.e. the vacuum.

There are several ways to look at a spin foam model. One is to consider it as giving the transition amplitudes for the gravitational field, consequently giving a precise formulation of the dynamics of the quantum theory (a problem still not fully resolved in the canonical loop quantum gravity approach), in the form of a sum over histories of its states, so we have a formulation which seems a precise implementation of the idea of a gravitational path integral \cite{HarHaw, Haw}. In this sense spin foam models represent a new form of the covariant approach to quantum gravity. Correspondingly we would like to interpret them as realizing the sum over geometries proposed in this approach as the way to realize a quantum gravity theory. And indeed this can be made more precise, as we will see in more detail in the following, since one can look at a spin foam as giving a quantum 4-geometry just like a spin network gives a quantum 3-geometry. In fact one can think of the 2-complex of a spin foam as being the dual 2-skeleton of a 4-dimensional triangulation, so that it has a vertex at the centre of each 4-simplex, an edge intersecting each tetrahedron, and a face intersecting each triangle (for a more detailed description and some picture of the dual 2-skeleton of a triangulation, see section ~\ref{sec:OW}). Then (analogously to what we described in section ~\ref{sec:3geo} for spin networks) we can think that the representation of $G$ labelling a face gives an area to the triangle this face intersects, and that the intertwiner labeling an edge gives a volume to the tetrahedron the edge intersects. This idea will be made precise and proven in the following. 
We note also that, seeing the 2-complex as the dual 2-skeleton of a triangulation, the sum over 2-complexes in (~\ref{eq:spinfoam}) is interpretable as a sum over triangulations, a crucial step, necessary to recover triangulation independence and the full degrees of freedom of the gravitational field in 4-dimensions when the theory itself is not topological. We will come back to this problem.
Another way to look at a spin foam model, suitable in a simplicial context, is to regard it as a state sum model \cite{B} for a (triangulated) manifold of the kind used in statistical mechanics. Consider a 4-dimensional triangulated manifold $M$, and assign a set of states $S$ to each $n$-simplex $s_{n}$, with vertices $0, 1,...,n$, for each $n\leq 4$ and the same set for any simplex of the same dimension. There are then maps $\partial_{i}:S(s_{n})\rightarrow S(s_{n-1})$, for $i=0,...,n$, by means of which a state on a $n$-simplex specify a state on anyone of its faces, the $i$-th map corresponding to the $i$-th ($n-1$)-face. Of course the states on intersecting simplices are related, because the boundary data must match. Then we specify a weight ($\mathbb{C}$-number) giving an amplitude to each state $w:S(s_{n})\rightarrow\mathbb{C}$. Finally we use all this information, which together specifies a {\it configuration} $c$, to construct a partition function on the triangulated manifold of the form: 
\be
Z(M)\,=\,\sum_{c}\prod_{s} w(s(c))
\ee where the sum is over all the possible configurations and the product is over all the simplices (of every dimension) of the triangulation.

We recognize that, again with the identification of the 2-complex of a spin foam with the 2-skeleton of a triangulation, and so of 4-simplices with vertices, tetrahedra with edges and triangles with faces, this state sum is exactly analogous to a spin foam model (~\ref{eq:spinfoam}) if we think of the sum over representations in this as the sum over configurations above. Of course if we want a state sum model for quantum gravity we should at the end get rid of the dependence on the particular triangulation chosen, and there are several ways in which this can be done, as we will see.

Whatever point of view we decide to take, the crucial point is that a
spin foam model aims to be a non-perturbative (since it is not based
on any kind of perturbative expansion, with the exception of the
formulation based on the field theory over a group, where, as we will
see, the spin foam model is given by the perturbative expansion of the
auxiliary field theory \footnote{However, even in this case it is a
perturbative expansion of a very unusual kind since it is not about a fixed
background geometry, but about ``nothingness'', as we shall see.}) and background independent (with the geometry and the metric being emerging concepts and not a priori fixed structures) formulation of quantum gravity.  

\subsection{Spin foam models for discrete topological quantum field theories: the Ponzano-Regge-Turaev-Viro state sum for 3d quantum gravity} \label{sec:PRTV}
The framework of spin foam models is very versatile and there exist spin foam models for many different kinds of theories. There exist spin foam formulation for topological field theories \cite{PonzReg, T-V, CrYet}, lattice gauge theories, both abelian and non-abelian \cite{Reis1, OePf}, and gravity \cite{BC,Reis2, Reis3,Iwa1, Iwa2,  BC2}. We cannot describe in detail all of them, and in the later sections we will concentrate only on the Barrett-Crane spin foam model for euclidean quantum gravity, but now we want to say something about a model which was very influential for this line of research, showing how one can arrive at spin foam and spin foam models from the very different (with respect to loop quantum gravity) point of view of discrete topological quantum field theories, with a striking convergence of results.
Consider a tetrahedron in a triangulated 3-manifold. Classically its geometry is characterized by its edge lengths (up to rescaling and reflection), and can be studied with the discretized version of general relativity given by Regge calculus \cite{Regge, Will}. Now let us make the simple quantization ansatz that the edge lengths are quantized as $l=j+\frac{1}{2}$ with $j\in\mathbb{N}/2$. Now we can associate to each \lq\lq quantized" tetrahedron an amplitude ($\mathbb{C}$-number) given by the classical 6j-symbol of Racah-Wigner calculus, a well-known object in recoupling theory of SU(2), given by the six variables $j$ associated to the edges of the tetrahedron and thought as irreducible representations of that group. 

\be
\includegraphics[width=4.5cm]{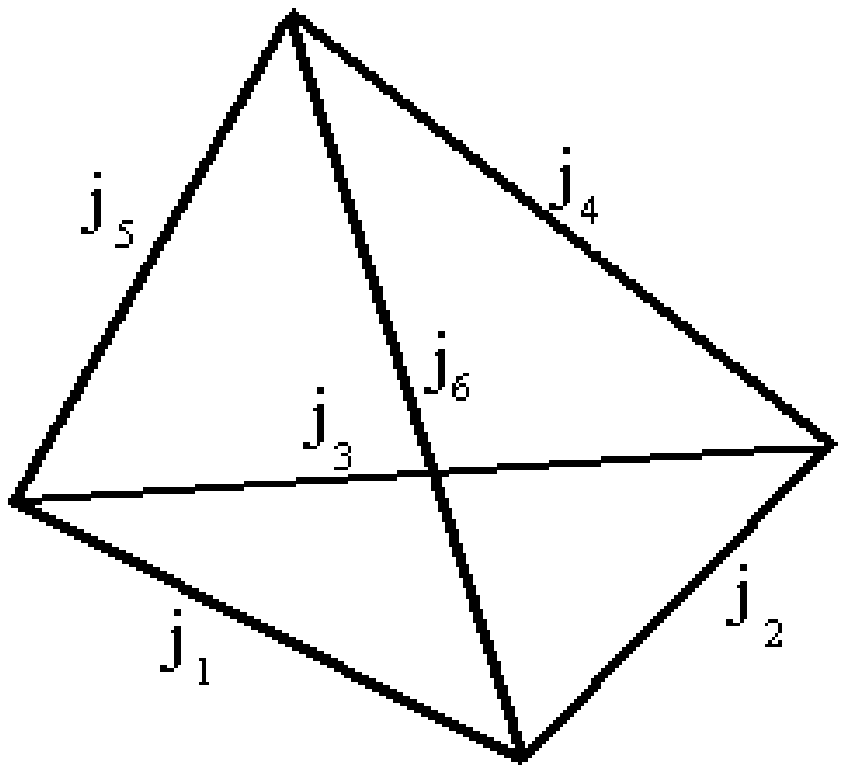} \leftrightarrow \left\{ \matrix
    { j_1 & j_2 & j_3 \cr j_4 & j_5 & j_6
\cr}
     \right\} \nonumber
\ee
\vspace{1cm}
\begin{center}
{\small Fig.6 - A tetrahedron and the corresponding 6j-symbol}
\end{center}

This association is meaningful also because the 6j-symbol has the same symmetries as a geometrical tetrahedron, which in fact was used for long time as a visual tool in recoupling theory calculations. Now it acquires however a deeper and unexpected significance; indeed Ponzano and Regge \cite{PonzReg} proved (at first heuristically, but this was then made rigorous \cite{Roberts}) that the asymptotic (semiclassical) value of the 6j-symbol for a tetrahedron in the limit of large $j$ (but the approximation turns out to be quite good also for general values of the $j$) is given by:
    \be \left\{ \matrix
    { j_1 & j_2 & j_3 \cr j_4 & j_5 & j_6
\cr}
     \right\}\sim_{j\rightarrow\infty}\sqrt{\frac{1}{V}}
\left(e^{iS_{R}+i\phi}+e^{-iS_{R}-i\phi}\right) \nonumber \ee where
$V$ is the volume of the tetrahedron, $\phi$ is a constant phase and $S_{R}=\sum_{e} (j_{e}+\frac{1}{2}) \theta_{e}$ is Regge action for
discrete gravity, with the sum being over the edges of the
tetrahedron, the $\theta$ being the dihedral angle of the edge $e$,
i.e. the angle between the two outward normals of the two faces
incident on that edge. So the 6j-symbol has the form of an amplitude
in a semiclassical path integral formulation of quantum gravity (the
reason for having both the exponentials with + and - sign is the
reflection ambiguity in characterizing the geometry with edge
lengths). 
This led Ponzano and Regge to propose a quantum gravity model in which one triangulates a 3-dimensional manifold, assigns a 6j-symbol to each tetrahedron in the triangulation, multiplies all the 6j-symbols for all the tetrahedra forming the manifold, and then sums over all the possible irreducible representations of SU(2), obtaining a kind of discretized path integral for quantum gravity, the only problem being that this sum diverges. This idea was then made rigorous much later by Turaev and Viro \cite{T-V}, who used the same kind of construction but with the quantum deformation of SU(2) at a root of unity, instead, obtaining a truncation of the sum over representation, so a finite model, and a new topological invariant for 3-manifolds given by the partition function:
\be
Z(M)\,=\,\sum_{j}\prod_{v}\frac{1}{N_{\Delta}}\prod_{e}dim_{j}\prod_{tet}\{6j\}_{q}
\ee
where  $N_{\Delta}$ is a normalization factor depending on the
    triangulation $\Delta$ chosen for the manifold $M$ (see \cite{T-V} for its definition), $dim_{j}$ is
    the (quantum) dimension of the representation space of the irrep
    $j$ (multiplied bt $(-1)^{2j}$), and $\{6j\}_{q}$ is the quantum deformed analogue of the
    classical 6j-symbol, with $q$ a root of unity. The topological
    invariance of the theory is reflected in the independence of $Z$
    from the particular triangulation chosen (this also implies that
    it is not necessary to sum over triangulations). 
This partition function was then proved \cite{giapp, ArchWill} to give
    in the asymptotic limit 3d Euclidean gravity with cosmological
    constant, and can be thought of as giving a quantization of BF-theory
    \cite{Horow,BBRT}, a topological field theory, classically equivalent to
    3d gravity (and also equivalent to two copies of Chern-Simons
    theory, proved at first by Witten \cite{Witt}, and then, more rigorously by Reshetkin and Turaev \cite{R-T}, to give a quantization of
    pure 3d gravity). The limit $q\rightarrow 1$, corresponding to
    the cosmological constant going to zero, reproduces the
    Ponzano-Regge model. It is apparent that the
    Ponzano-Regge-Turaev-Viro model is a state sum model of the kind
    described in the previous section, but it is also interpretable as
    a spin foam model. In fact we can consider the 2-skeleton of the
    triangulation (having a vertex for each tetrahedron, an edge
    intersecting each triangle, and a face intersecting each edge of
    the tetrahedron itself - see Fig.2), and think of each irrep of
    SU(2) (or of its quantum deformation) as assigned to the faces of
    it, and consequently the intertwiners assigned to the edges, and
    then the partition function above gives exactly a spin foam model with
    the structure (~\ref{eq:spinfoam}). Consequently it represents a
    spin foam model for 3d quantum gravity.

This construction was generalized to the 4-dimensional case by Crane and Yetter \cite{CrYet, CrKauYet}, giving a discrete topological field theory corresponding in the continuum to BF theory \cite{Kawa}, and an invariant of 4-manifolds. 

\section{The Barrett-Crane spin foam model for quantum gravity} \label{sec:BC}
We now turn to an analysis of a specific spin foam model for Euclidean
quantum gravity, trying to be as complete as possible, and pointing
out the basic ideas as well as the connections of the model with
classical gravity.
 
\subsection{The quantum 4-simplex} \label{sec:BC'} 
An appealing way to look at a spin foam, about which we will say a bit more in the following, is 
as a kind of  \lq\lq Feynman diagram" for spacetime encoding the
informations about interactions of spin network states, so that edge
and face amplitudes can be thought of as \lq\lq propagators" and the
\lq\lq interactions" occur at the vertices. So in a spin foam model
the crucial element is the vertex amplitude which encodes all the
information about the interactions and the dynamical content of the
theory. Thus the question to answer for a construction of a spin foam
model for quantum gravity is: what is the correct form of the vertex
amplitude? Thinking of the spin foam as embedded in a triangulated
4-manifold, as a coloured dual 2-skeleton of it, this is translated
into: what is the quantum amplitude for a 4-simplex? or how to
describe a quantum 4-simplex. More precisely, the problem to solve is
how to translate the geometrical information necessary to
characterize completely a 4-simplex at the classical level into the
quantum domain of algebra and representation theory, obtaining a
characterization of a quantum 4-simplex. Barrett and Crane \cite{BC}
answered this question precisely. 
A geometric 4-simplex in Euclidean space is given by the embedding of
an ordered set of 5 points $(0, 1, 2, 3, 4)$ in $\mathbb{R}^{4}$ (its
subsimplices are given by subsets of this set) with embedding
determined by the 5 position coordinates $x_{0}, x_{1}, x_{2}, x_{3},
x_{4} \in \mathbb{R}^{4}$ and required to be non-degenerate (the
points should not lie in any hyperplane). Each triangle in it
determines a bivector (i.e. an element of $\wedge^{2}\mathbb{R}^{4}$) constructed out of the displacement vectors for the edges, taking the wedge product of two of them. Barrett and Crane proved that, classically, a geometric 4-simplex in Euclidean space is completely and uniquely characterized (up to parallel translation and 
inversion through the origin) by a set of 10 bivectors $b_{i}$, each corresponding to a triangle in the 
4-simplex and satisfying the following properties:
\begin{itemize}
\item the bivector changes sign if the orientation of the triangle is changed;
\item each bivector is simple, i.e. given by a wedge product of two vectors;
\item if two triangles share a common  edge, the sum of the two bivectors is simple;
\item the sum (considering orientations) of the 4 bivectors corresponding to the faces of a 
tetrahedron is zero; this corresponds \lq\lq physically" to the fact that the 4 faces close among themselves forming a closed polyhedron;
\item the assignment of bivectors is non-degenerate; this means that for six triangles sharing the same vertex, the six corresponding bivectors are linearly independent;  
\item the bivectors (thought of as operators) corresponding to triangles meeting at a vertex of a 
tetrahedron satisfy the inequality $tr b_{1}[b_{2},b_{3}] > 0$, i.e. the tetrahedron has non-zero volume.
\end{itemize}
Now the problem is to find the corresponding quantum description, i.e. a characterization of a quantum 4-simplex. 
The crucial observation now is that bivectors can be thought of as being elements of the Lie 
Algebra so(4), because of the isomorphism between
$\wedge^{2}\mathbb{R}^{4}$ and $so(4)$, so we associate to each
triangle in the triangulation an $so(4)$ element. Then we turn these
elements into operators choosing a (different) representation of
$so(4)$ for each
of them,
i.e. considering the splitting $so(4)\simeq su(2) \oplus su(2)$, a
pair of spins $(j,k)$, so that we obtain bivector operators acting on
the Hilbert space given by the representation space chosen. Each
tetrahedron in the triangulation is then associated with a tensor in
the product of the four spaces of its triangles.
 
The point is to translate the conditions above into conditions on the representations of this 
algebra. 

The corresponding conditions on the representations were found to be
the following (more details about how to prove this will be given in
the next section):
\begin{itemize}
\item different orientations of a triangle correspond to dual representations;
\item the representations of the triangles are \lq\lq simple representations" of SO(4) of the form 
$(j,j)$, i.e. representations of class 1 with respect to the subgroup
SO(3) \cite{V-K}, characterized by the vanishing of the second Casimir
of the group (which for a representation $(j,k)$ is given by $C_{2}=\sqrt{j(j+1)}-\sqrt{k(k+1)}$); classically this corresponds to imposing equality of the self-dual and anti-self-dual parts of the bivectors, that is exactly requiring them to be simple: $\langle b, *b\rangle=\langle b^{+}, b^{+}\rangle - \langle b^{-},b^{-}\rangle =0$; 
\item given two triangles, if we decompose the pair of representations into its Clebsch-Gordan 
series, the tensor for the tetrahedron is decomposed into summands which are non-zero only 
for simple representations;
\item the tensor for the tetrahedron is invariant under SO(4).   
\end{itemize}  

At the quantum level, we allow for degenerate tetrahedra, so we admit representations such that the volume of the tetrahedron they are assigned to is zero.

Moreover Barrett and Crane proposed a form of the intertwiner to be used for labelling each tetrahedron, that was shown to be unique in \cite{Reis}. They called a spin network labeled by only simple representations of SO(4) and the intertwiners they propose a \lq\lq relativistic spin network". Now it is easy to construct an amplitude for a quantum 4-simplex. We can consider the relativistic spin network corresponding to a 4-simplex, whose graph is the 1-complex dual to the boundary of the 4-simplex, having five 4-valent vertices (corresponding to the five tetrahedra in the boundary of the 4-simplex) with each of the ten edges connecting two different vertices (corresponding to the ten triangles of the 4-simplex each shared by two tetrahedra), labelled as above, and evaluate it using the graphical calculus developed for spin networks or, equivalently, the recoupling theory for SO(4). This just means taking the product of the five intertwiners with the indices suitably contracted to get a number $\mathcal{B}_{BC}$ (amplitude)(see Fig.12). 
To get the partition function for the whole manifold we then multiply
all the amplitudes for the 4-simplices in it and sum over the
representation used in the labelling. There can be also amplitudes for
lower dimensional simplices, not determined uniquely in \cite{BC} and
we will see in section ~\ref{sec:OW} how these amplitudes can be derived, and, in particular, what is the origin of the amplitude for the tetrahedra. The proposed state sum (spin foam model) for gravity (for fixed triangulation $\Delta$) is then:
\be
Z(\mathcal{M}, \Delta)_{BC}\,=\,\sum_{J}\prod_{triangles}A_{tr}\,\prod_{tetrahedra}A_{tet}\,\prod_{4-simpl}\mathcal{B}_{BC}.
\ee

Barrett and Crane also suggest using the quantum deformation of SO(4)
at a root of unity in order to obtain a finite sum over representations, but the results
of \cite{P-R, Per}, to be discussed in section ~\ref{sec:matrix} show
that the sum above is finite even in the undeformed case. In fact, a particular choice of the amplitudes for the tetrahedra, that can be motivated by an analysis of the gluing between 4-simplices \cite{OW}, as we will see in section ~\ref{sec:BC1}, and also obtained naturally in the context of field theories over group manifolds \cite{P-R}(see section ~\ref{sec:matrix}), has the effect of making the sum over representations finite for any given triangulation.

We note that the formalism of \lq\lq relativistic spin networks" (or \lq\lq simple spin networks"), whose evaluation is interesting also from the purely mathematical point of view since it gives an invariant of the isotopy class of labelled embedded graphs, was then developed and generalized to higher-valent vertices in \cite{Yett}.

\subsection{Quantum geometry in the Barrett-Crane model} \label{sec:geom}
In this section we describe how a spin foam in the Barrett-Crane model
determines the 4-geometry of the triangulated manifold in which it is
embedded. Of course the most important feature of the model is the
Barrett-Crane description of the quantum 4-simplex, given above. Accordingly, we give now more details about this description, following \cite{Baez}. Then we discuss the characterization of a quantum tetrahedron in 4 dimensions, as it arises from the Barrett-Crane model, following \cite{BaezBarr}, to which we refer for more details.

\subsubsection{Quantum 4-geometry from spin foams}
We already anticipated and discussed a bit the basic idea about the encoding of a 4-geometry in a spin foam: spin foam faces give areas to the triangles they intersect and spin foam edges give volume to the tetrahedra they intersect. It is suspected (and expected) that spin foam vertices give 4-volume to the 4-simplices they are in, but this was not yet rigorously proven. Consider now a triangulated 4-manifold $M$ and a principal bundle $P$ with base $M$ and structure group $Spin(4)$. Let us take a spin foam with this group embedded in $M$, and try to understand how it gives a geometry to $M$. As a start we can consider 3-dimensional (triangulated) submanifolds $S$ of $M$. Considering the splitting $Spin(4)\simeq SU(2)\times SU(2)$, there is a corresponding splitting of the space of connections on $P\mid_{S}$ modulo gauge transformations and in the space of (square integrable) functions of these connections:
\be
L^{2}(\mathcal{A}_{S}/\mathcal{G}_{S})\simeq L^{2}(\mathcal{A}^{+}_{S}/\mathcal{G}^{+}_{S})\otimes L^{2}(\mathcal{A}^{-}_{S}/\mathcal{G}^{-}_{S}).
\ee
On this space we can define various interesting geometrical operators. We have seen in section ~\ref{sec:BC} that in the Barrett-Crane spin foam model the faces of the spin foam $F$ (and correspondingly the triangles of the triangulation) are labelled with pairs of SU(2) spins, and the edges (correspondingly, the tetrahedra in the triangulation) with intertwiners (in turn characterized by another pair of spins). Moreover this spin foam determines a state $F\mid_{S}\in L^{2}(\mathcal{A}_{S}/\mathcal{G}_{S})$ (see sections ~\ref{sec:spnet} and ~\ref{sec:spfo}). So, analogously to what we described in section ~\ref{sec:3geo}, a spin foam determines a 3-geometry by means of its induced state. In particular, there is an area operator $\hat{A}^{+}_{t}$ acting on $L^{2}(\mathcal{A}_{S}/\mathcal{G}_{S})$ for any triangle $t$ in the triangulation of $S$, obtained by tensoring the area operator defined on $L^{2}(\mathcal{A}^{+}_{S}/\mathcal{G}^{+}_{S})$ with the identity on $L^{2}(\mathcal{A}^{-}_{S}/\mathcal{G}^{-}_{S})$; we can define its expectation value in the spin foam to be:
\be
\langle F, \hat{A}^{+}_{t}\,F\rangle\,=\,\langle F\mid_{S}, \hat{A}^{+}_{t}\,F\mid_{S}\rangle
\ee     
and in the same way we can define a 3-volume operator $V^{+}_{T}$ on $L^{2}(\mathcal{A}_{S}/\mathcal{G}_{S})$ for any tetrahedron $T$, with expectation value
\be
\langle F, \hat{V}^{+}_{T}\,F\rangle\,=\,\langle F\mid_{S}, \hat{V}^{+}_{T}\,F\mid_{S}\rangle,
\ee     
and the above quantities are independent of the particular choice of submanifold $S$ in which the triangle $t$ and the tetrahedron $T$ are. 
Let us now focus on a particular submanifold given by the boundary of a (flat) 4-simplex, and try to describe the subspace $\chi\subset L^{2}(\mathcal{A}_{S}/\mathcal{G}_{S})$ (called the \lq\lq Hilbert space of a quantum 4-simplex") of corresponding states. Consider a classical 4-simplex in Euclidean $\mathbb{R}^{4}$, with vertices $0, 1, 2, 3, 4$, and the vertex $0$ at the origin, then fix 4 vectors $e_{1}, e_{2}, e_{3}, e_{4}\in\mathbb{R}^{4}$ corresponding to the edges $01, 02, 03, 04$. Now take a bivector (an element of $\Lambda^{2}\mathbb{R}^{4}$) for each triangle $abc$ in the 4-simplex as:
\be
E_{abc}\,=\,(e_{c}\,-\,e_{b})\,\wedge\,(e_{b}\,-\,e_{a})
\ee
with $e_{0}=0$. The conditions that these bivectors satisfy are: total antisymmetry in the indices, the {\it closure constraints}
\be
E_{abc}\,-\,E_{abd}\,+\,E_{acd}\,-\,E_{bcd}\,=\,0\;\;\;\;\;\;a\,<\,b\,<\,c\,<\,d \label{eq:close}
\ee
for each tetrahedron in the 4-simplex, and the {\it simplicity constraints}
\be
E_{abc}\,\wedge\,E_{a'b'c'}\,=\,0
\ee
holding for triangles $abc$ and $a'b'c'$ sharing at least one edge (so either one edge or all of them); in particular, the case $a'b'c'=abc$ implies that $E_{abc}$ can be written as a wedge product of two vectors in $\mathbb{R}^{4}$. Now using a metric on $\mathbb{R}^{4}$ we can identify $\Lambda^{2}\mathbb{R}^{4}$ with $so(4)^{*}\simeq so(3)\oplus so(3)$, in turn interpretable as a classical phase space equipped with a Poisson structure. We quantize this phase space obtaining the {\it Hilbert space of a quantum bivector}
\be
\mathcal{H}^{+}\otimes\mathcal{H}^{-}\simeq\bigoplus_{j^{+},j^{-}}j^{+}\otimes j^{-}.
\ee
Then we consider the triangulation corresponding to the boundary of the 4-simplex as given by an \lq\lq exploded complex", i.e. a disjoint union of 5 tetrahedra (so that each triangle is oriented counterclockwise), so counting each triangle twice and working with 20 bivectors, and take the Hilbert space 
\be
\bigotimes_{T}\,\left(\,\mathcal{H}^{+}\otimes\mathcal{H}^{-}\,\right)^{\otimes 4},
\ee
where the sum is over the 5 tetrahedra of the 4-simplex. Now we consider the $so(4)$ operators $\hat{E}_{i}(T)$ one for each triangle in the tetrahedron $T$, and pick out, from the Hilbert space above, states $\psi$ for which
\be
\left( \hat{E}_{0}(T)\,+\,\hat{E}_{1}(T)\,+\,\hat{E}_{2}(T)\,+\,\hat{E}_{3}(T)\right)\psi\,=\,0,
\ee
imposing in this way the closure constraints (the difference in signs with respect to (~\ref{eq:close}) is due to the difference in orientations of the triangles in the exploded complex and in the original one).     
This singles out the subspace
\be
\bigotimes_{T}\,\mathcal{T}^{+}\,\otimes\,\mathcal{T}^{-}.
\ee
Now we use the splitting $so(4)=so(3)\oplus so(3)$, giving $\hat{E}_{i}(T)=\hat{E}_{i}^{+}(T)+\hat{E}_{i}^{-}(T)$, and then consider the area operators
\be
\hat{A}_{i}^{\pm}(T)\,=\,\frac{1}{2}\sqrt{\hat{E}_{i}^{\pm}(T)\cdot \hat{E}_{i}^{\pm}(T)},
\ee and pick out states such that
\be \hat{A}_{i}^{\pm}(T)\,\psi\,=\,\hat{A}_{i'}^{\pm}(T')\,\psi
\ee
whenever the i-th triangle of the tetrahedron $T$ coincides with the i'-th triangle of the tetrahedron $T'$.
This characterizes the subspace
\be
\bigoplus_{j}\bigotimes_{T} Inv\left(j_{0}\otimes j_{1}\otimes j_{2}\otimes j_{3}\right) \simeq L^{2}(\mathcal{A}_{S}/\mathcal{G}_{S})      
\ee
where the $j$s are $so(4)$ ($Spin(4)$) representations given by pairs of $su(2)$ spins.
Finally we have to impose the quadratic simplicity constraints
\be
\left( \hat{E}_{i}(T)\wedge \hat{E}_{i'}(T)\right)\,\psi\,=\,0 \;\;\;\;\Leftrightarrow\;\;\;\;\left(\hat{E}_{i}^{+}(T)\cdot \hat{E}_{i'}^{+}(T)\right)\psi=\left(\hat{E}_{i}^{-}(T)\cdot \hat{E}_{i'}^{-}(T)\right)\psi .
\label{eq:simpl} \ee    
These constraints can be expressed in terms of left-handed and right-handed triangle areas as
\be
\hat{A}_{i}^{+}(T)\,\psi\,=\,\hat{A}_{i}^{-}(T)\,\psi
\ee
for the case $i=i'$, and in terms of left-handed and right-handed \lq\lq parallelogram areas" (they are proportional to the areas of the parallelogram with vertices given by midpoints of the edges of the tetrahedron that are contained in either the i-th or i'-th face of it, but not in both) (see ~\cite{Baez}) 
\be
\hat{A}_{ii'}^{\pm}(T)\,=\,\frac{1}{4}\sqrt{\left(\hat{E}_{i}^{\pm}(T)+ \hat{E}_{i'}^{\pm}(T)\right)\cdot\left(\hat{E}_{i}^{\pm}(T)+ \hat{E}_{i'}^{\pm}(T)\right)}
\ee
as
\be
\hat{A}_{ii'}^{+}(T)\,\psi\,=\,\hat{A}_{ii'}^{-}(T)\,\psi
\ee
for the case $i\not= i'$.
In the case $i=i'$ the only solution of the constraints is given by the states $\psi\in L^{2}(\mathcal{A}_{S}/\mathcal{G}_{S})$ for which the representations assigned to the triangles are of the form $j\otimes j$ (simple representations). For the case $i\not= i'$ a solution is the Barrett-Crane intertwiner $Inv_{BC}$, then shown to be unique by Reisenberger, as we said in section ~\ref{sec:BC}. We conclude that {\it the Hilbert space of a quantum 4-simplex} is the (infinite dimensional) subspace of $L^{2}(\mathcal{A}_{S}/\mathcal{G}_{S})$ given by:
\be
\chi\,=\,\left\{ \psi\in L^{2}(\mathcal{A}_{S}/\mathcal{G}_{S})\,:\,\left( \hat{E}_{i}(T)\wedge \hat{E}_{i'}(T)\right)\,\psi\,=\,0 \;\,\,\,\,\,\,\,\,\forall i, i', T\right\}
\ee
and constructed explicitely by Barrett and Crane as
\be
\chi\,=\,\bigoplus_{j}\bigotimes_{T}\,Inv_{BC}^{T}\left( (j_{1},j_{1})\otimes(j_{2},j_{2})\otimes(j_{3},j_{3})\otimes(j_{4},j_{4})\right)_{T}.
\ee  

\subsubsection{The quantum tetrahedron in 4 dimensions}
We have seen in the previous section that the Hilbert space of a
quantum 4-simplex is constructed by quantizing the tetrahedra forming its
boundary, by means of bivectors assigned to their faces and suitable
constraints imposed on them. The result is that a quantum tetrahedron
is characterized uniquely by 4 parameters, i.e. the 4 irreducible
simple representations of $so(4)$ assigned to the 4 triangles in it,
which in turn are interpretable as the (oriented) areas of the
triangles. 
Even if rigorous, this result is geometrically rather puzzling, since the geometry of a tetrahedron is classically determined by its 6 edge lengths, so imposing only the values of the 4 triangle areas should leave 2 degrees of freedom, i.e. a 2-dimensional moduli space of tetrahedra with given triangle areas. For example, this is what would happen in 3 dimensions, where we have to specify 6 parameters also at the quantum level (see section ~\ref{sec:PRTV}). So why does a tetrahedron have fewer degrees of freedom in 4 dimensions than in 3 dimensions, at the quantum level, so that its quantum geometry is characterized by only 4 parameters? The answer was given in \cite{BaezBarr}, to which we refer for more details. The essential difference between the 3-dimensional and 4-dimensional case is represented by the simplicity constraints (~\ref{eq:simpl}) that have to be imposed on the bivectors in 4 dimensions. These have a geometrical significance in that they imply that the all 4 faces of the tetrahedron lie in a common hyperplane, so that after having imposed them we are in some sense back to the 3-dimensional situation. In fact, constraints (~\ref{eq:simpl}) for $i=i'$ (second condition in ~\ref{sec:BC'}) imply that the bivectors come from triangles, being given by wedge products of the \lq\lq edge vectors", while for $i\not= i'$ (third condition in ~\ref{sec:BC'}) they imply that any pair of faces of the tetrahedron meet on an edge (which is a very strong condition in $\mathbb{R}^{4}$, where in general two planes intersect each other only in a point). At the quantum level these additional constraints reduce significantly the number of degrees of freedom for the tetrahedron, as can be shown using geometric quantization \cite{BaezBarr}, leaving us at the end with a 1-dimensional state space for each assignment of simple irreps to the faces of the tetrahedron, i.e. with a unique quantum state up to normalization, as we have just seen above. There is another interesting way to look at this fact. Looking back at the procedure we used in the previous paragraph, we can say that what is necessary for characterizing uniquely (up to normalization) a quantum state for a tetrahedron in 4 dimensions is basically to consider two copies of it, with faces labelled by SU(2) spins and then to impose that the two copies carry the same geometrical information (equality of their respective triangle areas and parallelogram areas). This suffices to identify a state. Then the question is: what is the classical geometry corresponding to this state?   
Analyzing the commutation relations of the quantum operators
corresponding to the triangle and parallelogram area operators, it
turns out (see \cite{Baez, BaezBarr}) that while the 4 triangle area
operators commute with each other and with the parallelogram areas
operators (among which only two are independent), the last ones  have
non-vanishing commutators among themselves. This implies that we are
free to specify 4 labels for the 4 faces of the tetrahedron, giving 4
triangle areas, and then only one additional parameter, corresponding
to one of the parallelogram areas (see section ~\ref{sec:BC1}), so
that only 5 parameters determine the state of the tetrahedron itself,
the other one being completely randomized, because of the uncertainty
principle. 
Consequently, we can say that a quantum tetrahedron does not have a unique metric geometry, since there are geometrical quantities whose value cannot be determined even if the system is in a well-defined quantum state. In the context of the Barrett-Crane spin foam model, this means that a complete characterization of two glued 4-simplices at the quantum level does not imply that we can have all the informations about the geometry of the tetrahedron they share. This is a very interesting example of the kind of quantum uncertainty relations that we can expect to find in a quantum gravity theory, i.e. in a theory of quantum geometry.

\subsection{The evaluation of a simple spin network} \label{sec:eval}
Now we want to give more details and some explicit formulas for the evaluation of a simple spin network, one example of which, as we saw, gives the quantum amplitude for a 4-simplex.
The first one was given \cite{Barr}, for a classical group, and in 4 dimensions (the group is SO(4)). We start by assigning a variable $h_{k}\in SU(2)$ to each vertex of the (underlying 1-complex of the) spin network, then to each edge $e$ connecting the vertices $e(0)$ and $e(1)$ we associate a real function $(-1)^{2j_{e}}Tr \rho^{j_{e}}(h_{e(0)}h_{e(1)}^{-1})$, where $\rho^{j_{e}}$ is the representation matrix for the representation $j_{e}$ of SU(2) associated to the edge $e$. Then the evaluation of the spin network is given by integrating over a copy of SU(2) for each vertex the product of the functions associated to all the edges of the graph:
\be
I\,=\,(-1)^{\sum_{e}2j_{e}}\int_{h\in SU(2)\times....\times SU(2)}\prod_{e}Tr \rho^{j_{e}}(h_{e(0)}h_{e(1)}^{-1}). \label{eq:eval}
\ee 
The integration is done with the Haar measure on SU(2), normalized to one, and ensures that the the tensor product of representations at each vertex contains the trivial representation, and implies some restrictions on the edge labels. This formula can be proven \cite{Barr} to arise from the evaluation of an ordinary SU(2) spin network, taking the product of two copies of it, and then normalizing it in a sensible way, as can be expected from the splitting $Spin(4)=SU(2)\times SU(2)$ and the restriction to simple representations $(j,j)$, with $j\in irrep(SU(2))$.
There is also a very nice geometrical interpretation for the formula we gave. Noting that $SU(2)\simeq S^{3}$, we can think of each variable $h_{k}$ as a unit vector in $\mathbb{R}^{4}$, so that:
\be
Tr \rho^{j_{e}}(h_{e(0)}h_{e(1)}^{-1})\,=\,\frac{\sin(2j_{e}+1)\phi}{\sin\phi} \label{eq:dihe}
\ee
where $\phi$ is the angle between the two vectors $h_{e(0)}$ and $h_{e(1)}$. Consider then a compact triangulated 3-manifold and its dual 1-skeleton (one vertex for each tetrahedron and one edge for each triangle). We regard the $h_{k}\in S^{3}$ at each vertex of the graph as the normal to a 3-dimensional hyperplane associated to a tetrahedron in the triangulation. In the case of the graph dual to the boundary of a 4-simplex (giving the amplitude for a quantum 4-simplex) the five hyperplanes (corresponding to the five vertices of the graph) determine the geometry of the 4-simplex uniquely up to rescaling and translation, and in the same way this is determined uniquely by the ten dihedral angles $\phi$, so that the evaluation $I$ is given by averaging over these geometries.

We now show how simple spin networks can be evaluated as Feynman graphs in an internal space (a group manifold), and the formula (~\ref{eq:eval}) can be generalized to different groups (and higher spacetime dimensions), following \cite{F-K}. This is also relevant for the results to be described in the next sections regarding the derivation of the Barrett-Crane model and the generalized matrix models.

The definition of simple representations we used in describing the Barrett-Crane model can be generalized to an arbitrary group $G$: consider an irreducible representation $J$ of a (semisimple, compact) group $G$ and consider a subgroup $H$ of $G$; if the representation space $V^{J}$ contains vectors invariant under $H$ and if all the operators $\rho^{J}(h)$, $h\in H$, are unitary, then $J$ is a {\it representation of class 1 with respect to $H$}. The crucial property is that these representations can be realized as functions on the homogeneous space $X=G/H$. In particular, if $G=SO(D)$, then $H=SO(D-1)$ and $X=G/H=S^{D-1}$. Now choose an orthonormal basis $P^{J}_{i}(x)$ ($x\in X$) in the representation space $J$ realized in the space of functions on $X$, so that the matrix elements of the group operators are given by 
\be
\rho^{J}(g)^{l}_{m}\,=\,\int_{X}dx\,\bar{P}^{J}_{m}(x)P^{J l}(xg), 
\ee       
so that they are realized as integrals over $X$. Then take the special set of intertwiners (for k-valent vertices), representing the generalization of the Barrett-Crane ones, given by:
\be
\iota(v)^{l_{n+1}...l_{k}}_{m_{1}...m_{n}}=\int_{X}dx\,\bar{P}^{J_{1}}_{m_{1}}(x)...\bar{P}^{J_{n}}_{m_{n}}(x) P^{J_{n+1}l_{n+1}}(x)...P^{J_{k}l_{k}}(x).
\ee
We then introduce the functions
\be
G^{J}(x,y):=\sum_{i}\bar{P}^{J}_{i}(x)P^{J i}(y)
\ee
 satisfying the \lq\lq propagator" property:
\be 
\int_{X}dz\, G^{J}(x,z)\,G^{J}(z,y)\,=\, G^{J}(x,y).
\ee 
Now, the evaluation of the simple spin network with underlying graph $\Gamma$ is given by a Feynman graph with the following Feynman rules: - associate to each edge $e$ of the graph $\Gamma$ the propagator $G^{J_{e}}(x,x')$, where $x$ and $x'$ refer to the two vertices connected by the edge $e$; - take the product of all the propagators and integrate over one copy of $X$ for each vertex. The resulting formula is:
\be
I\,=\,\prod_{v}\int_{X}dx_{v}\,\prod_{e}\,G^{J_{e}}(x,x')
\ee 
The construction can be generalized to the case of open spin networks. So we see that the structure is exactly that of a Feynman diagram on the homogeneous space $X$, in which one has a Green function for each edge and integrates over the possible positions of the vertices. In the particular case of SO(D), as we said, the simple representations are representations of class 1 with respect to SO(D-1), they can be realized as functions $L^{2}(S^{D-1})$ and are characterized by a single integer parameter $N$ (see \cite{V-K}). Moreover we have an explicit formula for the propagators, that are given by:
\be
G^{(D)}_{N}(x,y)=\frac{D+2N-2}{D-2}\,C_{N}^{\frac{D-2}{2}}(x\cdot y),
\ee
where $C_{N}^{\frac{D-2}{2}}$ is the Gegenbauer polynomial.
Of course, this reduces to (~\ref{eq:eval}) for $D=4$. 

\subsection{The asymptotics of the Barrett-Crane amplitude} \label{sec:asym}
Now we want to describe some results suggesting that the Barrett-Crane model can indeed be considered as an important step in the construction of a quantum theory of gravity, and since it is really related to general relativity  in more than one aspect.

The first result was obtained in \cite{BarrWill}, where it was shown
that the asymptotics of the Barrett-Crane amplitude for a 4-simplex
gives the cosine of the Regge action for a triangulated 4-manifold, so
contains the amplitude for a semiclassical path integral formulation of discrete general relativity. This is the 4-dimensional analogue of the Ponzano-Regge result discussed in section ~\ref{sec:PRTV}.
The proof goes as follows. We start from the evaluation formula (~\ref{eq:eval}) and its form in terms of dihedral angles using (~\ref{eq:dihe}). We then write $\sin(2j+1)\phi$ in terms of exponentials, and treat the $\sin\phi$ factors as part of the integration measure since they are not affected by the limit $j\rightarrow\infty$ in which we are interested. Introducing a variable $\epsilon_{kl}=\pm 1$ we obtain the expression:
\be
I\,=\,\frac{(-1)^{\sum_{k<l}2j_{kl}}}{(2i)^{10}}\sum_{\epsilon=\pm 1}\int_{h\in SU(2)^{5}}\left( \prod_{k<l}\frac{\epsilon_{kl}}{\sin\phi_{kl}}\right)\,\exp\left(i\sum_{k<l}\epsilon_{kl}(2j_{kl}+1)\phi_{kl}\right)
\ee
Now we compute the stationary points in the action
$S=\sum\epsilon_{kl}(2j_{kl}+1)\phi_{kl}$, with the angles $\phi$
constrained to satisfy $\sum
A_{kl} d\phi_{kl}=0$ (Schl\"{a}fli differential identity), where the $A_{kl}$ are the areas of the triangles of one of the geometric 4-simplices determined by the $\phi$ (i.e. fixing a particular scale). 

The result is that, for each stationary phase point the $\epsilon_{kl}$ are all positive or all negative; the $\phi$ are the dihedral angles for a geometric 4-simplex with triangle areas $A_{kl}=2j_{kl}+1$; the integrand is $\exp{i\mu S_{E}}$, where $S_{E}=\sum A_{kl}\phi_{kl}$ is the Regge calculus version of the Einstein action for a 4-simplex, and $\mu=\pm 1$.
Since each stationary phase point occurs $2^{5}$ times, the final asymptotic formula (considering only non-degenerate simplices) is:
\be
I\,\simeq\,-\frac{(-1)^{\sum 2j_{kl}}}{2^{4}}\left( \sum_{\sigma} P(\sigma)\,\cos\left( S_{E}(\sigma)\,+\,\kappa\frac{\pi}{4}\right)\right)
\ee
where the sum is over the (finite) set of metric 4-simplices $\sigma\subset\mathbb{R}^{4}$ modulo isometries, such that the areas of the triangle $kl$ are given by $2j_{kl}+1$, and $P(\sigma)$ is a non-oscillatory and explicitely computable function.

This result is very important, but it is worth noting that it relies
on the assumption that the dominant contribution to the value of the
Barrett-Crane amplitude, at least in the limit of large \lq\lq spins",
is given by non-degenerate configurations, i.e. configurations in
which the ten spins the ten areas of the 4-simplex, and that this
assumption has still to be verified by explicit (analitical or
numerical) calculations (see \cite{Num1, Num2} for the first numerical
work on spin foam models), and preliminary numerical results tend to disprove it\footnote{Baez, private communication}. However, the failure of this assumption would not be necessarily a problem for the Barrett-Crane model, because it may well be the case that large 4-simplices tend to be degenerate, but the whole partition function is dominated by configurations with a huge number of very small, non-degenerate 4-simplicies, and not by those with a small number of degenerate ones. Only extensive calculations can then settle this issue.

\subsection{Barrett-Crane model, Plebanski action, BF theory and general relativity} \label{sec:BFGR}
The second important result suggesting that the Barrett-Crane spin foam model is indeed related to general relativity at the classical level was obtained in \cite{DP-F}, where it was shown that the Barrett-Crane constraints on the bivectors can be thought of as the quantum analogue of the constraints that reduce BF theory to general relativity in the Plebanski formulation of it, which is equivalent at the classical level with the usual Einstein-Hilbert one. We now review briefly the relationship between the Plebanski action, BF
theory and general relativity. The so(4)-Plebanski action \cite{Pleb,C-D-J} (without cosmological constant) is 
given by:
\be 
S\,=\,S(\omega,B,\phi)\,=\,\int_{\mathcal{M}}\left[ B^{IJ}\,\wedge\,F_{IJ}(\omega)\,-\frac{1}{2}\phi_{IJKL}\,B^{KL}\,\wedge\,B^{IJ}\right]
\ee
where $\omega$ is an so(4)-valued connection 1-form, $\omega=\omega_{\mu}^{IJ}X_{IJ}dx^{\mu}$, $X_{IJ}$ are the generators of so(4), $F=d\omega$ is the corresponding two-form curvature, $B$ is an so(4)-valued 2-form, $B=B_{\mu\nu}^{IJ}X_{IJ}dx^{\mu}\wedge dx^{\nu}$, and $\phi_{IJKL}$ is a Lagrange multiplier. The associated equations of motion are:
\bea
\frac{\delta S}{\delta \omega}\rightarrow \mathcal{D}B\,=\,dB\,+\,[\omega,B]\,=\,0 \\
\frac{\delta S}{\delta B}\rightarrow F^{IJ}(\omega)\,=\,\phi^{IJKL}B_{KL} \\
\frac{\delta S}{\delta \phi}\rightarrow B^{IJ}\,\wedge\,B^{KL}\,=\,e\,\epsilon^{IJKL} \label{eq:constrB}
\eea
where $e=\frac{1}{4!}\epsilon_{IJKL}B^{IJ}\wedge B^{KL}$.

Thus it is evident that this theory is like a BF topological field
theory (which would be given by the first term in the above action alone), with a type of source term 
and with a non-linear constraint on the B field. In turn the relation with gravity arises because the 
constraint (~\ref{eq:constrB}) is satisfied if and only if there exists a real tetrad field 
$e^{I}=e^{I}_{\mu}dx^{\mu}$ so that one of the following equations holds:
\bea &I&\;\;\;\;\;\;\;\;\;\;B^{IJ}\,=\,\pm\,e^{I}\,\wedge\,e^{J} \label{eq:degsol} \\ 
&II&\;\;\;\;\;\;\;\;\;\;B^{IJ}\,=\,\pm\,\frac{1}{2}\,\epsilon^{IJ}  _{KL}\,e^{K}\,\wedge\,e^{L}.   
\eea

If we restrict the field B to be always in the sector II (with the plus sign), and substitute the 
expression for B in terms of the tetrad field into the action, we obtain:
\be 
S\,=\,\int_{\mathcal{M}}\,\epsilon_{IJKL}\,e^{I}\,\wedge\,e^{J}\,\wedge\,F^{KL}
\ee
which is just the action for general relativity in the first order Palatini formalism.

This restriction on the B field is always possible classically, so the
two theories do not differ at the 
classical level, but they are different at the quantum level, since in the quantum theory one cannot
 avoid interference between different sectors. In fact in a partition function for the Plebanski action we have to integrate over all the possible values of the $B$ field, so considering all the 4 sectors. Another way to see it is the existence in the Plebanski action of a $\mathbb{Z}_{2}\times\mathbb{Z}_{2}$ symmetry $B\rightarrow -B$, $B\rightarrow *B$ responsible for this interference. This is discussed in \cite{DP-F}.

It was shown in \cite{DP-F} that a discretization of the constraints (~\ref{eq:constrB}) which give 
gravity from BF theory prove that they are the classical 
analogue of the Barrett-Crane constraints. 

The discretization goes as follows. Consider a triangulation of our 4-manifold. We take the $B$ field to be constant inside each 4-simplex ($dB^{IJ}=0$). We can associate an so(4)-element, and consequently a bivector, to each triangle (T) in the 4-simplex by means of the integral
\be
B^{IJ}(T)\,=\,\int_{T}\,B_{\mu\nu}^{IJ}\,dx^{\mu}\wedge dx^{\nu}.
\ee 
Note that with this discretization the first of the classical Barrett-Crane constraints on the 4-simplex is automatically satisfied.
Then we see that by Stokes' theorem the sum of bivectors assigned to the faces of a tetrahedron of a 4-simplex is zero, so we have the closure constraint of Barrett and Crane. Coming to the simplicity constraint, it is possible to prove that the constraint  (~\ref{eq:constrB}) implies:
\be
V(S, S')\,=\,\epsilon_{IJKL}B^{IJ}(S)\,B^{KL}(S')\,=\,\delta_{ij}[B^{(+)i}(S)B^{(+)j}(S')\,-\,B^{(-)i}(S)B^{(-)j}(S')]
\ee
where $S$ and $S'$ are two arbitrary 2-surfaces, $V(S,S')$ is the 4-volume spanned by them, and $B^{(+)i}$ and $B^{(-)i}$ are the self-dual and anti-self-dual parts of the $B$ field. This in turn implies:
\be
\delta_{ij}[B^{(+)i}(T)B^{(+)j}(T)\,-\,B^{(-)i}(T)B^{(-)j}(T)]\,=\,0
\ee 
for each triangle in the 4-simplex, i.e. the simplicity constraint of Barrett and Crane, and
\be
\delta_{ij}[B^{(+)i}(T)B^{(+)j}(T')\,-\,B^{(-)i}(T)B^{(-)j}(T')]\,=\,0
\ee
for each couple of triangles sharing an edge, i.e., together with the previous one, the last of the Barrett-Crane constraints. 

Consequently, we can look at the Barrett-Crane state sum model as a
quantization of the Plebanski action, and so strongly related (even if
somewhat different) to gravity. This conclusion is indeed confirmed by
the results to be discussed in section ~\ref{sec:OW}. 

Moreover, in \cite{BaezBarr} it was shown that the four bivectors assigned to the faces of a tetrahedron satisfy automatically a \lq\lq chirality constraint", implying that they belong to the geometric sector II above and not to the non-geometric I. Consequently we can say that the Barrett-Crane model gives a quantization of this geometric sector, and does not \lq\lq see" the non-geometric one.

Recently
\cite{LivOr} it was also
shown that the Barrett-Crane model arises quite naturally from a spin
foam quantization of the most general BF-type action for gravity,
proposed in \cite{CMPR}, in the sense that a discretization of the
constraints reducing BF theory to gravity in this new action shows their equivalence
with the Barrett-Crane constraints on bivectors, and that a
translation of them into conditions on the representations of the Lie
algebra also gives the Barrett-Crane quantum constraints. This suggests the uniqueness of the Barrett-Crane constraints, in the sense that they arise whenever we express gravity in terms of 2-forms.   

We can identify the crucial fact which allows the Barrett-Crane spin foam quantization of gravity (with Plebanski action) with the structure of the theory as a constrained BF topological one. Consequently it is very interesting, also from the point of view of the generalization of the Barrett-Crane model to higher dimensions, that this formulation of general relativity as a constrained BF theory holds true in any dimension (and in both Euclidean and Lorentzian signature) \cite{F-K-P}.
In fact, consider the action
\be
S(A, B, \Phi)\,=\,\int d^{D}x\,B^{\mu\nu}_{IJ}\,F^{IJ}_{\mu\nu}\,+\,\frac{1}{2}\Phi^{IJKL}_{\mu\nu\rho\sigma}B^{\mu\nu}_{IJ}B^{\rho\sigma}_{KL},
\ee
where, for any dimension $D\geq 4$, $A_{\mu}^{IJ}$ is an SO(D) (or SO(D-1,1)) gauge field whose curvature is $F^{IJ}_{\mu\nu}$, $B^{IJ}_{\mu\nu}$ is an SO(D) (or SO(D-1,1)) bivector field, and $\Phi^{IJKL}_{\mu\nu\rho\sigma}$ is a Lagrange multiplier field.

In order to ensure the relation with gravity, $\Phi$ should be chosen to be completely anti-symmetric in one set of indices (for convenience in the spin foam quantization, the internal ones) and with the anti-symmetrization on the other set (the spacetime ones) vanishing.
The variation of this action with respect to $\Phi$ gives:
\be
\epsilon^{[M]IJKL}B_{IJ}^{\mu\nu}B^{\rho\sigma}_{KL}\,=\,\epsilon^{[\alpha]\mu\nu\rho\sigma}c^{[M]}_{[\alpha]}, \label{eq:con}
\ee
with coefficients $c$ given by $c^{[M]}_{[\alpha]}=\frac{1}{(D-4)!4!}\epsilon^{[M]IJKL}B^{\mu\nu}_{IJ}B^{\rho\sigma}_{KL}\epsilon_{[\alpha]\mu\nu\rho\sigma}$, and $[M]$ and $[\alpha]$ are cumulative anti-symmetric indices of lenght $D-4$.

It is then proven \cite{F-K-P} that in any dimension $D>4$ a non-degenerate $B$ field satisfies the constraints (~\ref{eq:con}) if and only if there exists a frame field $e^{\mu}_{I}$ such that
\be
B^{\mu\nu}_{IJ}\,=\,\pm\,e\,e^{[\mu}_{I}\,e^{\nu]}_{J}, \label{eq:higsol}
\ee
where $e$ is the (absolute value of the) determinant of $e^{\mu}_{I}$.
We note that the other solutions of the constraints present for $D=4$ (see equation (~\ref{eq:degsol})) are instead absent for $D>4$.

Substituting these solutions (~\ref{eq:higsol}) into the action, we find:
\be
S(A, B(e))\,=\,\pm\,\int d^{D}x\,e\,e^{[\mu}_{I}\,e^{\nu]}_{J}\,F^{IJ}_{\mu\nu}
\ee
which is exactly the standard Palatini action in the first order formalism of general relativity. Thus, we can say that in any dimension general relativity can be formulated as a constrained BF theory (with {\it quadratic} constraints) (incidentally, the same was proposed for supergravity 
\cite{Eza,L-S,L-S2}).

Consequently a spin foam quantization of general relativity, following
Barrett and Crane, in any dimension, seems viable. This can be attempted using the
results of \cite{F-K} described in section ~\ref{sec:eval} and a very
general formalism developed in \cite{FK} (a sketch of this is given
in \cite{OW}).

\subsection{A derivation of the Barrett-Crane model} \label{sec:OW}
We will now show how the Barrett-Crane spin foam model can be derived \cite{OW} from a discretization of
 the SO(4) BF theory, imposing the constraints that reduce this theory to gravity 
(the Barrett-Crane constraints) at the quantum level, i.e. at the level of the representations of 
SO(4) used, and not starting from a discretization of the Plebanski action, i.e. from a constrained
 action at the classical level. This, which is the starting point of the Reisenberger model \cite{Reis2}, is much more complicated because of the
 non-linearity of the constraint on the B-field (similar problems exist for the discretization of the 
BF theory with a cosmological constant, see \cite{O'L}). 

\subsubsection{Constraining of the BF theory and the state sum for a single 4-simplex} \label{sec:1BC}
Consider a piecewise linear 4-dimensional simplicial manifold, which is given by a 
triangulation of the manifold $M$. Consider also the complex 
which is dual to the triangulation, having a vertex for each 4-simplex of the triangulation, an 
edge (dual link) for each tetrahedron connecting the two different 4-simplices that share it, and 
a (dual) face for each triangle in the triangulation (see Fig.7).

\begin{center} 

\includegraphics[width=5cm]{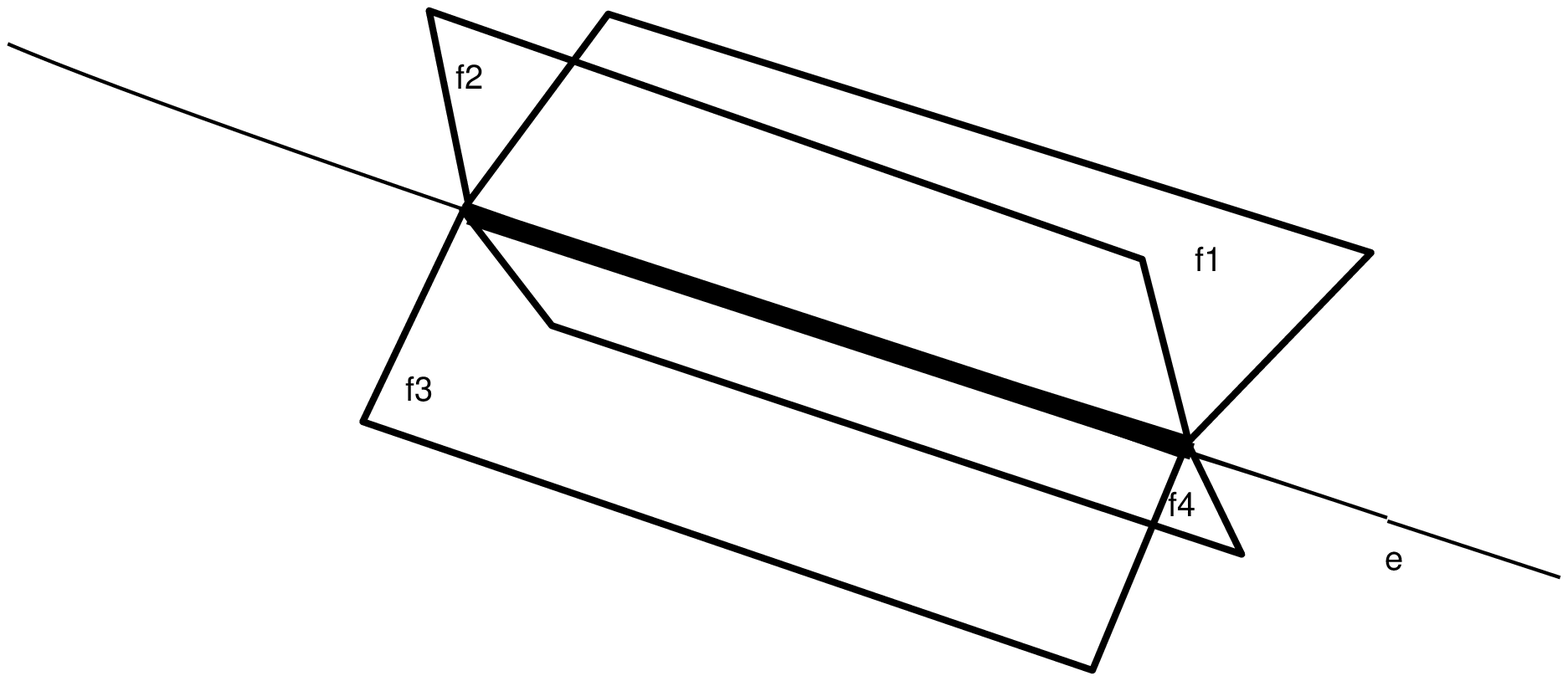}

{\normalsize Figure 7 - A dual edge e with the four dual faces meeting on it}
\end{center}

The 2-dimensional surface bounded by the 
dual links connecting the 4-simplices that share the same triangle is called a dual plaquette. It is 
easy to see that the correspondence between a triangle in the original triangulation and a dual 
plaquette is 1-1 (see Figure 8). 

\begin{center}
\includegraphics[width=5cm]{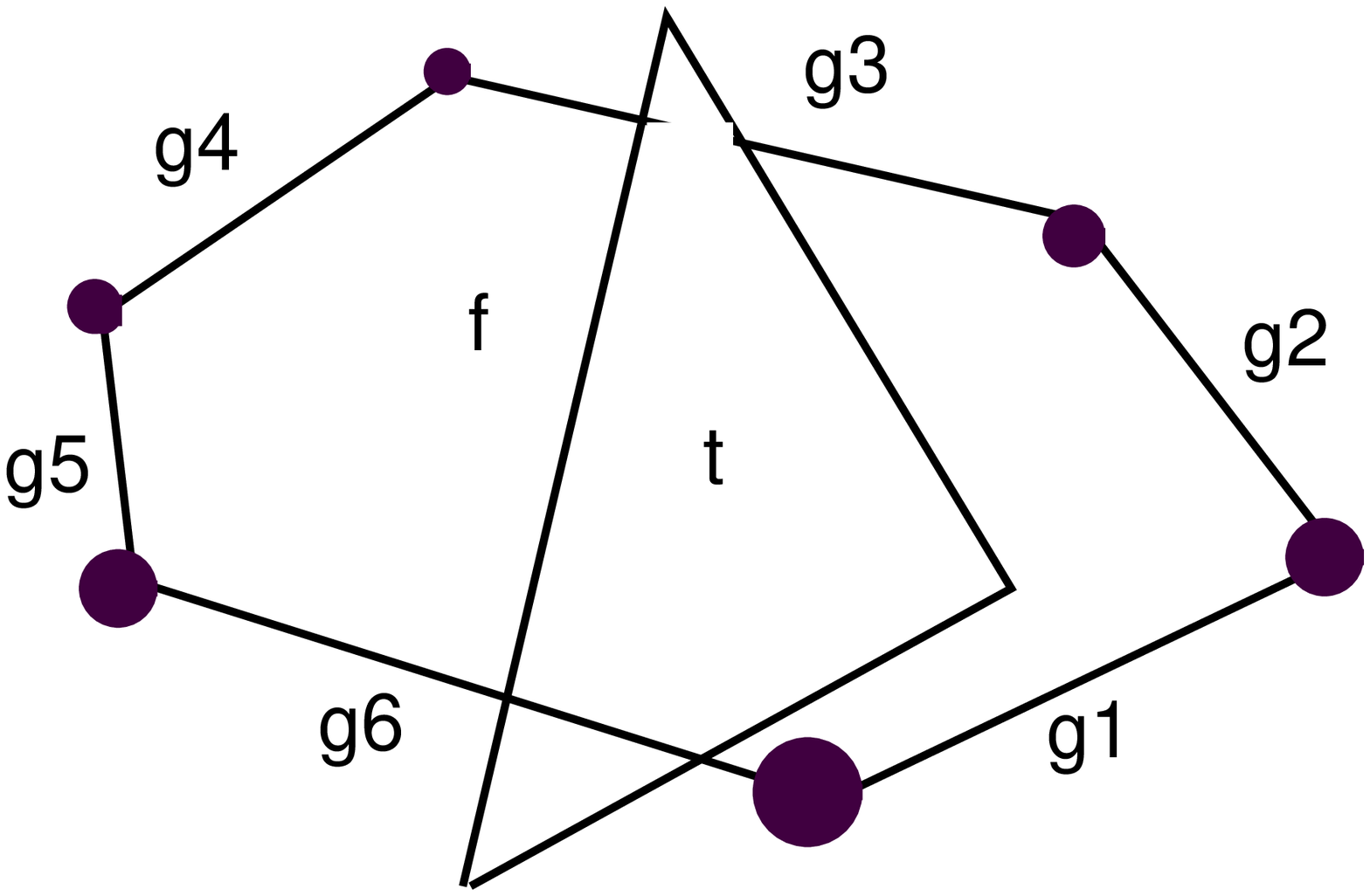}

{\normalsize Figure 8 - The dual plaquette f for the triangle t}
\end{center}

We introduce a dual link variable $g_{e}=e^{i\omega(e)}$ for 
each dual link $e$, through the holonomy of the $so(4)$-connection $\omega$
along the link. Consequently the product of dual link variables along the boundary
$\partial f$ of a dual plaquette $f$ leads to a curvature located at the center of  
the dual plaquette, i.e. at the center of the triangle t:
\be
\prod_{e\in \partial f} g_{e} \equiv e^{i\,F(t)}.
\ee

We then approximate the 2-form field $B$ with a distributional field $B(t)$ with values on the 
triangles of the original triangulation.
Note that this gives an exact theory for a topological field theory like the BF one, but it 
represents only an approximation for a non-topological theory like gravity. Nevertheless this 
approximation would be better and better when we refine the
triangulation, or sum over all the possible different triangulations,
which would be the next step after constructing a spin foam model for
a given triangulation.

Using this discretization procedure \cite{Baez2,Kawa,OW,FK}, we have \cite{OW} the following expression for the discretized
partition function of $Spin(4)$ BF theory:
\be
Z_{BF}(Spin(4))\,=\,\int_{Spin(4)}dg\,\prod_{\sigma}\sum_{J_{\sigma}}\,\Delta_{J_{\sigma}}\,\chi_{J_{\sigma}}(\prod_{e}
g_{e}) \ee
where the first product is over the plaquettes in the dual complex (remember the 1-1 correspondence between triangles and plaquettes), the sum is over (the highest weight of) the 
representations of $Spin(4)$, $\Delta_J$ being their dimension, and $\chi_{J_{\sigma}}(\prod g)$ is the
character (in the representation $J_{\sigma}$) of the product of the
group elements assigned to the boundary edges of the dual plaquette
$\sigma$.

The partition function for the SO(4) BF theory is
consequently obtained by considering only the representation for which
the components of the vectors $J_{\sigma}$ are all integers. 

We see that the integral over the connection
field corresponds to the integral over group elements assigned to the
links of the dual lattice, while the integration over the $B$ field
is replaced by a sum over representations of the group. The meaning of
this can be understood by recalling from the previous sections that, roughly speaking, the $B$ field
turns into a product of tetrad fields after the imposition of the
constraints reducing BF theory to gravity, so giving the geometrical
information about our manifold, and that this information, in spin
foam models, has to be encoded into the algebraic language of
representation theory, and be given by the representations of the
gauge group labelling our spin foam. 

A generic 4-simplex has 5
tetrahedra and 10 triangles in it (see Fig.9).
 Each dual link goes from a
4-simplex to a neighbouring one through the shared tetrahedron, so we
have 5 dual links coming out from a 4-simplex.

\begin{center}

\includegraphics[width=6cm]{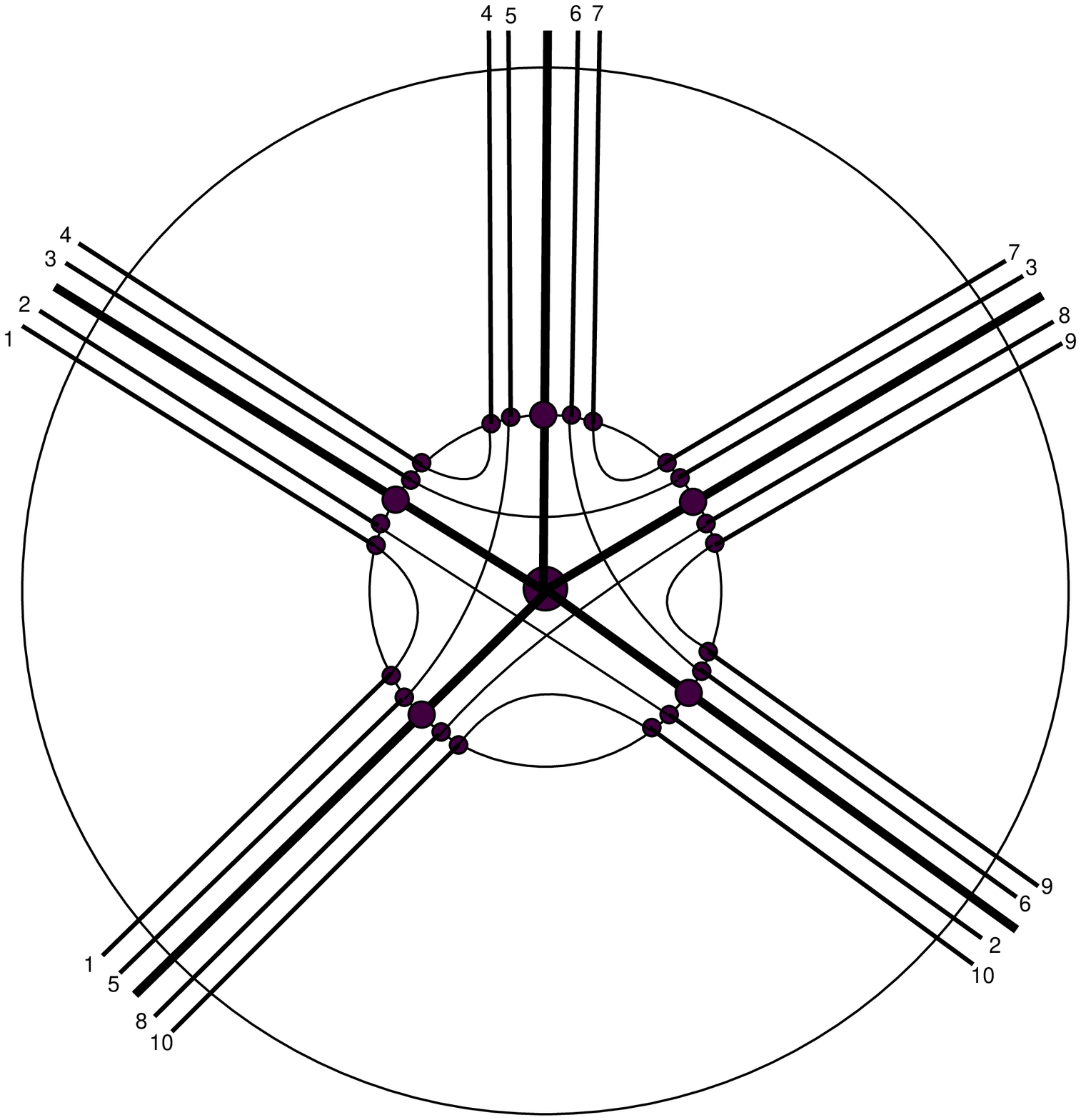}

{\normalsize Fig.9 - Schematic representation of a 4-simplex; the thick lines represent the 5 tetrahedra and the thin lines the triangles}
\end{center}

 We can assign two dual
link variables to each dual link dividing it into two segments going
from the centre of each 4-simplex to the centre of the boundary
tetrahedron, i.e. we assign one group element $g$ to each of them (see
Fig.10).

\begin{center}

\includegraphics[width=5cm]{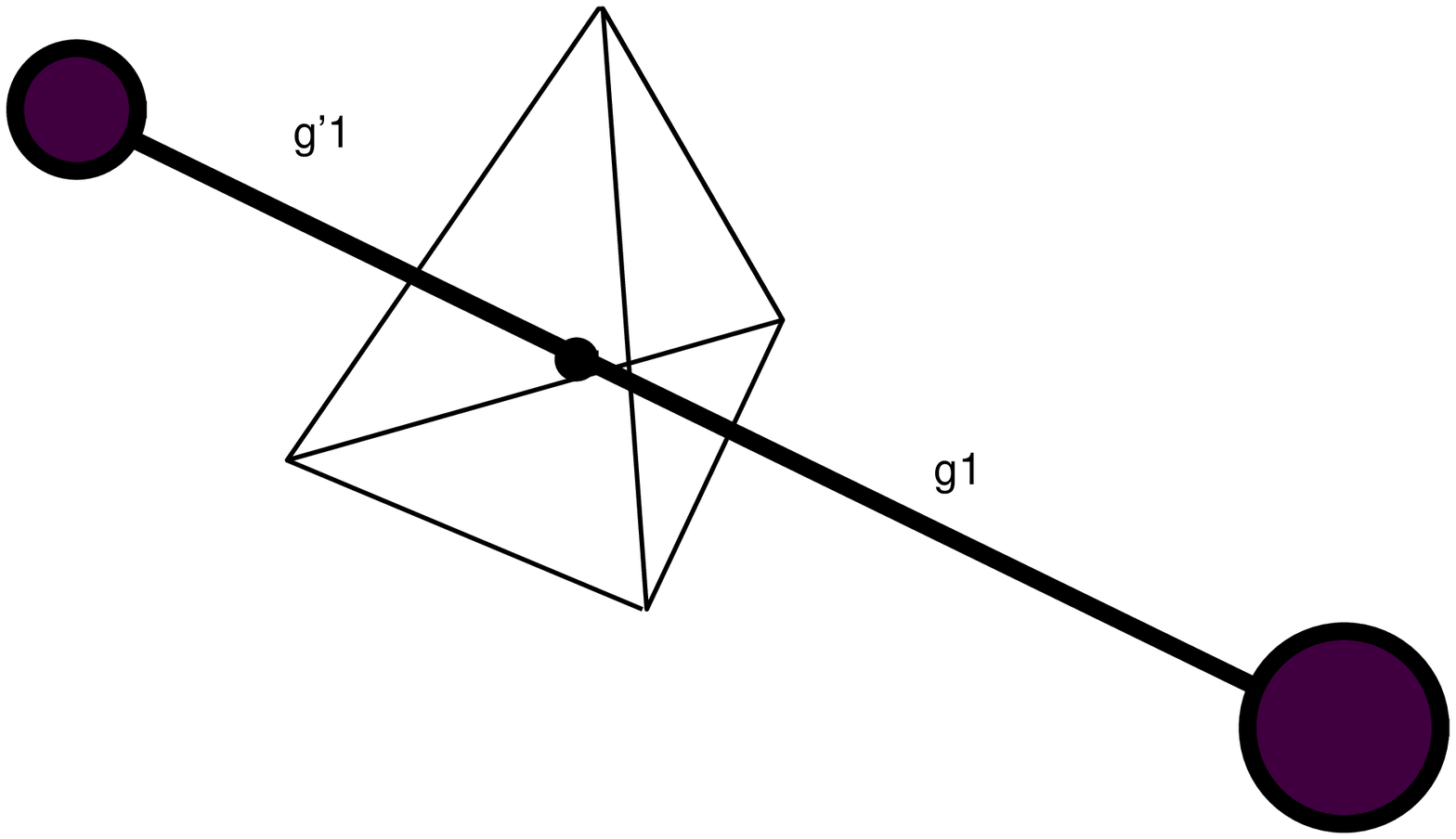} 

{\normalsize Fig.10 - The dual link corresponding to the tetrahedron on which 2 4-simplices meet}
\end{center}
Consider now a dual plaquette. It is given by a number, say, m of dual
links each divided into two segments, so there are 2m dual link
variables on the boundary of each plaquette. When a tetrahedron sharing the triangle to which the plaquette corresponds is on the boundary of the manifold, the plaquette results in being truncated by the boundary, and there will be edges exposed on it (not connecting 4-simplices). To each of these exposed edges we also assign a group variable.

We now make use of the character decomposition formula which
decomposes the character of a given representation of a product of
group elements into a product of (Wigner) D-functions in that representation:
\be 
\chi_{J_{\sigma}}\left(\prod_{\tilde{l}\in \partial\tilde{P}}
g_{e}(\tilde{l})\right)\,=\,\sum_{\{k\}}\prod_{i}D_{k_{i}k_{i+1}}^{J_{\sigma}}\left(g_{e_{i}}\right),\;\;\;{\it
with}\;\;\;k_{1}=k_{2m+1},
\ee
where the product on the $i$ index goes around the boundary of the
dual plaquette surrounding the triangle labeled by $J_{\sigma}$, and
there is a D-function for each group element assigned to a dual link and to the edges exposed on the boundary .
We choose real representations of Spin(4) (this is always possible).

Consider now a single 4 simplex. Note that in this case all the tetrahedra are on the boundary of the manifold, which is
 given by the interior of the 4-simplex.
Writing down explicitly all the products of D-functions and labelling
 the indices appropriately, we can write down the partition function for the
Spin(4) BF theory on a manifold consisting of a single 4-simplex in the
following way:
\bea 
\lefteqn{Z_{BF}(Spin(4))=} \nonumber \\ &=&\sum_{\{J_{\sigma}\},\{k_{e}\}}\left(\prod_{\sigma}\Delta_{J_{\sigma}}\right)\prod_{e}\int_{Spin(4)}dg_{e}\,D_{k_{e1}m_{e1}}^{J_{1}^{e}}D_{k_{e2}m_{e2}}^{J_{2}^{e}}D_{k_{e3}m_{e3}}^{J_{3}^{e}}D_{k_{e4}m_{e4}}^{J_{4}^{e}}\;\left(\prod_{\tilde{e}}D_{il}^{J}\right)\;\;\;\;. 
\eea
The situation is now as follows: we have a contribution for each of the 5 edges of the dual complex, corresponding to 
the tetrahedra of the triangulation, each of them made of a product of the 4 D-functions for the 4 representations 
labelling the 4 faces incident on an edge, corresponding to the 4 triangles of the tetrahedron. There is an extra product 
over the faces with a weight given by the dimension of the representation labelling that face, and the indices of the 
Wigner D-functions refer one to the centre of the 4-simplex, one end of the dual edge, and the other to a tetrahedron 
on the boundary, the other end of the dual edge. There is also an additional product of D-functions, one for each group element assigned to an edge exposed on the boundary, and not integrated over because we are working with fixed connection on the boundary.  

Now we want to go from BF theory to gravity (Plebanski) theory by
imposing the Barrett-Crane constraints on the BF partition
function. These are quantum constraints on the representations of SO(4) which
are assigned to each triangle of the triangulation, so they can be
imposed at this \lq\lq quantum" level.
The constraints are essentially two: the simplicity constraint,
saying that the representations by which we label the triangles are to
be chosen from the simple representations of SO(4) (Spin(4)), and the closure
constraint, saying that the tensor assigned to each tetrahedron has to
be an invariant tensor of SO(4) (Spin(4)). As we have chosen real representations, there is no need to impose the 
first constraint of section ~\ref{sec:BC}, and the third one will be imposed automatically in the following. We can 
implement the second constraint
at this level by requiring that all the representation functions have
to be invariant under the subgroup SO(3) of SO(4), so realizing these representations in the space of harmonic 
functions over the coset ${SO(4)}/{SO(3)}\simeq S^{3}$, which was proven in \cite{F-K-P, F-K} to be a 
complete characterization of the simple representations of SO(D) for
any dimension D, as we discussed previously. We then implement the fourth 
constraint
by requiring that the amplitude for a tetrahedron is invariant under a
general SO(4) transformation.
We note that these constraints have the effect of breaking the topological invariance of the theory.
Moreover, from now on we can replace the integrals over Spin(4) with integrals over SO(4), and the sum with a sum 
over the SO(4) representations only. 

Consequently we write:
\bea 
\lefteqn{Z_{BC}=\sum_{J_{\sigma},\{k_{e}\}}\left(\prod_{\sigma}\Delta_{J_{\sigma}}\right)}
\nonumber \\ && \prod_{e}\int_{SO(4)}dg_{e}\int_{SO(3)}dh_{1}\int_{SO(3)}dh_{2}\int_{SO(3)}dh_{3}\int_{SO(3)}dh_{4}\int_{SO(4)}dg'_{e}
\nonumber \\ && D_{k_{e1}m_{e1}}^{J_{1}^{e}}(g_{e}h_{1}g'_{e})D_{k_{e2}m_{e2}}^{J_{2}^{e}}(g_{e}h_{2}g'_{e})D_{k_{e3}m_{e3}}^{J_{3}^{e}}(g_{e}h_{3}g'_{e})D_{k_{e4}m_{e4}}^{J_{4}^{e}}(g_{e}h_{4}g'_{e}) \left(\prod_{\tilde{e}}D\right)\;\;\;\;\;\;\;\;\nonumber \\ 
&=& \sum_{J_{\sigma},\{k_{e}\}}\left(\prod_{\sigma}dim_{J_{\sigma}}\right)\prod_{e}\,A_{e}\left(\prod_{\tilde{e}}D\right).  
\eea
Let us consider now the amplitude for each edge $e$ of the dual
complex (corresponding to a tetrahedron of the 4-simplex):
\bea
\lefteqn{A_{e}=\int_{SO(4)}dg_{e}\int_{SO(3)}dh_{1}\int_{SO(3)}dh_{2}\int_{SO(3)}dh_{3}\int_{SO(3)}dh_{4}\int_{SO(4)}dg'_{e}\,}
\nonumber \\ 
&& D_{k_{e1}m_{e1}}^{J_{1}^{e}}(g_{e}h_{1}g'_{e})D_{k_{e2}m_{e2}}^{J_{2}^{e}}(g_{e}h_{2}g'_{e})D_{k_{e3}m_{e3}}^{J_{3}^{e}}(g_{e}h_{3}g'_{e})D_{k_{e4}m_{e4}}^{J_{4}^{e}}(g_{e}h_{4}g'_{e})
\eea
for a particular tetrahedron (edge) made out of the triangles
1,2,3,4. Performing explicitely all the integrals, the amplitude for a
single tetrahedron on the boundary turns out to be:
\bea 
A_{e}\,=\,\sum_{simple\;I,L}\frac{1}{\Delta_{J_{1}}\Delta_{J_{2}}\Delta_{J_{3}}\Delta_{J_{4}}}C_{k_{1}k_{2}k_{3}k_{4}}^{J_{1}J_{2}J_{3}J_{4}I}C_{m_{1}m_{2}m_{3}m_{4}}^{J_{1}J_{2}J_{3}J_{4}L}
\nonumber
\\
=\frac{1}{\Delta_{J_{1}}\Delta_{J_{2}}\Delta_{J_{3}}\Delta_{J_{4}}}B_{k_{1}k_{2}k_{3}k_{4}}^{J_{1}J_{2}J_{3}J_{4}}B_{m_{1}m_{2}m_{3}m_{4}}^{J_{1}J_{2}J_{3}J_{4}},
\eea
where the $B$'s are the (un-normalized) Barrett-Crane intertwiners,
defined in \cite{BC} (there are two different normalization currently in use,
see \cite{DP-F-K-R, P-R}), and shown to be unique up to scaling in 
\cite{Reis}, $\Delta_{J}$ is the dimension of the representation $J$ of $SO(4)$, and all the representations for faces and edges in the
sum are now constrained to be simple. Considering the usual decomposition $Spin(4)\simeq SU(2)\times SU(2)$ a representation $J$ of $SO(4)$ corresponds (as we mentioned in section~\ref{sec:BC}) to a pair of $SU(2)$ representations $(j,k)$, so that its dimension is $(2j+1)(2k+1)$. This means that a simple representation $J$ of $SO(4)$ would have dimension $(2j+1)^{2}$, where $j$ is the corresponding $SU(2)$ representation. 

Explicitely, the (un-normalized) Barrett-Crane intertwiners are given by \cite{BC,DP-F-K-R, P-R}:
\be
B_{k_{1}k_{2}k_{3}k_{4}}^{J_{1}J_{2}J_{3}J_{4}}\,=\,\sum_{simple\;I}C_{k_{1}k_{2}k_{3}k_{4}}^{J_{1}J_{2}J_{3}J_{4}I}\,=\,\sum_{simple\;I}\sqrt{\Delta_{I}}C_{k_{1}k_{2}k}^{J_{1}J_{2}I}C_{k_{3}k_{4}k}^{J_{3}J_{4}I}
\ee
where $I$ labels the representation of $SO(4)$ that can be thought as assigned to the tetrahedron whose triangles are instead labeled by the representations $J_{1},...,J_{4}$, $C_{k_{1}k_{2}k_{3}k_{4}}^{J_{1}J_{2}J_{3}J_{4}I}$ is an ordinary $SO(4)$ intertwiner between four representations, and finally $C_{k_{1}k_{2}k}^{J_{1}J_{2}I}$ are Wigner $3j$-symbols.

Note that the simplicity of the representations labelling the
tetrahedra (the third of the Barrett-Crane constraints) comes 
automatically, without the need to impose it explicitly.
We note also that because of the restriction to the simple representations of the 
group, the result we end up with is independent of having started from the Spin(4) or the SO(4) BF partition 
function.

We see that each tetrahedron on the boundary of the 4-simplex contributes with two Barrett-Crane
intertwiners, one with indices referring to the centre of the 4-simplex
and the other indices referring to the centre of the tetrahedron
itself (see Fig.11).
\begin{center}

\includegraphics[width=5.5cm]{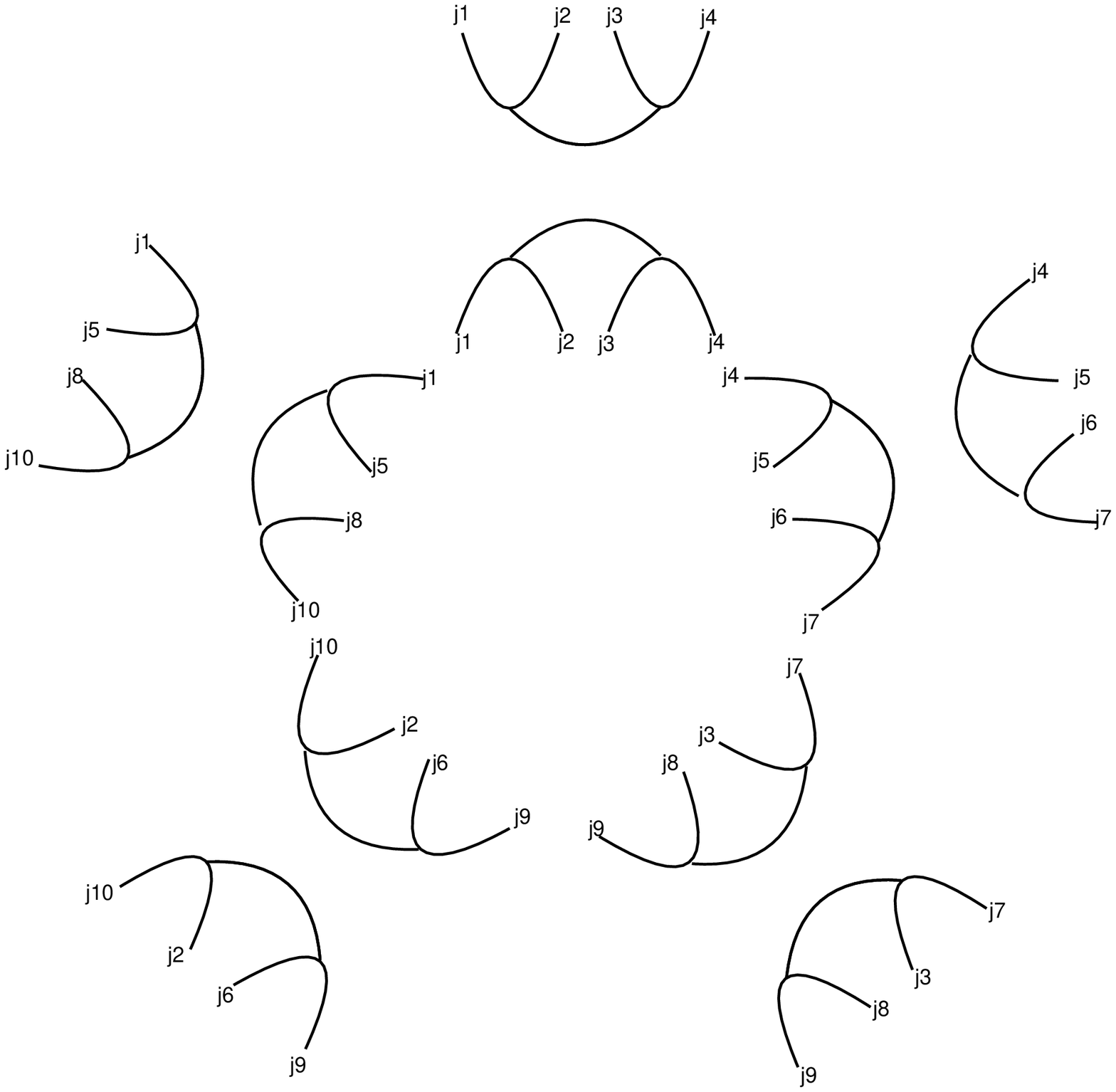}

{\normalsize Fig.11 - Diagram of a 4-simplex, indicating the two Barrett-Crane intertwiners assigned to each tetrahedron}
\end{center}
The partition function for this theory (taking into account all the
different tetrahedra) is then given by:
\bea 
\lefteqn{Z_{BC}=\sum_{\{J\},\{k\},\{n\},\{l\},\{i\},\{m\}}\Delta_{J_{1}}...\Delta_{J_{10}}\frac{1}{(\Delta_{J_{1}}...\Delta_{J_{10}})^{2}}}
\nonumber
\\
&&
B_{k_{1}k_{2}k_{3}k_{4}}^{J_{1}J_{2}J_{3}J_{4}}B_{l_{4}l_{5}l_{6}l_{7}}^{J_{4}J_{5}J_{6}J_{7}}B_{n_{7}n_{3}n_{8}n_{9}}^{J_{7}J_{3}J_{8}J_{9}}B_{h_{9}h_{6}h_{2}h_{10}}^{J_{9}J_{6}J_{2}J_{10}}B_{i_{10}i_{8}i_{5}i_{1}}^{J_{10}J_{8}J_{5}J_{1}} \nonumber
\\ && B_{m_{1}m_{2}m_{3}m_{4}}^{J_{1}J_{2}J_{3}J_{4}}B_{m_{4}m_{5}m_{6}m_{7}}^{J_{4}J_{5}J_{6}J_{7}}B_{m_{7}m_{3}m_{8}m_{9}}^{J_{7}J_{3}J_{8}J_{9}}B_{m_{9}m_{6}m_{2}m_{10}}^{J_{9}J_{6}J_{2}J_{10}}B_{m_{10}m_{8}m_{5}m_{1}}^{J_{10}J_{8}J_{5}J_{1}}\left(\prod_{\tilde{e}}D\right).\;\;\;\;\;
\eea

Now the product of the five Barrett-Crane intertwiners with indices
$m$ gives just the Barrett-Crane amplitude for the 4-simplex to which the
indices refer, given by a 15j-symbol constructed out of the 10 labels of the triangles and the 5 labels of the 
tetrahedra (see Fig.12), 
\begin{center}
\includegraphics[width=5.5cm]{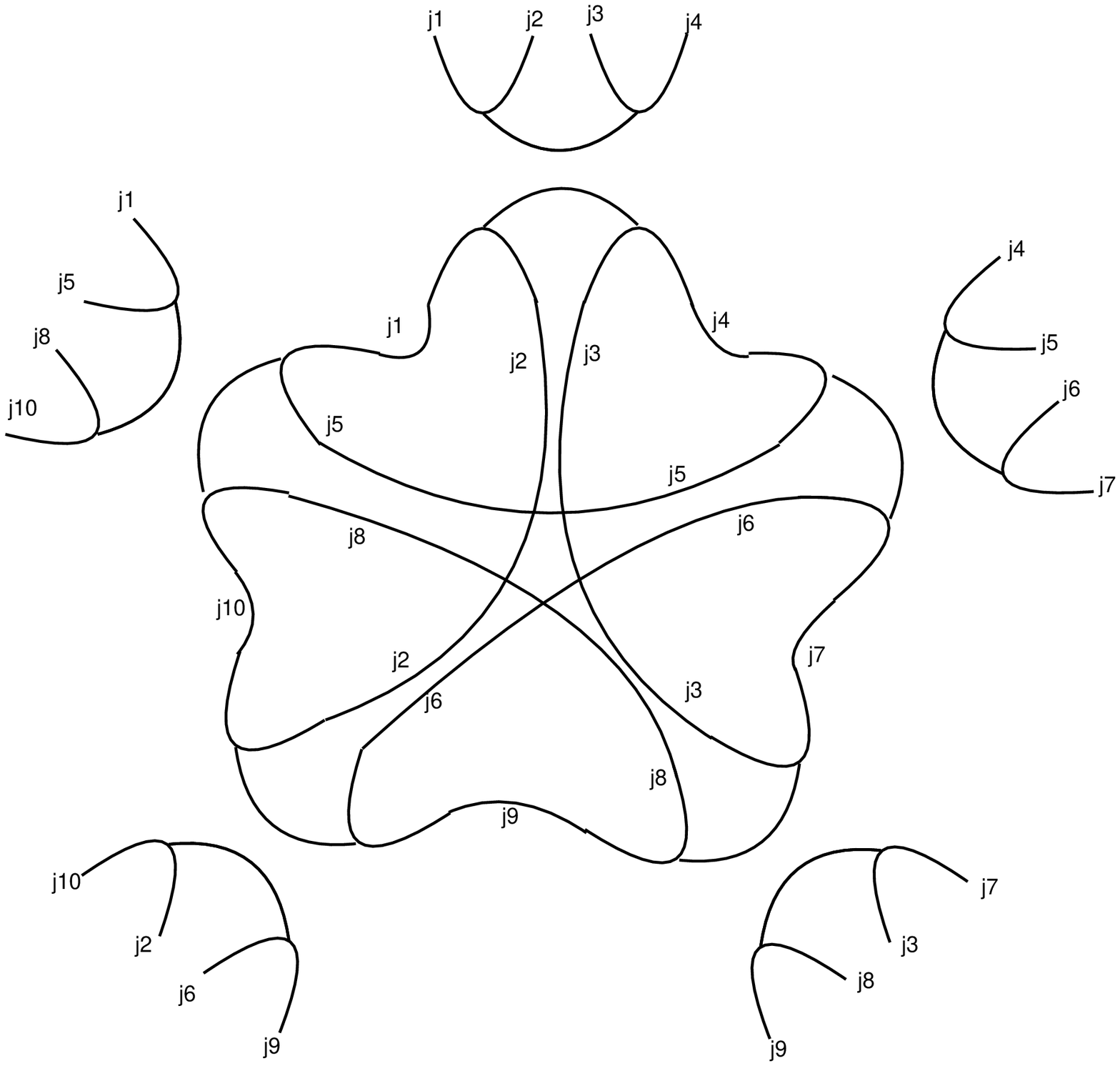}

{\normalsize Fig.12 - Schematic representation of the Barrett-Crane amplitude for a 4-simplex}
\end{center}
so that we can write down explicitly the state sum
for a manifold consisting of a single 4-simplex as:
\be
Z_{BC}=\sum_{\{j_{f}\},\{k_{e'}\}}\prod_{f}\Delta_{j_{f}}\prod_{e'}\frac{B_{k_{e'1}k_{e'2}k_{e'3}k_{e'4}}^{j_{e'1}j_{e'2}j_{e'3}j_{e'4}}}{\Delta_{j_{e'1}}\Delta_{j_{e'2}}\Delta_{j_{e'3}}\Delta_{j_{e'4}}}\prod_{v}\mathcal{B}_{BC}\;\left(\prod_{\tilde{e}}D\right),
\ee
where it is understood that there is only one vertex, $\mathcal{B}_{BC}$ is the Barrett-Crane amplitude for a 
4-simplex, and the notation $e'i$ means that we are referring to the i-th face (in some given ordering) of the 
tetrahedron $e'$, which is on the boundary of the 4-simplex, or equivalently to the i-th 2-simplex of the four which are
 incident to the dual edge (1-simplex) $e'$ of the spin foam (dual 2-complex), which is open, i.e. not ending on any 
other 4-simplex. Also the D-functions for the exposed edges are constrained to be in the simple representation.

\subsubsection{Gluing 4-simplices and the state sum for a general manifold with boundary} \label{sec:BC1} 
Now consider the problem of gluing two 4-simplices together along a common tetrahedron, say, 1234. This important issue was not addressed in the original paper \cite{BC} and of course it is necessary to settle it in order to construct properly a partition function for the whole manifold.

Having the state sum for a single 4-simplex, we consider two adjacent 4-simplices separately, so considering the
common tetrahedron in the interior twice (as being in the boundary of two different 4-simplices), and glue 
them together along it.  

The gluing is done by multiplying the two single partition functions, and
imposing that the values of the spins and of the projections (the $k_{e'i}$'s) of the
common tetrahedron are of course the same in the two partition
functions (this comes from the integration over the group elements assigned to the exposed edges that are being glued and become part of the interior, and thus have to be integrated out).

Everything in the state sum is unaffected by the gluing, except for the
common tetrahedron, which now is in the interior of the manifold.
In this naive sense we could say that this way of gluing is local,
because it depends only on the parameters of the 
common tetrahedron, i.e. it should be determined only by the two
boundary terms which are associated with it when it is 
considered as part of the two different 4-simplices that are being glued.

What exactly happens for the amplitude of this interior tetrahedron
is:
\bea
\sum_{\{m\}}\frac{B_{m_{1}m_{2}m_{3}m_{4}}^{J_{1}J_{2}J_{3}J_{4}}}{\Delta_{J_{1}}\Delta_{J_{2}}\Delta_{J_{3}}\Delta_{J_{4}}}\frac{B_{m_{1}m_{2}m_{3}m_{4}}^{J_{1}J_{2}J_{3}J_{4}}}{\Delta_{J_{1}}\Delta_{J_{2}}\Delta_{J_{3}}\Delta_{J_{4}}}=\sum_{\{m\},I,L}\frac{C_{m_{1}m_{2}m_{3}m_{4}}^{J_{1}J_{2}J_{3}J_{4}I}C_{m_{1}m_{2}m_{3}m_{4}}^{J_{1}J_{2}J_{3}J_{4}L}}{\left(\Delta_{J_{1}}\Delta_{J_{2}}\Delta_{J_{3}}\Delta_{J_{4}}\right)^{2}}\nonumber
\\ = \sum_{I,L}\frac{\sqrt{\Delta_{I}\Delta_{L}}C_{m_{1}m_{2}m}^{J_{1}J_{2}I}C_{m_{3}m_{4}m}^{J_{1}J_{2}I}C_{m_{1}m_{2}n}^{J_{1}J_{2}L}C_{m_{3}m_{4}n}^{J_{3}J_{4}L}}{\left(\Delta_{J_{1}}\Delta_{J_{2}}\Delta_{J_{3}}\Delta_{J_{4}}\right)^{2}}
\nonumber \\ = \sum_{I,L}\frac{\Delta_{I}\delta_{IL}\delta_{mn}}{\left(\Delta_{J_{1}}\Delta_{J_{2}}\Delta_{J_{3}}\Delta_{J_{4}}\right)^{2}} = \sum_{I}\frac{\Delta_{I}}{\left(\Delta_{J_{1}}\Delta_{J_{2}}\Delta_{J_{3}}\Delta_{J_{4}}\right)^{2}} \label{eq:edam}
\eea
where we have used the orthogonality between the intertwiners, and $I$ labels the interior edge (tetrahedron).

We see that the result of the gluing is the insertion of an
amplitude for the tetrahedra (dual edges) in the interior of the
triangulated manifold, and of course the disappearance of the boundary terms $B$ since the tetrahedron is not anymore 
part of the boundary of the new manifold (see Fig.13).
\begin{center}

\includegraphics[width=6cm]{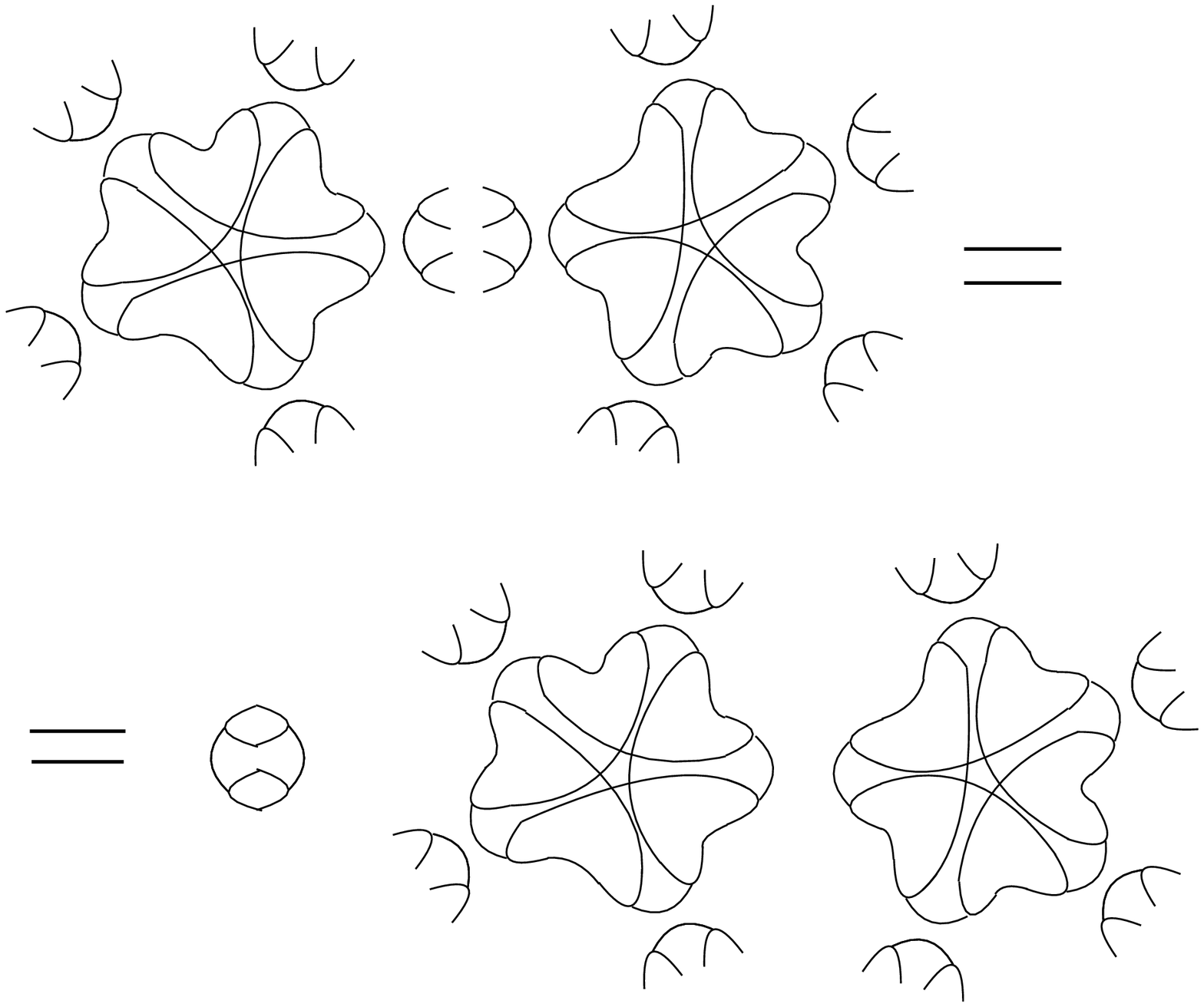}

{\normalsize Fig.13 - The gluing of two 4-simplices along a common tetrahedron}
\end{center}
We can now write down explicitly the state sum for a manifold with
boundary which is then constructed out of an arbitrary number of 4-simplices, and
has some tetrahedra on the boundary and some in the interior:

\be
Z_{BC}=\sum_{\{j_{f}\},\{k_{e'}\},\{J_{e}\}}\prod_{f}\Delta_{j_{f}}\prod_{e'}\frac{B_{k_{e'1}k_{e'2}k_{e'3}k_{e'4}}^{j_{e'1}j_{e'2}j_{e'3}j_{e'4}}}{\Delta_{j_{e'1}}\Delta_{j_{e'2}}\Delta_{j_{e'3}}\Delta_{j_{e'4}}}\prod_{e}\frac{\Delta_{J_{e}}}{\left(\Delta_{j_{e1}}\Delta_{j_{e2}}\Delta_{j_{e3}}\Delta_{j_{e4}}\right)^{2}}\prod_{v}\mathcal{B}_{BC}
\;\left( \prod_{\tilde{e}}D\right),\label{eq:final}
\ee
where the $\{e'\}$ and the $\{e\}$ are the sets of boundary and interior edges of the spin foam, respectively, while the $\tilde{e}$ are the remaining exposed edges.

It is important to note that the number of parameters which determine the gluing and that in the end characterize the 
tetrahedron in the interior of the manifold is five (4 labels for the faces and one for the tetrahedron itself), which is 
precisely the number of parameters necessary in order to determine  a first quantized geometry of a tetrahedron 
\cite{BaezBarr}, as we have seen. 

Moreover, the partition function with which we ended, apart from the boundary terms, is the one obtained in 
\cite{P-R} by studying a generalized matrix model that we will describe in the following, and shown to be
finite at all orders in the sum over the 
representations \cite{P-R,Per}.
 
Several comments are opportune at this point.

The gluing procedure used above is consistent with the formalism developed for general spin foams \cite{Baez,Baez2}, and discussed in section ~\ref{sec:spfo}, saying that when we glue two manifolds $\mathcal{M}$ and $\mathcal{M}'$ along a common 
boundary, the partition functions associated to them and to the composed manifold satisfy 
$Z(\mathcal{M})Z(\mathcal{M}')=Z(\mathcal{M}\mathcal{M}')$, as is easy to verify; see \cite{Baez,Baez2}
 for more details. 

Also, this result suggests that the correct way of deriving a state sum  that implements the 
Barrett-Crane constraints from a generalized matrix model (we will discuss this topic in section ~\ref{sec:matrix}) is like in \cite{P-R}, i.e. imposing the constraints on the 
representations only in the interaction term of the field over a group manifold, because this derivation leads to the 
correct edge amplitudes coming from the gluing, and these amplitudes
are not present in \cite{DP-F-K-R}.
 
Regarding the regularization issue, it seems that even starting from a discretized action in which the sum over the 
representations is not convergent, we end up with a state sum which
is finite at all orders, according to the results of
\cite{P-R,Per}. Anyway another way to regularize completely the state sum model, making it finite at all orders, is to use a quantum group at a root of unity so that the 
sum over the representations is automatically finite due to the finiteness of the number of representations of any such quantum group; in this case we have only to replace the elements of the state sum coming from the recoupling 
theory of SO(4) (intertwiners and 15j-symbols) with the corresponding objects for the quantum deformation of it.     

The structure of the state sum and the form of the boundary terms is a very close analogue of that discovered in 
\cite{CCM1,CCM2} for SU(2) topological field theories in any 
number of dimensions, the difference being the group used, of course, and the absence of any constraints on the 
representations so that the topological invariance is maintained.

Let us stress that the fact that, starting from classical BF theory,
quantizing, and imposing the constraints that are the quantum analogue
of those giving the Plebanski action from the BF one at the classical
level, as discussed in section ~\ref{sec:BFGR}, we obtained precisely
the Barrett-Crane spin foam model, represents a strong confirmation of
the conjecture in \cite{DP-F} that this model gives a quantization of
the Plebanski action.
   
Of course, in order to obtain a complete spin foam model for gravity we should implement a sum over triangulations or over dual 2-complexes, in some form, and both the implementation itself and the issue of its convergence are still open problems, not completely solved yet;  we note that a way to  implement naturally a sum over spin foams giving also a sum over topologies is given by the generalized matrix models that we will discuss in section ~\ref{sec:matrix}. 

Finally, the results reviewed in sections ~\ref{sec:eval} and
~\ref{sec:BFGR} allow for a generalization of this procedure for
deriving the Barrett-Crane spin foam model to any dimension \cite{OW}.
 
\section{Barrett-Crane state sum from a generalized matrix model} \label{sec:matrix}
We want now to describe a recent development which can lead to much progress in the future, i.e. the derivation of spin foam models, and in particular the Barrett-Crane model for quantum gravity, from generalized matrix models or, in other words, from field theories over a group manifold. For a comparison of this approach to quantization with the canonical one and the spin foam procedure in the simple case of 2d BF theory, see \cite{LivPerRov}. Matrix models were invented more than 15 years ago to give a precise formulation of 2d quantum gravity (or \lq\lq zero-dimensional string theory") \cite{D, ADF, KKM, BKKM, DS}, then generalized to 3 dimensions by Boulatov \cite{Boul}, whose model gives the Ponzano-Regge-Turaev-Viro (spin foam) formulation of 3d quantum gravity  (still a topological theory), and to 4 dimensions by Ooguri \cite{Oog}, giving the Crane-Yetter spin foam formulation of 4d BF theory (see section ~\ref{sec:PRTV})\cite{CrYet, CrKauYet}. It could be expected that imposing in a proper way the Barrett-Crane constraints in the Ooguri model would give in the end the Barrett-Crane state sum instead of the Crane-Yetter one, so giving quantum gravity in 4 dimensions instead of a topological theory. This is indeed the case, as we will see. The first derivation of the Barrett-Crane model from a field theory over a group manifold was given in \cite{DP-F-K-R}, and an alternative one was proposed in \cite{P-R}. A general formalism for deriving any kind of spin foam model from a generalized matrix model was developed in \cite{R-R1, R-R2}, proving the power and versatility of the formalism itself, and further study on this topic is in \cite{DP1, DP2}. We will focus on the model of \cite{P-R}, since it reproduces exactly the Barrett-Crane state sum in the form given in section ~\ref{sec:OW}.

Before turning to a more detailed description of a field theory giving the Barrett-Crane state sum, we would like to point out
what are the most interesting aspects of this approach. First of all
the purely algebraic and combinatorial nature of the model, and its
background independence, are made manifest, since the field theory
action is defined in terms of integrals over a group of scalar
functions of group elements only, while its mode expansions involves
only the representation theory of the group used, and its Feynman
graphs (describing \lq\lq interactions") require only notions of
combinatorial topology to be constructed. Then these Feynman graphs
are interpreted as encoding the geometry of a 4-dimensional (Euclidean) spacetime, but no geometric notion enters in their definition. Moreover, this approach permits a natural implementation of a suitable sum over spin foams, i.e. the key ingredient missing in the other derivations and necessary, as we stressed in section ~\ref{sec:BC} and ~\ref{sec:OW}, for a complete definition of a spin foam model for 4-dimensional quantum gravity.    
 
\subsection{The model} \label{sec:mod}   
Let us now consider the generalized matrix model of \cite{P-R}. The
idea is to construct a spin foam model, or a state sum, as the
Feynman expansion of a quantum field theory defined over a group
manifold. In the case we are interested in, the group is $SO(4)$ and
the (scalar) field is a real function over $SO(4)^{4}$, i.e. a function of 4 group elements $\phi(g_{1}, g_{2}, g_{3}, g_{4})$ with $g_{i}\in SO(4)$ (sometimes we will write it simply as $\phi(g_{i})$). It is required (but alternatives can be studied \cite{DP-F-K-R}) to be invariant under any permutation of its arguments. 
We then define the following projector operators:
\be
P_{g}\,\phi(g_{1},g_{2},g_{3},g_{4})\,=\,\int_{SO(4)}d\gamma\,\phi(g_{1}\gamma,g_{2}\gamma,g_{3}\gamma,g_{4}\gamma) 
\ee
and
\be
P_{h}\,\phi(g_{1},g_{2},g_{3},g_{4})\,=\,\int_{SO(3)}dh_{1}...dh_{4}\,\phi(g_{1}h_{1},g_{2}h_{2},g_{3}h_{3},g_{4}h_{4})
\ee
(all the integrals here and in the following are in the normalized Haar measure).
The first projector $P_{g}$ imposes on the field the invariance under the action of $SO(4)$ on its 4 arguments (closure constraint of section~\ref{sec:BC}); the second one $P_{h}$ projects the field over the subspace of fields that are constant on the orbits of $SO(3)$ in $SO(4)$; if we expand the field in modes, i.e. in terms of a sum over the irreducible representations of $SO(4)$, as we will do later, the projection amounts to considering only the representations in which there is a vector invariant under $SO(3)$, i.e. the simple representations (simplicity constraint of section~\ref{sec:BC}).

Taking 10 arbitrary group elements, the action for this field is:
\bea
\lefteqn{S[\phi]\,=\,\frac{1}{2}\,\int\,dg_{1}...dg_{4}\,\left[P_{g}\phi(g_{1}, g_{2}, g_{3}, g_{4})\right]^{2}\,+} \nonumber \\ &+&\,\frac{\lambda}{5!}\,\int \,dg_{1}...dg_{10}\,\left[P_{g}P_{h}\phi(g_{1}, g_{2}, g_{3}, g_{4})\right]\,\left[P_{g}P_{h}\phi(g_{4}, g_{5}, g_{6}, g_{7})\right]\, \nonumber \\ &+&\left[P_{g}P_{h}\phi(g_{7}, g_{3}, g_{8}, g_{9})\right]\,\left[P_{g}P_{h}\phi(g_{9}, g_{6}, g_{2}, g_{10})\right]\,\left[P_{g}P_{h}\phi(g_{10}, g_{8}, g_{5}, g_{1})\right],\;\;\;\;\;\;\;\;\;\;    \label{eq:act}
\eea
so it is given by a quadratic kinetic term without derivatives and a potential term of fifth order, again with no derivative, with a coupling constant $\lambda$.
The partition function of the theory is given, as in the usual quantum field theory literature, by an integral over the field values of the exponential of the action, and can be re-expressed in terms of its Feynman graphs (as an expansions in powers of $\lambda$) as:
\be
Z\,=\,\int\,\mathcal{D}\phi\,e^{-\,S[\phi]}\,=\,\sum_{\Gamma}\frac{\lambda^{v[\Gamma]}}{sym[\Gamma]}\,Z[\Gamma], \label{eq:Z}
\ee
where $v[\Gamma]$ and $sym[\Gamma]$ are the number of vertices and the order of symmetries of the Feynman graph $\Gamma$.
So we are interested in finding the form of the amplitude for a generic Feynman graph, i.e. the Feynman rules of the theory.
It is then convenient to write the action (~\ref{eq:act}) as:
\be
S[\phi]=\frac{1}{2}\int dg_{i}d\tilde{g}_{i}\phi(g_{i})\mathcal{K}(g_{i},\tilde{g}_{i})\phi(\tilde{g}_{i})+\frac{\lambda}{5!}\int dg_{ij}\mathcal{V}(g_{ij})\phi(g_{1j})\phi(g_{2j})\phi(g_{3j})\phi(g_{4j})\phi(g_{5j}),
\ee
with $\phi(g_{1j})=\phi(g_{12},g_{13},g_{14},g_{15})$ and so on. $\mathcal{K}$ and $\mathcal{V}$ are the kinetic operator (whose inverse, in the space of gauge invariant fields, is the propagator of the theory) and the vertex (or potential) operator respectively.
In the present case, the kinetic term in \lq\lq coordinate space" is given by:
\be
\mathcal{K}(g_{i},\tilde{g}_{i})\,=\,\sum_{\sigma}\int d\gamma \,\prod_{i}\delta\left( g_{i}\,\gamma\,\tilde{g}_{\sigma(i)}^{-1}\right), \label{eq:K}
\ee 
which corresponds to a projector in to the space of gauge invariant fields, and is such that it is equal to its inverse, so that the propagator is $\mathcal{K}$ itself. In the formula above the product is over the four arguments of the field and the sum is over the possible permutations of them.

\begin{center}
\includegraphics[width=6cm]{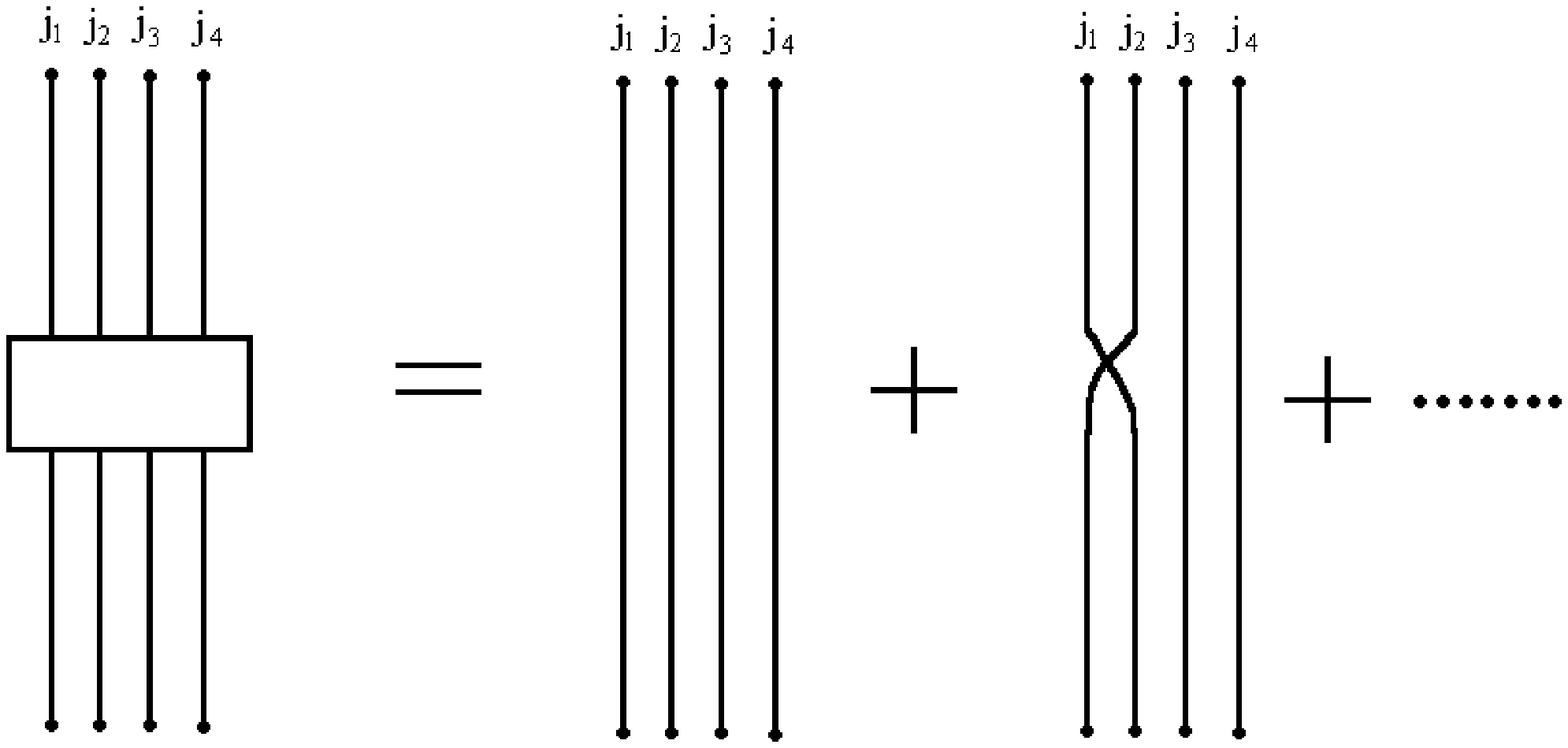}

{\small Fig.14 - The propagator of the theory; each of the four strands carries a (simple) representation of the group and the box stands for a symmetrization of the four arguments}

\end{center}

The vertex operator is:
\be
\mathcal{V}(g_{ij})\,=\,\frac{1}{5!}\,\int d\beta_{i}d\tilde{\beta}_{i}dh_{ij}\,\prod_{i<j}\,\delta\left( g_{ji}^{-1}\tilde{\beta}_{i}h_{ij}\beta_{i}^{-1}\beta_{j}h_{ji}\tilde{\beta}_{j}^{-1}g_{ij}\right) \label{eq:V}
\ee
where $\beta$ and $\tilde{\beta}$ are $SO(4)$ integration variables, and $h_{ij}\in SO(3)$. Again the product is over the arguments of the fields entering  in the interaction term.

\begin{center}
\includegraphics[width=5cm]{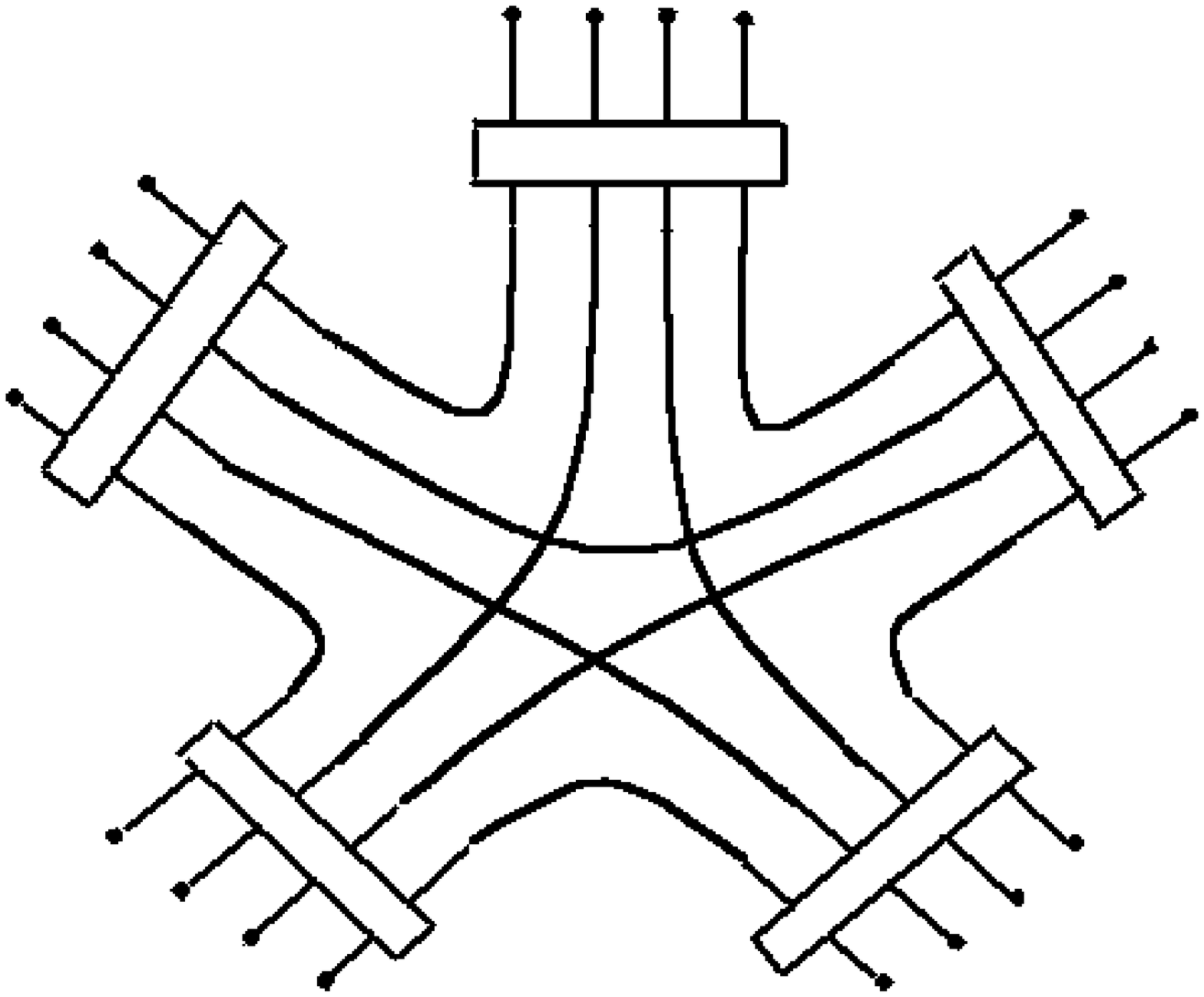}

{\small Fig.15 - The vertex of the theory; it has the combinatorial structure of a 4-simplex and has a (simple) representation of the group at each open end}

\end{center}

The propagator (see Fig.14) can be represented by four straight parallel lines,
each representing a delta function in (~\ref{eq:K}), while the vertex
operator (~\ref{eq:V}) (see Fig.15) has the structure of a 4-simplex, with 5
vertices and 4 lines coming out of each of them, and connecting it
with the others according to the deltas in the expression
above. Everywhere at the open ends of propagators and vertices are the
four group variables that are the arguments of the field. 
All the possible Feynman graphs are obtained (see Fig.16) connecting a number a
vertices (~\ref{eq:V}) with the propagators (~\ref{eq:K}),
constructing in this way what is called a \lq\lq fat graph".

\begin{center}
\includegraphics[width=7cm]{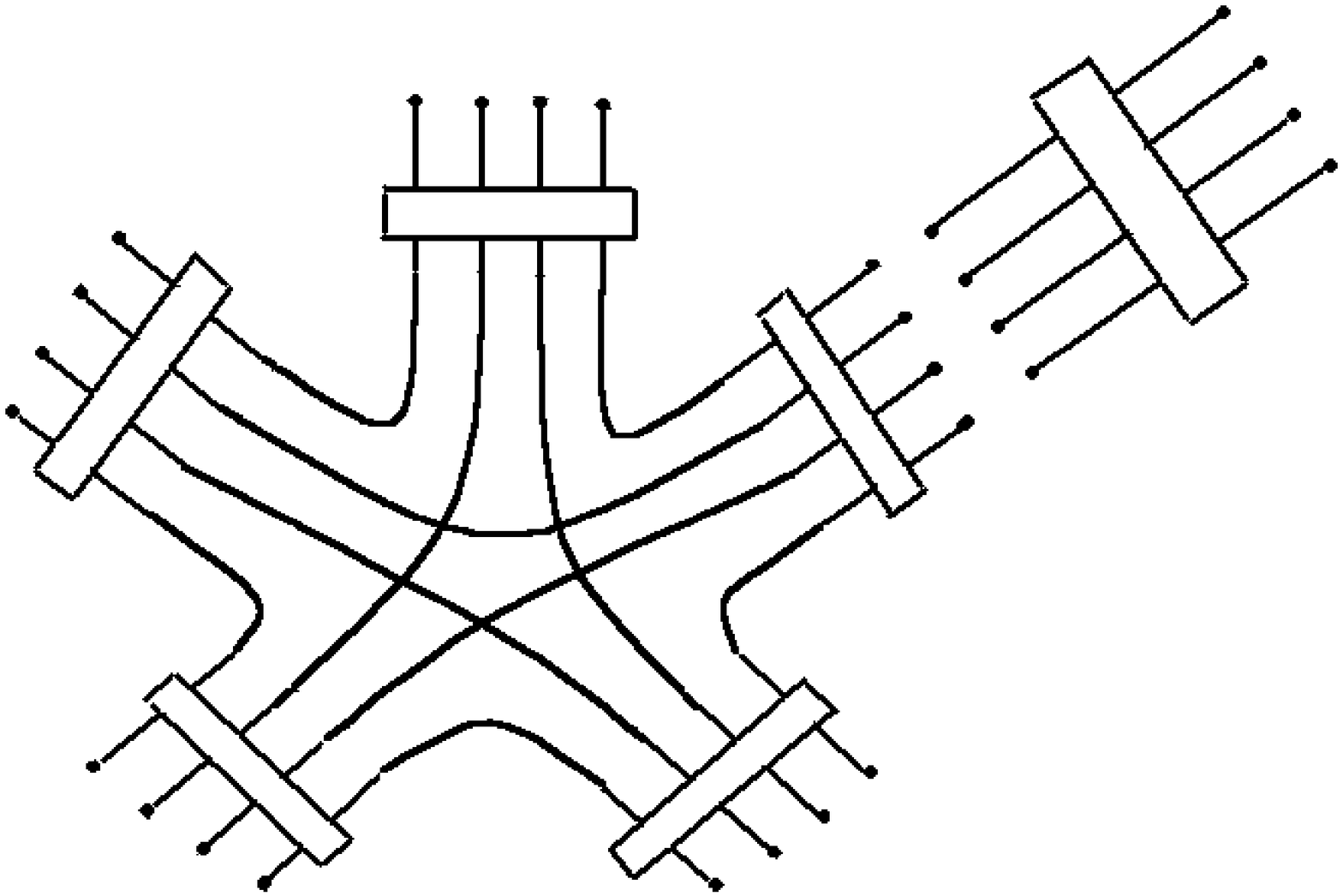}

{\small Fig.16 - The construction of a Feynman graph of the theory, connecting a propagator with a vertex}

\end{center}

Each of the strands of a propagator is connected to a strand in one of
the five \lq\lq open sites" of the vertex, in such a way that the
orientations in vertices and propagators match. Moreover, because of
the symmetry of the field, each propagator really corresponds to many
different terms given by the different ways in which its four carried
indices can be permutated. 
Now, each strand of the fat graph can go through several propagators
and several vertices, and at the end closes on itself, and
consequently forms a cycle. This cycle is labelled by the representation assigned to the strand that encoles it, which is asimple representation of $SO(4)$ because of the projection $P_{h}$ we imposed above. The abstract set formed by these
cycles, the edges and the vertices is a 2-complex, moreover, since
each face (cycle) is labelled by a simple representation of $SO(4)$,
this is a labelled 2-complex, i.e. a spin foam. So we see that there is
a 1-1 correspondence between the Feynman graphs of our field theory
and spin foams. 
This means that the sum in (~\ref{eq:Z}) can be interpreted as a sum over spin foams (labelled 2-complexes), the amplitude $Z(\Gamma)$ as an amplitude for each spin foam, and the partition function itself defines a spin foam model. 
Knowing this, we can interpret the vertices as referring to 4-simplices, and the propagators to the tetrahedra they share.

Now we want to determine the exact expression for $Z(\Gamma)$. In order
to do this, and compare easily the results with the spin foam
formalism, it is convenient to expand the field in modes (which means using the Peter-Weyl decomposition), passing to \lq\lq momentum space". Because of the Peter-Weyl theorem we can
expand any $\mathcal{L}^{2}(G)$ function $\phi(g)$ over the group
$G=SO(4)$ in terms of matrices $D^{\Lambda}_{\alpha\beta}(g)$ of the
irreducible representations $\Lambda$ of $SO(4)$ (repeated indices are
summed over), so we can expand our field $\phi(g_{1},g_{2},g_{3},g_{4})$ as:
\be
\phi(g_{1},g_{2},g_{3},g_{4})=\sum_{J_{1},J_{2},J_{3},J_{4}}\phi^{J_{1}J_{2}J_{3}J_{4}}_{\alpha_{1}\beta_{1}\alpha_{2}\beta_{2}\alpha_{3}\beta_{3}\alpha_{4}\beta_{4}}D^{J_{1}}_{\alpha_{1}\beta_{1}}(g_{1})...D^{J_{4}}_{\alpha_{4}\beta_{4}}(g_{4}).
\ee   
Inserting this expansion into the action (~\ref{eq:act}), using standard results from representation theory as in section~\ref{sec:OW} (for more details see \cite{P-R}) and redefining the field components:
\be
\Phi^{J_{1}J_{2}J_{3}J_{4}\Lambda}_{\alpha_{1}\alpha_{2}\alpha_{3}\alpha_{4}}\,\equiv\,\frac{\phi^{J_{1}J_{2}J_{3}J_{4}}_{\alpha_{1}\beta_{1}\alpha_{2}\beta_{2}\alpha_{3}\beta_{3}\alpha_{4}\beta_{4}}\,C^{J_{1}J_{2}J_{3}J_{4}\Lambda}_{\beta_{1}\beta_{2}\beta_{3}\beta_{4}}}{\left( \Delta_{J_{1}}\Delta_{J_{2}}\Delta_{J_{3}}\Delta_{J_{4}}\right)^{\frac{3}{2}}},
\ee
(the $\Delta_{J}$ are the dimensions of the representations $J$ and the $C$'s are $SO(4)$ intertwiners, as in section~\ref{sec:OW}), 
we have for the kinetic term of the action the expression:
\be
\sum_{J_{1},J_{2},J_{3},J_{4}}\,\Phi^{J_{1}J_{2}J_{3}J_{4}\Lambda}_{\alpha_{1}\alpha_{2}\alpha_{3}\alpha_{4}}\Phi^{J_{1}J_{2}J_{3}J_{4}\Lambda}_{\beta_{1}\beta_{2}\beta_{3}\beta_{4}}\,\left( \Delta_{J_{1}}\Delta_{J_{2}}\Delta_{J_{3}}\Delta_{J_{4}}\right)^{2}\,\delta_{\alpha_{1}\beta_{1}}...\delta_{\alpha_{1}\beta_{1}},
\ee
and for the potential term:
\be
\frac{1}{5!}\sum_{J}\sum_{\Lambda}\Phi^{J_{1}J_{2}J_{3}J_{4}\Lambda_{1}}_{\alpha_{1}\alpha_{2}\alpha_{3}\alpha_{4}}\Phi^{J_{4}J_{5}J_{6}J_{7}\Lambda_{2}}_{\alpha_{4}\alpha_{5}\alpha_{6}\alpha_{7}}\Phi^{J_{7}J_{3}J_{8}J_{9}\Lambda_{3}}_{\alpha_{7}\alpha_{3}\alpha_{8}\alpha_{9}}\Phi^{J_{9}J_{6}J_{2}J_{10}\Lambda_{4}}_{\alpha_{9}\alpha_{6}\alpha_{2}\alpha_{10}}\Phi^{J_{10}J_{8}J_{5}J_{1}\Lambda_{5}}_{\alpha_{10}\alpha_{8}\alpha_{5}\alpha_{1}}\,\mathcal{B}_{J_{1}...J_{10}}^{BC},
\ee
where $\mathcal{B}^{BC}$ is the Barrett-Crane amplitude given by a product of five Barrett-Crane intertwiners as in section~\ref{sec:OW}.
From these expressions we can read off the formulae for the propagator
\be
\mathcal{P}_{\alpha_{1}\beta_{1}...\alpha_{4}\beta_{4}}\,=\,\frac{\delta_{\alpha_{1}\beta_{1}}...\delta_{\alpha_{4}\beta_{4}}}{\left( \Delta_{J_{1}}\Delta_{J_{2}}\Delta_{J_{3}}\Delta_{J_{4}}\right)^{2}}
\ee
and for the vertex amplitude given by the Barrett-Crane amplitude, as we said. Using these formulae, and taking into account the extra sum over the $\Lambda$s, labeling the intertwiners of the 4 representations $J_{1},J_{2},J_{3},J_{4}$, we have for each Feynman graph $J$ the amplitude:
\be
Z(J)\,=\,\sum_{J,\Lambda}\,\prod_{f}\Delta_{J_{f}}\,\prod_{e}\frac{\Delta_{\Lambda_{e}}}{\left( \Delta_{J_{e_{1}}}\Delta_{J_{e_{2}}}\Delta_{J_{e_{3}}}\Delta_{J_{e_{4}}}\right)^{2}}\,\prod_{v}\mathcal{B}^{BC}_{v}
\ee
where the sum is over the simple (because of the projection operators $P_{h}$ imposed above) representations of $SO(4)$ and the products are over the faces, edges and vertices of the Feynman graph, or equivalently over the faces, edges and vertices of the corresponding spin foam.
Moreover, we see that we have obtained precisely the Barrett-Crane state sum model as derived in section~\ref{sec:OW}, with a precise matching of the face, edge and vertex amplitudes. In addition using this procedure we have also implemented in a natural way a sum over spin foams, with weights as in (~\ref{eq:Z}), necessary to restore the full covariance of the theory and get rid of the dependence on the particular triangulation chosen. In fact, under suitable conditions \cite{DP1,DP2}, a fat graph can be put in correspondence with a 4-dimensional triangulated manifold, so that the sum over Feynman graphs is a sum over triangulations.
More precisely \cite{DP1, DP2}, the fat graphs representing Feynman
diagrams of our field theory are in direct correspondence with
2-complexes, as we have seen, that can be in turn interpreted as the
dual 2-skeleton of a triangulation; then the crucial question is
whether this triangulation is a manifold or not, i.e. whether it is
true that each point of it has a closed neighbourhood homeomorphic to
an n-disc $D^{n}$ ($n=4$ in our case). Precise conditions for the
answer to be positive can be found, and translated into conditions on
the fat graphs, and there is reasonable hope that the field theories
leading to the Barrett-Crane model are rich enough to encode such
conditions.

Let us note that, because at each joining of a vertex with a
propagator we sum over all the possible ways of pairing the 4 indices
$\alpha_{i}$, we have obtained also a sum over triangulations with the
same number of simplices but with different topology, so we have also
a natural implementation of a sum over topologies.

Of course this does not mean that it is the \lq\lq right" sum over topologies which is needed in quantum gravity. Much more work has indeed to be done to understand what triangulations are included in the sum, to distinguish precisely between different triangulations of the same manifold and those corresponding really to different manifolds, to avoid overcounting of topologies, and so on, i.e. to understand precisely the physical meaning of the field theory described above.  Moreover, not only there is currently no real understanding of the sum over triangulations generated  by the field theory over the group, but it is also not clear that a sum over triangulations is the correct way of getting rid of the dependence on the triangulation, the other possible way being the refinement of it. Actually there is not even general agreement on the idea that a quantum gravity theory has to implement a sum over topologies, which can be appealing but raises a huge number of questions and difficulties.   
 
\subsection{Generalizations, finiteness and physical interpretation}\label{sec:propfield}
Let us now point out some important features and discuss the physical interpretation of the generalized matrix model we have just described.

First of all we note that it is easy to modify the field theory so that only {\it oriented} 2-complexes are generated , in fact it is enough to require invariance of the field under only even permutations of its arguments and to modify slightly the kinetic term in the action \cite{DP-F-K-R}. Similarly other symmetry properties can be considered leading to different models.
Also, it is possible to construct similar models in higher dimensions, replacing $SO(4)$ with $SO(D)$ and $SO(3)$ with $SO(D-1)$, so that the representations used are realized on the homogeneous space $SO(D)/SO(D-1)\simeq S^{D-1}$, and so obtaining the higher dimensional analogue of the Barrett-Crane model.

A key feature of the model we have described is that the sum over
colorings of the 2-complex, i.e. over the representations used, is
convergent at all orders, according to \cite{P-R, Per}, without using
any quantum deformation of the group, i.e. without any truncation in
the number of representations, or any other regularization technique. This is due basically to the presence of
the edge amplitudes (~\ref{eq:edam}), that we can interpret, following
the results of section~\ref{sec:OW} \cite{OW}, as coming from the
gluing of 4-simplices along tetrahedra, in the partition function
(~\ref{eq:final}). 

More precisely, the following upper bound for the amplitude $Z(J)$ of
an arbitrary 2-complex $J$ was found \cite{Per}:

\be
\mid Z(J)\mid \leq \prod_{f\in J}\sum_{j_{f}}\left(\Delta_{j_{f}}\right)^{-1}\,=\,(\zeta(2)-1)^{F_{J}}\approx(0.6)^{F_{J}},
\ee
where the product is over the faces of the 2-complex, the sum is over
their associated representations, $\zeta$ is the Riemann zeta
function, and $F_{J}$ is the total number of faces in the 2-complex
$J$.

This is a surprising and striking result, since, as we have seen in the previous sections, this sum is interpretable as a sum over geometries for a fixed triangulation of the manifold and a fixed topology, so we have a rigorously defined discrete version of a path integral formulation for Euclidean quantum gravity, or sum over geometries, which moreover is finite at all orders, without any need for regularization. This parallel with path integrals has however to be taken just as an analogy, at the present stage of development of the theory, since the exact relation between sums over spin foams and path integrals over equivalence classes of 4-metrics is not understood.
Of course, there is another potential source of divergences in the model, that is the sum over 2-complexes, that includes a sum over topologies, but the possible convergence of this latter sum was not yet studied in detail (see \cite{DP-F-K-R} for a brief discussion). Beacuse of this potential difficulty, all we can really say is that the field theory defined above is perturbatively finite at all orders, but more work is needed to check that the sum defining the perturbation expansion in Feynman diagrams makes sense.  

Regarding the physical interpretation of the field theory above, note that the field $\Phi^{J_{1}J_{2}J_{3}J_{4}\Lambda}$ depends on 5 parameters, corresponding to the assignment of simple representations to the faces of a tetrahedron and of an intertwiner to the bulk of it. As we discussed extensively in sections~\ref{sec:BC}, ~\ref{sec:geom}, ~\ref{sec:OW}, this determines the geometry of the tetrahedron at the quantum level, so that we can consider  $\Phi^{J_{1}J_{2}J_{3}J_{4}\Lambda}$ as a quantum amplitude for a particular geometry of a tetrahedron. It is the amplitude for an elementary quantum of space. Consequently the field $\phi(g_{1},g_{2},g_{3},g_{4})$ can be seen as a kind of Fourier transform of that amplitude. This suggests a very interesting interpretation for the quantum theory defined by the action (~\ref{eq:act}): it is the second quantization (in the sense of quantum field theory) of a quantum theory of geometry of these quanta of space, in which these are created and annihilated. Four dimensional spacetime is thus interpretable as the Feynman history of creation, destruction and interaction of elementary quanta of space. A similar picture can be made more precise and indeed corresponds to a way to deal with the field theory above \cite{Mikov}. 

\subsection{Quantum gravity observables}
A very difficult (conceptual and technical) problem in quantum gravity
(but the situation in the classical theory is not much better) is the
definition and computation of the observables of the theory (see \cite{Rovel}), which are required
to be fully gauge (diffeomorphism) invariant. The field theory over
a group permits us to define and compute a natural set of observables for
the theory, namely the n-point functions representing transition
amplitudes between eigenstates of geometry, i.e. spin networks (better
s-knots), meaning states with fixed number of quanta of geometry, as
we explained in section ~\ref{sec:spnet}, and as was done first in \cite{Mikov} and then in \cite{P-R-Obs}. Clearly these are purely quantum observables and have no
analogue in the classical theory. 

We start the discussion of these observables, following
\cite{P-R-Obs}, from their definition in the canonical loop quantum
gravity theory, and then turn to their realization in the field theory
over group framework. We stress that the results presented here are very recent and not yet rigorously extablished, so that the presentation will be rather heuristic as well.

The kinematical state space of the canonical theory is given, as we
have seen in section~\ref{sec:spnet}, by a Hilbert space of s-knot states, solutions of the
gauge and diffeomorphism constraints, $\mid s\rangle$, including the
vacuum s-knot $\mid 0\rangle$. We can formally define a projection $P:\mathcal{H}_{diff}\rightarrow\mathcal{H}_{phys}$
from this space to the physical state space of the solutions of the
Hamiltonian constraint, $\mid s\rangle_{phys}=P\mid s\rangle$. The
operator $P$ is assumed to be real, meaning that $\langle s_{1}\cup
s_{3}\mid P\mid s_{2}\rangle=\langle s_{1}\mid P\mid s_{2}\cup
s_{3}\rangle$, and to represent physically the invariance under
exchange of past and future boundaries, which is sensible in the
Euclidean case. The $\cup$ stands for the disjoint union of two s-knots, which is another s-knot. The quantities
\be
W(s,s')\equiv \,\,_{phys}\langle s\mid s'\rangle_{phys}\,=\,\langle s\mid P\mid s'\rangle
\ee
are fully gauge invariant (invariant under the action of all the
constraints) objects and represent transition amplitudes between
physical states.
 
We can introduce in $\mathcal{H}_{phys}$ the operator
\be
\phi_{s}\mid s'\rangle_{phys}\,=\,\mid s\cup s'\rangle_{phys}
\ee
with the properties of being self-adjoint (because of the reality of
$P$) and of satisfying $[\phi_{s},\phi_{s'}]=0$, so that we can define
\be
W(s)\,=\,\,_{phys}\langle 0\mid \phi_{s}\mid 0\rangle_{phys}
\ee
and
\be
W(s,s')\,=\,\,_{phys}\langle 0\mid \phi_{s}\phi_{s'}\mid
0\rangle_{phys}\,=\,W(s\cup s').
\ee
In this way we have a field-theoretic definition of the $W$'s as
n-point functions for the field $\phi$. 

Before going on to the realization of these functions in the context of
the field theory over a group, we point out in which sense they encode
the dynamics of the theory, an important property in our background
independent context.

Consider the linear space of (linear combination of) spin networks
(with complex coefficients), say, $\mathcal{A}$, with elements $A=\sum
c_{s} s$. Defining on
$\mathcal{A}$ an algebra 
product (with values in the algebra itself) as $s\cdot s'=s\cup s'$, a star operation giving, for each
s-knot $s$, an s-knot $s^{*}$ with the same underlying graph and the
edges labelled by dual representations, and the norm as $\mid\mid
A\mid\mid= sup_{s}\mid c_{s}\mid$, then $\mathcal{A}$ acquires the
stucture of a $C^{*}$-algebra (assuming that the product is continuous in a suitably chosen topology, and other important technicalities, which are actually not proven yet). Moreover, $W(s)$ is a linear functional
on this algebra, which turns out to be positive definite. 

This permits the application of the GNS construction \cite{Haag} to reconstruct,
from $\mathcal{A}$ and $W(s)$, a Hilbert space $\mathcal{H}$,
corresponding to the physical state space of our theory, including a
vacuum state $\mid 0\rangle$, and a representation $\phi$ of
$\mathcal{A}$ in $\mathcal{H}$ with $W(s)=\langle 0\mid\phi\mid
0\rangle$.

Clearly $\mathcal{H}$ is just the Hilbert space $\mathcal{H}_{phys}$ of the
canonical theory, so this means that there is the possibility of
defining the physical state space, annihilated by all the canonical
constraints, without making use of the projection operator $P$. The
full (dynamical) content of the theory is given by the $W$ functions,
and these, as we are going to see now, can be computed in a fully
covariant fashion.

Using the field theory described above, its n-point functions are
given, as usual, by:

\be
W(g_{1}^{i_{1}},...,g_{n}^{i_{1}})=\int\mathcal{D}\phi\,\,\phi(g^{i_{1}}_{1})...\phi(g^{i_{n}}_{n})\,e^{-S[\phi]},
\ee
where we have used a shortened notation for the four arguments of the
fields $\phi$ (each of the indices $i$ runs over the four arguments of the field).

Expanding the fields $\phi$ in \lq\lq momentum space'', as in section
~\ref{sec:mod}, we have \cite{P-R-Obs} the following explicit (up to a
rescaling depending on the representations $J_{i}$)
expression in terms of the ``field components''
$\Phi_{J_{1}J_{2}J_{3}J_{4}\Lambda}^{\alpha_{1}\alpha_{2}\alpha_{3}\alpha_{4}}$:
\be
W_{J^{1}_{1}J^{2}_{2}J^{3}_{3}J^{4}_{4}\Lambda^{1}}^{\alpha^{1}_{1}\alpha^{2}_{2}\alpha^{3}_{3}\alpha^{4}_{4}}....._{J^{n}_{1}J^{n}_{2}J^{n}_{3}J^{n}_{4}\Lambda^{n}}^{\alpha^{n}_{1}\alpha^{n}_{2}\alpha^{n}_{3}\alpha^{n}_{4}}\,=\,\int\mathcal{D}\phi\,\,\phi_{J^{1}_{1}J^{2}_{2}J^{3}_{3}J^{4}_{4}\Lambda^{1}}^{\alpha^{1}_{1}\alpha^{2}_{2}\alpha^{3}_{3}\alpha^{4}_{4}}...\phi_{J^{n}_{1}J^{n}_{2}J^{n}_{3}J^{n}_{4}\Lambda^{n}}^{\alpha^{n}_{1}\alpha^{n}_{2}\alpha^{n}_{3}\alpha^{n}_{4}}\,e^{-S[\phi]}.
\ee
However, the W functions have to be invariant under the gauge group
$G$ to which the $g$'s belong, and this requires all the indices $\alpha$ to be 
suitably paired (with the same representations for the paired indices)
and summed over. Each independent choice of indices and of their pairing defines an
independent W function. If we associate a 4-valent vertex to each
$\phi_{J^{1}_{1}J^{2}_{2}J^{3}_{3}J^{4}_{4}\Lambda^{1}}^{\alpha^{1}_{1}\alpha^{2}_{2}\alpha^{3}_{3}\alpha^{4}_{4}}$
with label $\Lambda^{1}$ at the vertex and $J^{i}_{i}$ at the i-th edge,
and connect all the vertices as in the chosen pairing, we see that we
obtain a 4-valent spin network, so that independent  n-point functions
$W$ are labelled by spin networks with n vertices.

To put it differently, to each spin network $s$ we can associate a
gauge invariant product of field operators $\phi_{s}$

\be
\phi_{s}\,=\,\sum_{\alpha}\,\prod_{n}\,\phi_{J^{1}_{1}J^{2}_{2}J^{3}_{3}J^{4}_{4}\Lambda^{1}}^{\alpha^{1}_{1}\alpha^{2}_{2}\alpha^{3}_{3}\alpha^{4}_{4}}.
\ee
This provides us with a functional on the space of spin networks
\be
W(s)\,=\,\int\mathcal{D}\phi\,\phi_{s}\,e^{-S[\phi]}
\ee
that we can use, if positive definite, to reconstruct the full Hilbert space
of the theory, using only the field theory over the group, via the GNS
construction. This is
very important in light also of the difficulties in implementing the
Hamiltonian constraint in the canonical loop quantum gravity approach.

The transition functions between spin networks can be easily computed using a perturbative
expansion in Feynman diagrams. As we have seen above, this turns out
to be given by a sum over spin foams $\sigma$ interpolating between the n spin
networks representing their boundaries, for example:

\be
W(s,s')\,=\,W(s\cup s')\,=\,\sum_{\sigma/\partial\sigma=s\cup
s'}A(\sigma).
\ee  

Given the interpretation of spin foams as quantum 4-geometries,
discussed in detail in the last sections, it is clear that this
represents an implementation of the idea of constructing a
quantum gravity theory as a sum over geometries, as sketched in the
introduction.

For a different way of using the field theoretical techniques
in this context, giving rise to a Fock space of spin networks on which creation and annihilation operators constructed from the field act, see \cite{Mikov}. The main differences between the approach in \cite{Mikov} and the one we described above are the mentioned construction of a Fock space of spin network states and the fact that the spacetime is considered to be discrete from the beginning (the continuum approximation corresponds to the limit in which the states have high occupation number), while in the approach above, as in loop quantum gravity, spin networks are considered as embedded in a continuum manifold, with the discreteness arising when computing the spectrum of geoemtric operators, like area and volume. Morever, in \cite{Mikov}, the perturbative expression for the transition amplitude between states in which one sums over spin foams only for a given number of vertices, and does not sum over this number, is considered as physical meaningful on its own. It would represent the transition amplitude between spin networks for a given number of time steps, each of these corresponding to a 4-simplex of planckian size in the triangulation. In this approach, then, time would be discrete as well, the time variable would be given by the number of 4-simplices, and the transition amplitudes of the theory would be finite. More work is needed however to understand to which extent this idea is viable. 

\section{Lorentzian Barrett-Crane spin foam model} \label{sec:lor}
We want now to discuss briefly the extensions of results 
described in the previous sections to the Lorentzian case, leading to
intriguing spin foam models of Lorentzian quantum gravity.
Again we concentrate on the Barrett-Crane model in 4d, but we point
out that a spin foam model of Lorentzian quantum gravity in 3d,
generalising the Ponzano-Regge-Turaev-Viro state sum, was constructed
by Freidel \cite{Freidel}, using methods similar to those used in
section ~\ref{sec:OW}. Moreover, a very general class of models of
causal evolution of spin networks, sharing many features of spin foam
models, was constructed by Markopoulou and Smolin \cite{M-S, M, M-S2};
the exact relationship between these models and the Lorentzian
Barrett-Crane one, however, is still to be clarified.

As far as the Barrett-Crane model is concerned, the general idea and
interpretation of the model and its construction are analogous to
those in the Euclidean signature. In a sense, then, the difficulties
of the Lorentzian case are mainly technical difficulties, related to
the non-compactness of the Lorentz group as opposed to the rotation
group that is used, being the local gauge group of gravity, in the
construction of the spin foam models. But the conceptual difficulties
are not less important, since causality has to play a major role in
spin foam quantum gravity, as it does in classical gravity, and
both its correct implementation and its consequences are not well
understood at present.

\subsection{The Lorentzian quantum 4-simplex}
The characterization of a quantum 4-simplex embedded in
$\mathbb{R}^{4}$ endowed with a Minkowski metric is based on the same
kind of analysis we reviewed in section ~\ref{sec:BC'} \cite{BC2}. A classical
geometric 4-simplex is completely determined by a set of 10 bivectors
assigned to its 10 triangles, and satisfying the same set of
constraints that we gave above. Again the bivectors $b_{i}$ can be put
in 1-1 correspondence with the elements $L$ of the Lie algebra of the
Lorentz group (using $b=*L$, $*$ being the Hodge star operation). The quantization then is obtained by a choice of a
representation for each of these Lie algebra elements, so replacing
them with operators acting on some representation space. The classical
geometric conditions on bivectors turn into quantum conditions for
these representations. The irreducible representations of
$sl(2,\mathbb{C})$ (which is the double covering of the Lorentz group
itself) are labelled by a pair of numbers $(j, \rho)$, where $j$ is a
(half-)integer, and $\rho$ is a real number, and, in particular, the
simplicity constraint on bivectors translates again into the vanishing
of the second Casimir of the algebra, $C_{2}=2j\rho$, forcing the
representations we use to be in the two series $(j, 0)$ and $(0,\rho)$.

Moreover, in this Lorentzian context we should be able to distinguish
spacelike, null and timelike bivectors. This distinction is given by
the sign of $\langle b, b\rangle$, where $\langle , \rangle$ represents
the inner product in the space of bivectors. We have $\langle b, b\rangle=-
\langle L, L\rangle$ and the quantization of $\langle L, L\rangle$ is given by $C_{1}+1$, where $C_{1}$ is the first Casimir of
the algebra, i.e. $C_{1}=j^{2}-\rho^{2}-1$. Consequently, simple
representations of the type $(j,0)$ correspond to timelike bivectors,
while those of the type $(0,\rho)$ correspond to spacelike ones.

Note that this means that we are incorporating a non-perturbative and
background independent notion of local causality in our spin foam
model of quantum spacetime.

In order to get an amplitude for a 4-simplex we have to construct the
graph dual to its boundary, as in section ~\ref{sec:BC'}, and label
each edge (dual to a triangle) with a simple representation, and each
vertex (dual to a tetrahedron) with an intertwiner, obtaining a
Lorentzian relativistic spin network. The evaluation of this spin
network gives then the desired amplitude. Note that, in general, the
representation labelling the tetrahedra (the intertwiners) can be both
a
simple timelike or a simple spacelike one even if the representations
labelling its triangles (edges) are all spacelike or all timelike,
corresponding geometrically to the fact that spacelike (timelike)
bivectors do not add necessarily to spacelike (timelike) bivectors.

The actual form of the Barrett-Crane model in the Lorentzian case
differ depending on the way we choose to realize our simple
representations, with a consequently different geometrical
interpretation of the resulting model. In fact \cite{V-K} the simple
representations of the Lorentz group can be realized in the space of
square integrable functions on hyperboloids in Minkowski
space. These are: $Q_{1}$, or Lobachevskian space, characterized by
$x^{\mu}x_{\mu}=1$ and $x^{0}>0$, the positive null cone $Q_{0}$, with
$x^{\mu}x_{\mu}=0$ and $x^{0}>0$, and the imaginary Lobachevskian
space $Q_{-1}$, with $x^{\mu}x_{\mu}=-1$. 

The relevant Barrett-Crane models are those using the realizations on
$Q_{1}$, which involves only spacelike representations of the type
$(0,\rho)$, and $Q_{-1}$, which instead involves both timelike $(j,0)$
and spacelike $(0,\rho)$ representations. The fist case was developed
in \cite{BC2,P-R2}, while the second one appeared in \cite{P-R3}. In both cases one can give
an explicit formula for the evaluation of spin networks, and for the
amplitudes appearing in the spin foam partition function, as we will
see in the next section.  

The general expression for the evaluation of closed spin networks in
which only representations of the type $(0,\rho)$ are present is given
by the formula \cite{BC2}: 

\be
\int_{Q_{1}}dx_{2}...dx_{n}\,\prod_{i<j}\,K_{\rho_{ij}}(x_{i},x_{j})
\ee
where a variable $x_{i}\in Q_{1}$ is assigned to each vertex of the spin
network, a representation $(0.\rho_{ij})$ is assigned to each edge $ij$
linking two vertices $i$ and $j$, the integral is over only n-1 of the
(variables assigned to the) vertices (of course the integral is
independent of the choice of which vertex is not considered), and
finally the function $K_{\rho_{ij}}(x_{i},x_{j})$ is given by

\be
K_{\rho_{ij}}(x_{i},x_{j})\,=\,\frac{\sin{\rho_{ij} r}}{\rho_{ij}\,\sinh{r}}
\label{eq:Kp}
\ee
with $r$ the hyperbolic distance between $x_{i}$ and $x_{j}$. 
Moreover, in most of the interesting cases
the evaluation of these spin networks is finite, i.e. they are integrable,
as was proven in \cite{BB2}.
   
The geometric interpretation of the different models is the following.
The model obtained using $Q_{1}$ involves only tetrahedra lying in
spacelike hypersurfaces, thus having only spacelike
faces. Consequently only representations $(0,\rho)$ are present. The
model based on $Q_{-1}$, on the other hand, involves only timelike
tetrahedra, so that its faces can be either timelike or spacelike and
both kinds of simple representations appear.  

\subsection{Field theory over group formulation}
A field theory over group formulation  can be given also for the Lorentzian Barrett-Crane model  \cite{P-R2,P-R3}, using both realizations of the simple
representations on $Q_{1}$ and $Q_{-1}$. 

The procedure is the same as the one explained in section
~\ref{sec:matrix}; we consider a field $\phi(g_{i})=\phi(g_{1},g_{2},g_{3},g_{4})$
over 4 copies of $SL(2,\mathbb{C})$, assumed to be completely
symmetric in its four arguments, and we define the two following
projectors:
\be
P_{g}\phi(g_{i})\,=\,\int_{SL(2,\mathbb{C})}dg\,\phi(g_{1}g,g_{2}g,g_{3}g,g_{4}g)
\ee
imposing the Barrett-Crane closure constraint, and
\be
P_{h^{\pm}}\phi(g_{i})\,=\,\int_{U^{\pm}}..\int_{U^{\pm}}dh_{1}..dh_{4}\,\phi(g_{1}h_{1},g_{2}h_{2},g_{3}h_{3},g_{4}h_{4})
\ee
imposing the simplicity constraint, where $U^{\pm}$ are two subgroups
of $SL(2,\mathbb{C})$; more precisely, $U^{+}=SU(2)$ and
$U^{-}=SU(1,1)\times\mathbb{Z}_{2}$. 

Choosing the first sign in the
projector appearing in the action leads to the Barrett-Crane model
based on $Q_{1}$ and containing only spacelike simple representations
$(0,\rho)$, since $SL(2,\mathbb{C})/SU(2)\simeq Q_{1}$, while the
choice of the second sign in the projector leads to the Barrett-Crane model
on $Q_{-1}$, with both series of simple representations $(0,\rho)$ and
$(j,0)$, since the coset space
$SL(2,\mathbb{C})/SU(1,1)\times\mathbb{Z}_{2}$ is isomorphic to the
hyperboloid $Q_{-1}$ with the opposite points identified.

The action for the theory, analogous to that in
section~\ref{sec:matrix}, is 

\be
S^{\pm}[\phi]\,=\,\int
dg_{i}\,[P_{g}\phi(g_{i})]^{2}\,+\,\frac{\lambda}{5!}[P_{g}P_{h^{\pm}}\phi(g_{i})]^{5},
\ee
where $\lambda$ is an arbitrary real constant, and the fifth power stands for
\be
[\phi(g_{i})]^{5}\equiv\,\phi(g_{1},g_{2},g_{3},g_{4})\phi(g_{4},g_{5},g_{6},g_{7})\phi(g_{7},g_{3},g_{8},g_{9})\phi(g_{9},g_{6},g_{2},g_{10})\phi(g_{10},g_{8},g_{5},g_{1})
\ee
As was shown in section~\ref{sec:matrix} for the Euclidean case, we
can expand the partition function in terms of the Feynman graphs of
the theory (power series in $\lambda$) as in (~\ref{eq:Z}).

Carrying out the same steps as in section~\ref{sec:matrix}, with
particular care in the technical problems coming from the fact that
$SL(2,\mathbb{C})$ is not compact, we can compute the expressions for
the propagator and vertex of the theory, both in ``coordinate'' and in
``momentum'' space, and the amplitude for each Feynman diagram, which
in turn is, as before, in 1-1 correspondence with a combinatorial
2-complex $J$.

The general structure of this amplitude is of course the same spin
foam structure we encountered before, with amplitudes for faces,
edges, and vertices of the 2-complex, and an integral over the
continuous representations instead of a sum, but of course the particular
form of this amplitude and the representations involved are different
for the two cases based on $Q_{1}$ or $Q_{-1}$.
 
In the first case ($Q_{1}$) the amplitude is given by \cite{P-R2,
P-R3}:

\be
A^{+}(J)\,=\,\int_{\rho_{f}}d\rho_{f}\prod_{f}\rho_{f}^{2}\prod_{e}A_{e}^{+}(\rho_{1},\rho_{2},\rho_{3},\rho_{4})\prod_{v}A^{+}_{v}(\rho_{1},...,\rho_{10}) \label{eq:ampllor}
\ee
with
\be
A_{e}^{+}(\rho_{1},\rho_{2},\rho_{3},\rho_{4})\,=\,\int_{Q_{1}} dx_{1}dx_{2}
K^{+}_{\rho_{1}}(x_{1},x_{2})K^{+}_{\rho_{2}}(x_{1},x_{2})K^{+}_{\rho_{3}}(x_{1},x_{2})K^{+}_{\rho_{4}}(x_{1},x_{2})
\nonumber \label{eq:edg}
\ee
and 
\bean 
\lefteqn{A^{+}_{v}(\rho_{1},...,\rho_{10})\,=\,\int_{Q_{1}}dx_{1}...dx_{5}K^{+}_{\rho_{1}}(x_{1},x_{5})K^{+}_{\rho_{2}}(x_{1},x_{4})K^{+}_{\rho_{3}}(x_{1},x_{3})K^{+}_{\rho_{4}}(x_{1},x_{2})}
\\ && K^{+}_{\rho_{5}}(x_{2},x_{5})K^{+}_{\rho_{6}}(x_{2},x_{4})K^{+}_{\rho_{7}}(x_{2},x_{3})K^{+}_{\rho_{8}}(x_{3},x_{5})K^{+}_{\rho_{9}}(x_{3},x_{4})K^{+}_{\rho_{10}}(x_{4},x_{5}).
\eean

Note that all these expression correspond to the evaluation of a
relativistic spin network as defined in the previous section, with $K$
given by (~\ref{eq:Kp}), and
require the dropping of one of the integrals to be regularized. Of
course, only representations of the type $(0,\rho)$ appear.

For the model based on $Q_{-1}$ we have instead:
\be
 A^{-}(J)\,=\,\sum_{j_{f}}\int_{\rho_{f}}d\rho_{f}\prod_{f}(\rho_{f}^{2}+j_{f}^{2})\prod_{e}A_{e}^{-}(\rho_{1},...,\rho_{4};j_{1},...,j_{4})\prod_{v}A^{-}_{v}(\rho_{1},...,\rho_{10};j_{1},...,j_{10}), \label{eq:ampllor2}
\ee
where it is clear that each face of the 2-complex is labelled either
by a representation $(0,\rho)$ or by a representation $(j,0)$, so that
either $j_{f}=0$ or $\rho_{f}=0$ for each face $f$. 
In this case, we have:

\be
A_{e}^{-}(\rho_{1},...,\rho_{4};j_{1},...,j_{4})\,=\,\int_{Q_{-1}} dx_{1}dx_{2}
K^{-}_{j_{1}\rho_{1}}(x_{1},x_{2})K^{-}_{j_{2}\rho_{2}}(x_{1},x_{2})K^{-}_{j_{3}\rho_{3}}(x_{1},x_{2})K^{-}_{j_{4}\rho_{4}}(x_{1},x_{2})
\nonumber \label{edgeamplor}
\ee
and 
\bean 
\lefteqn{A^{-}_{v}(\rho_{1},...,\rho_{10};j_{1},...,j_{10})\,=\,\int_{Q_{-1}}dx_{1}...dx_{5}K^{-}_{j_{1}\rho_{1}}(x_{1},x_{5})K^{-}_{j_{2}\rho_{2}}(x_{1},x_{4})K^{-}_{j_{3}\rho_{3}}(x_{1},x_{3})K^{-}_{j_{3}\rho_{4}}(x_{1},x_{2})}
\\ && K^{-}_{j_{5}\rho_{5}}(x_{2},x_{5})K^{-}_{j_{6}\rho_{6}}(x_{2},x_{4})K^{-}_{j_{7}\rho_{7}}(x_{2},x_{3})K^{-}_{j_{8}\rho_{8}}(x_{3},x_{5})K^{-}_{j_{9}\rho_{9}}(x_{3},x_{4})K^{-}_{j_{10}\rho_{10}}(x_{4},x_{5}).
\eean
Again it is understood that one of the integrations has to be dropped,
and an explicit expression for the functions
$K^{-}_{j_{f}\rho_{f}}(x_{1},x_{2})$ (more complicated than the previous
one) can be given \cite{P-R3}.

The properties of the Euclidean model and its physical interpretation, discussed in section~\ref{sec:propfield}, hold also for the Lorentzian one, including the finiteness.
In fact it was proven \cite{CPR}, using the results of \cite{BB2}, that, for any non-degenerate and finite (finite number of simplices) triangulation, the amplitude (~\ref{eq:ampllor}) is finite, in the sense that the integral over the simple continuous representations of the Lorentz group converges absolutely. This convergence is made possible by the simplicity constraint on the representations and by the presence of the edge amplitude (~\ref{eq:edg}), as in the euclidean case. The analysis of the amplitude (~\ref{eq:ampllor2}) was not yet carried out. 

Since, as we have seen, the amplitude for a given 2-complex is interpretable as a term in a Feynman expansion of the field theory, its finiteness means that the field theory we are considering is finite order by order in perturbation theory, which is truly remarkable for a theory of Lorentzian quantum gravity. This result is even more remarkable if we think that the sum over the representations of the gauge group is a precise implementation of the sum over geometries proposal for quantum gravity we mentioned in the introduction, and that the path integral implementing this in the original approach was badly divergent in the Lorentzian case.  Of course, finiteness only does not mean necessarily correctness and in particular the choice of the edge amplitude ~\ref{eq:edg} has still to be justified by geometrical or physical considerations.

\section{Conclusions}
We want to conclude with a discussion of present limits and possible
developments of the subject, i.e. spin foam models in general and the
Barrett-Crane one in particular.
We start focusing on the latter.

A basic issue to be understood is the triangulation independence of the
model. As we have stressed, gravity in four and higher dimensions is not a topological theory (even if strongly related to
topological BF theory), and to use a fixed triangulation to represent
spacetime corresponds to a partial breaking of its degrees of freedom.
The full content should be recovered either by taking an infinite
refinement of the triangulation and going back to the continuum limit, or
by summing over all the possible triangulations, with suitable and
possibly physically motivated weights. The approach based on a
refinement is more conventional in lattice gauge theory; it is possible of
course also in this context and involves ideas and methods from
lattice gauge theories, statistical mechanics, and piecewise linear
topology, since a change to a different (finer) triangulation is obtained
by the application of the so-called Pachner moves and the issue of the
behaviour of a spin foam model under this is basically the issue of its
behaviour under coarse graining and renormalization procedures. On the
other hand, in some sense, a refinement towards a continuum limit is in
sharp contrast with one of the ideas on which the spin foam approach is
based, that is the fundamental discreteness of spacetime at the Planck
scale, so that a sum over triangulations would be a more sensible way to
define a full theory of quantum gravity. Even more natural in our case is
the idea of summing over 2-complexes, and letting a sum over triangulated
manifolds emerge out of it. As we have seen, the field theory over group
formalism provides a way to obtain this sum over 2-complexes in a
straightforward manner, and also gives a prescription for the weight to be
assigned to each 2-complex, but the geometrical and physical meaning of
this approach has still to be fully investigated and understood. The
development of alternative approaches would also be of much interest. A different suggestion for the solution of this problem of triangulation dependence is given in \cite{Crane}.
Moreover, the convergence of this (or any other) sum over 2-complexes
(triangulations) is a highly non-trivial issue, and a proof of finiteness
or a regularization of it would anyway require a great deal of work.   

A physically realistic theory of quantum gravity has clearly to be
formulated in the Lorentzian signature, so the Lorentzian version of the
Barrett-Crane spin foam model recently constructed is a promising
candidate for such a theory. It encodes a non-perturbative and algebraic
definition of local causality and of course presents all the features that
make spin foam models interesting models of quantum geometry.  On the other hand much more study is needed to put the construction of the model on a more solid basis, in particular regarding the form of the edge amplitudes and the sum over 2-complexes, and to improve the understanding of its physical meaning, e.g. the interpretation in terms of spacelike/timelike surfaces and the effects of local causality.   

The issue of causality in the evolution of spin network was studied in great generality in \cite{M-S, M, M-S2, M2}, where a general class of globally causal models for this evolution was outlined. These models have the structure of spin foam models, and it can be expected that a Lorentzian version of the Barrett-Crane model, probably with additional restrictions necessary to obtain global causality in addition to the local one, will result in a concrete realization of them, but the precise relationship between the Barrett-Crane model and these globally causal ones is not yet clear.  The construction in \cite{M-S, M, M-S2, M2} is also deeply related to the causal set formalism \cite{BLMS} for quantum gravity, with an interesting additional (see the introduction) convergence of ideas.

More generally, much remains to be understood about the quantum geometry emerging from the Barrett-Crane model, both in the Euclidean and Lorentzian signatures, in addition to what we discussed in section~\ref{sec:geom}. 

The crucial issue to settle is however the complete understanding
of the classical limit of the model. 

First of all we lack a clear picture of the emergence of a classical
background from the quantum structures on which the model is
based. The issue is how exactly the fundamental quantum, discrete, foamy and algebraic description of spacetime at the Planck scale as given by the spin foam model (which as we said is still to be clarified anyway) can be replaced in a suitably defined limit by the usual description in terms of a smooth manifold of fixed topology and given metric structure. Until this is extablished in a clear way, any judgement about the success and results of spin foam models can be only provisional. 
Probably the implementation of causality is necessary for this to happen, as suggested by the
results obtained in the context of dynamical triangulations \cite{Amb,
AmbLol}. 
Also the suggestion \cite{M-S, M, M-S2, K-S} that this is basically a
problem  of self-organization of a dynamical percolative system could
be investigated, after the connection between the Lorentzian
Barrett-Crane model and the causal evolution of spin networks cited
above is established. 
Another approach for studying the classical limit of spin foam models is the general and beautiful scheme for a purely algebraic coarse graining of spin foams being developed by Markopoulou \cite{Marko}, based on ideas and methods from the theory of Hopf algebras.  
Here the ambiguous nature of spin foam models as systems which are both quantum geometrical, statistical mechanical, and field theoretical will probably turn out to be essential.  

The connections with general relativity have also to be much strengthened, improving the results discussed in sections ~\ref{sec:asym} and ~\ref{sec:BFGR}, since of course the reduction of a spin foam model to classical gravity in a suitably defined limit is a necessary condition for being regarded as a good quantum gravity model.

Related to this,  the general problem of the perturbation of a spin
foam model needs to be studied and understood. A first work on this
topic is \cite{Baez4}, and a general procedure for extracting an
effective action for the system describing the perturbations of a spin foam model is
outlined in \cite{Smo}. A precise study is missing anyway, not
only for the Barrett-Crane model, but also for the simpler
Ponzano-Regge-Turaev-Viro model. This issue is of crucial importance, since there is no definite proof so far that the Barrett-Crane model possesses truly physical degrees of freedom and admits a sensible linearized approximation describing linearized gravity, maybe coupled with additional fields, i.e. we do not know yet if the model contains gravitons, as is compulsory for a good theory of quantum gravity.  

Let us say something now about possible extensions of the Barrett-Crane model (and of the other proposed spin foam models).

We already mentioned the importance of a precise definition and analysis of causal structures in the spin foam context, e.g. causal relations, causal propagation, horizons, and so on.

Another step in the development of the model is the inclusion of
matter (or gauge fields) in it (recent results on this issue are in
\cite{Mikov2, CraMatt, OriPf}). The construction of the generalized
matrix models gives a path to follow in this direction, since
the coupling of matter fields to gravity was already studied in the
context of 2-dimensional matrix models \cite{Amb,Ginsp1,Ginsp2}. 
The results of \cite{OePf}, where a spin foam model was obtained for a general Yang-Mills field on a lattice, provide another route to the construction of a model for quantum gravity coupled to other gauge fields. The possibility of a truly non-perturbative definition of a theory describing gravity (geometry) and other interactions in a unified manner, through a spin foam model with a large gauge group, is clearly intriguing, as is the idea of using the Barrett-Crane model to describe the lattice, i.e. a triangulated spacetime, and its geometry, and to study the spin foam formulation for the Yang-Mills field on this (quantized) lattice, obtaining again a unified description. 

As a possible solution (among many) to the problem of includig matter fields in the models, but also for its own interest and
relevance, a supersymmetric extension of the Barrett-Crane model, and
of its lower-dimensional (Ponzano-Regge-Turaev-Viro) and
higher-dimensional analogues, would be a very nice development. This extension would amount to the replacement of the (quantum)
group used for the assignment of labels to the spin foam with its
supersymmetric counterpart ($Osp(N,4)$ for $Spin(4)$ or $Osp(N,2)$ for
$SU(2)$, for example, in the Euclidean case).
In 3 dimensions a \lq\lq super-Turaev-Viro" model would be reliably related at the continuum level to a supersymmetric BF theory with $Osp(N,2)$ as gauge symmetry, and to 3-dimensional supergravity, but its relevance for the construction of a new topological invariant for 3-manifolds is not so clear. In higher dimensions, the costruction of supersymmetric spin foam models as a step in a quantization of supergravity is an intriguing possibility, but also no more than a speculation at present. 

In order to be considered seriously as consistent quantum gravity models, if not theories, spin foam models have to provide a framework in which quantum gravity effects can be not only described but also computed explicitely and in details. In other words, in addition to formal development of these models, there is a need to make contact with physical reality. On the road towards this goal, the first thing a quantum gravity theory is expected to give us is a microscopic derivation of black hole entropy, which is in agreement with the semiclassical Bekenstein-Hawking formula at zeroth order, but permits us also to compute the quantum corrections to it. Important results were already obtained in the canonical loop quantum gravity approach \cite{AshBaKra}, and a first attempt to perform this derivation in the context of a spin foam model in 3d is in \cite{BHETV}, but a complete and satisfactory procedure in the 4-dimensional case is still missing, and would be of paramount importance. 

Being ambitious, we would like also to ask our quantum gravity theory to make contact with experiment, i.e. to provide a (possibly) rich phenomenology that can be tested in one way or another. This would have been considered probably foolish until some years ago, but recent proposals for the possibility of a quantum gravity phenomenology \cite{AC} make not completely unmotivated, even if still rather optimistic, the hope ofobtaining physically testable predictions (coming maybe from the discreteness or the foamy structure of spacetime) from spin foam models of quantum gravity. 

In parallel with the development of spin foam models themselves, there is a need to understand better their relations with each other, first of all, and then with other approaches to non-perturbative quantum gravity, like loop quantum gravity. In fact we mentioned that different spin foam models for gravity were proposed \cite{Reis1,Reis2,Iwa1,Iwa2}, and in particular the Reisenberger model \cite{Reis2, Reis3} is of much interest, but the precise relation between them and the Barrett-Crane one is not at all clear. Moreover, also the relations between loop quantum gravity and spin foam models, e.g. the Barrett-Crane model, is still to be investigated and understood. The general picture is that the canonical space of a spin foam theory is associated with boundaries of spacetime, and so is spanned by spin networks, since the boundary of a spin foam is a spin network, as in loop quantum gravity, but now these spin networks are $SO(4)$ or $SO(3,1)$ ones with simple representations (at least in the Barrett-Crane model), while in loop quantum gravity they are based on $SU(2)$. Anyway, the details are missing and they can be expected to be of help both in solving the difficulties of the canonical approach in defining the correct form of the Hamiltonian constraint, thanks to the definition of the dynamics of the theory in terms of the spin foam vertex amplitudes, and in providing a better understanding of the physical content of spin foam geometry, through the use of a more familiar 3+1 canonical decomposition.

Finally we can ask what are the connections (if any) between spin foam models and other, at first, completely different approaches to quantum gravity, like string theory and non-commutative geometry.        

In \cite{Smo} (but interesting ideas in this direction are also in
\cite{Crane}) it was argued that the system describing the
perturbations of a spin foam model for quantum gravity will be given
by a kind of perturbative string theory, something that it is of
course expected on general grounds, since the only consistent
perturbative theories of quantum gravity so far constructed are string
theories, but no clear and precise picture of it was ever produced in
the context of any particular quantum gravity model. 
On the other hand, if true, a similar result would give a strong motivation for considering spin foam models as consistent quantum gravity theories and as candidates for a background independent formulation of non-perturbative string (M-)theory (for other suggestions on this, see \cite{Smo2, Smo3}), in additon to the speculated possible spin foam formulation of quantum supergravity. Moreover, encouragement for looking for such connections are the strong similarities between some ideas of the spin foam approach and those proposed to be a conceptual basis for a truely non-perturbative and background independent form of M-theory \cite{BanFis}.     
    
We can speculate that also deep connections between the spin foam approach and the non-commutative geometry programme \cite{Connes, Majid, Majid2} are possibly to be discovered, since the insistence on background independence and algebraic characterization of geometrical quantities, the use of quantum groups, the idea of a discreteness of quantum geometry, are all common to both. In particular the key role here may be played by quantum groups, which are at the very roots of the non-commutative geometry programme \cite{Majid2}, that can be used for labellings of the spin foams, like $SO(4)$ is used in the formulation of the Euclidean Barrett-Crane model. Thus, like in that case a smooth Euclidean geometry is expected to emerge from the spin foam, now the emerging spacetime would result in being endowed naturally with a non-commutative structure.    

\vspace{1cm}

As a conclusion, which is not a conclusion, we can just say that spin foam models in general, and the Barrett-Crane one in particular, represent a promising new approach to the construction of a quantum theory of gravity, providing some answers and posing new questions, making use of interesting ideas and pointing to a rather radical conceptual change in the way we look at the Nature of which we are part. Many aspects of these models are still obscure, and at the same time also their possible improvements and developments cannot be fully predicted, so that any opinion about them can just be provisional, and, while forced to be necessarily cautious, at the same time we have reasons to be optimistic. Much more work is certainly needed, but this is going to be interesting and worth doing. 
  
\section*{Acknowledgements}
The author is really grateful to  John C. Baez, Richard E. Livine, Hendryk Pfeiffer and Ruth M. Williams for discussions, careful reading of the manuscript and many helpful comments.

\end{document}